\documentclass[numberedappendix,11pt,a4paper,iop]{emulateapj}

\usepackage{amsmath,amssymb}
\usepackage{graphicx}
\usepackage{xcolor}
\usepackage[english]{babel}

\usepackage{floatrow}
\usepackage[caption=false]{subfig}

\newcommand{\sigmab}{\sigma_{\mbox{\tiny $B_\phi$}}}
\newcommand{\sigmaa}{\sigma_{\mbox{\tiny $A_\phi$}}}

\newcommand{\myvect}[1]{\mathbf#1}
\newcommand{\myhat}[1]{\mathbf#1}

\newcommand{\firstrev}[1]{{#1}}
\newcommand{\firstrevii}[1]{{#1}}
\newcommand{\secrev}[1]{{#1}}
\newcommand{\secrevii}[1]{{#1}}

\shorttitle{Meridional Circulation and the 11-yr Solar Cycle}
\begin{document}

\title{Estimating the deep solar meridional circulation using magnetic observations and a dynamo model: a variational approach}

\author{Ching Pui Hung$^{1,2}$, Laur\`ene Jouve$^{1,3}$, Allan Sacha Brun$^{1}$, Alexandre Fournier$^{2}$, Olivier Talagrand$^{4}$}

\affil{$^1$ Laboratoire AIM Paris-Saclay, CEA/IRFU Universit\'{e} Paris-Diderot CNRS/INSU, 91191 Gif-Sur-Yvette, France\\
\and $^2$ Institut de Physique du Globe de Paris, Sorbonne Paris Cit\'{e}, Universit\'{e} Paris Diderot UMR 7154 CNRS, F-75005 Paris, France\\
\and $^3$ Universit\'{e} de Toulouse, UPS-OMP, Institut de Recherche en Astrophysique et Plan\'{e}tologie, 31028 Toulouse Cedex 4, France\\
\and $^4$ Laboratoire de m\'{e}t\'{e}orologie dynamique, UMR 8539, Ecole Normale Sup\'{e}rieure, Paris Cedex 05, France}

\begin{abstract}
We show how magnetic observations of the Sun can be used in conjunction with an
 axisymmetric flux-transport solar dynamo model in order           
to estimate the large-scale meridional circulation throughout the
 convection zone. Our innovative approach rests on variational data assimilation,
 whereby the distance between predictions and observations (measured
 by an objective function) is
 iteratively \secrev{minimized} by means of an optimization algorithm
 seeking the  meridional flow which best accounts for the data.
 The minimization is
 performed using a quasi-Newton technique, which requires
 the knowledge of the sensitivity of the objective function
 to the meridional flow. That sensitivity is
 efficiently computed via the integration of the adjoint
 flux-transport dynamo model.
 Closed-loop (also known as twin) experiments using synthetic 
data demonstrate the validity
 and accuracy of this technique, for a variety of
  meridional flow configurations, ranging from unicellular and equatorially
 symmetric  to multicellular and equatorially asymmetric. 
  In this well-controlled synthetic context,
 we perform a systematic study of the behavior of our variational approach
 under different observational configurations, by varying their
 spatial density, temporal density, noise level, as well as the
 width of the assimilation window. We find that the method
 is remarkably robust, leading in most cases to a recovery of the true
  meridional flow to within better than $1$\%. These encouraging
 results are a first step towards using this technique to i) better
 constrain the physical processes occurring inside the Sun and
 ii) better predict  solar activity on decadal time scales.

\end{abstract}

\keywords{dynamo – magnetohydrodynamics (MHD) – methods: data analysis – Sun: activity – Sun: interior – Sun: magnetic fields}

\section{Introduction} \label{sec_intro}

    The magnetic activity of our Sun has been monitored for centuries now.
    \firstrev{Modern systematic} observations of sunspots started in the early 1600's. In addition,
    the analysis of the concentration of cosmogenic $^{10}$Be and $^{14}$C (found in
    ice cores and tree rings, respectively) makes it possible to trace
    back solar activity over the past 10,000 years [\citep{Beer98} and \citep{Usoskin13}].
    This monitoring has revealed a cyclic magnetic activity in our Sun: Sunspots emerge at mid-latitudes during
    the rising phase of the cycle; they reach a maximum number 3 to 5 years later when the polar field flips polarity,
    while emerging gradually closer to the equator as the cycle progresses. If plotted against time, the location in latitude
    of sunspots then produces the so-called butterfly diagram. A quantitative estimate of the cycle strength
    was proposed by Wolf in 1859. He introduced the so-called Wolf number as $R=k (10 g +s)$, with $g$ the number of sunspot groups,
    $s$ the total number of individual sunspots in all groups and $k$ a variable scaling factor that accounts for instruments
    or observation conditions. The time series of the Wolf number starts in 1749, making the solar cycle which peaked in June 1761
    Cycle number 1. At the time of writing, we have just passed the maximum of Cycle 24. The monthly smoothed Wolf number has indeed reached
    a moderate maximum peak of about 82 in April 2014, which will probably become the maximum of Cycle 24.
    This makes Cycle 24 the weakest cycle since Cycle 14, which peaked in 1906.
   This relative low value is likely connected  with the expected end of the current Gleissberg cycle \citep{Abreu08}.
   Predicting solar activity has become key for our technological society in which strong solar flares, coronal
  mass ejections (CMEs) or any violent event linked with solar activity can cause significant damage to satellites,
  air traffic or telecommunication networks \citep{2007AN....328..329B}. Consequently, a solar cycle panel, whose role is to produce predictions of forecoming
   solar activity, was created in 1997. This panel has provided us with estimates of the
   sunspot number for Cycles 23 and 24 \citep{joselyn1997panel,biesecker2007solar}. In the ensemble of~75 predictions
 of Cycle 24 maximum sunspot number listed by \cite{pesnell2012solar}, only~20 had anticipated such a moderate value
 of 82, taking
  into account the uncertainties provided for each prediction.
  This poor performance reveals the (expected) difficulties to produce reliable forecasts of magnetic activity for such
  a turbulent chaotic astrophysical system, especially if
   the prediction ignores the dynamics of the system and is entirely data-driven, as was the case for most
   predictions of Cycle 24, which relied on geomagnetic precursors or other statistical estimates \citep[see][for two reviews on the subject]{Hathaway09,pesnell2012solar}. 

    Recently, however, the use of numerical models in conjunction with observations,
   which goes by the generic name of {\sl data assimilation},
   has started to emerge in the solar physics community. What is data assimilation?
   Let us assume that some observations of the Sun are available over a finite time interval $[t_s,t_e]$ and that
   a numerical model governing the temporal evolution of the Sun over this interval is available. In a deterministic setting, the
dynamic trajectory of the Sun is then entirely controlled by a set of initial conditions (at $t=t_s$) and, possibly, by a set of static control parameters.

    In a first attempt of combining data and model,
  one can simply use the observations made at the Sun surface as boundary conditions
    to impose on the numerical model between $t_s$ and $t_e$.
    Such a strategy was followed by \cite{Dikpati06, Choudhuri07, Upton14} for the predictions of the amplitude and timing
    of the solar cycle and by \citet{Cheung12} for eruptive events originating from active regions. This strategy
    is, however, not optimal, in particular because it does not \firstrev{combine·
the uncertainties affecting the observations and the physical model}.

    More advanced techniques exist to remediate this problem, which can be classified into two categories: variational and sequential.
   Both share the same goal, which is to provide 
\secrev{a least-variance statistically optimal fit}
   fit to the \secrev{unknown reality} (in a generalized least-squares sense) over the time window $[t_s,t_e]$
   over which observations are available (which we will refer to 
   indifferently as the observation or sampling window in the following). 
\secrev{Variational assimilation provides a globally optimal fit over the whole time window. Sequential assimilation provides an optimal fit at the end of the window.}
The {\sl variational} approach is rooted in
   the mathematical theory of control and
   aims at correcting the initial conditions (and possibly the set of static control parameters) by making use of all
   the data available over the entire $[t_s, t_e]$ interval. In contrast, the sequential approach rests on
   estimation (or filtering) theory. In that case, the stream of observations is assimilated sequentially,
   each time a new observation becomes available at, say,
   $t=t_o \in [t_s, t_e]$. 
The sequential and variational approaches \secrev{lead to} the same results \secrev{at the end of the assimilation window} if the dynamics of the system is
linear.
More generally, and regardless of their respective merits, both
approaches illustrate the same philosophy of combining data with numerical models.
Both can lead to the production of a forecast for $t>t_e$ and
 are indeed used on a daily basis for the best-known problem of weather forecast, which requires
several tens of millions of data to be assimilated every day into physical models
of the atmosphere (and ocean), to first initialize a
state (or ensemble of states) of the atmosphere (and ocean) and subsequently generate weather forecasts \citep[see e.g.][]{kalnay2003amd}. 

The problem at hand may involve some nonlinearities, for instance, in the dynamics or in the
relationships linking the state of the system to the available observations. If so, both sequential and variational approaches need be adapted. In practice, this amounts to performing a linearization at some stage in the analysis.
In the case of the sequential approach, this leads to a class of methods known as
the extended Kalman filter (EKF) and the ensemble Kalman filter, the latter commonly known as the EnKF \citep{evensen2009dae}.
 In the variational framework, the most popular approach is the so-called 4D-Var approach, whose
 efficiency rests on the implementation of the so-called {\sl adjoint} model \citep[see ][for a recent review]{talagrand2010variational, Fournier10}.

Regardless of the assimilation approach followed, the first question
 one may ask when seeking to apply data assimilation to the prediction of the solar cycle is: What type of numerical model
of the origin of solar magnetism should be used? Indeed, in spite of some recent encouraging progress made
by 3D magnetohydrodynamic (MHD) simulations to self-consistently produce "solar-like"
magnetic features \citep{Ghizaru10,Kapyla12,Warnecke14,Augustson14},  some of the ingredients needed to account for
all the properties of the solar cycle remain to be understood. In particular, full 3D MHD simulations are not able to self-consistently produce, through an internal dynamo mechanism, sunspots emerging at the solar surface [refer to \citep{Nelson13, FanFang14} for a first step in that direction].
The first attempts to apply data assimilation to solar physics have instead used simplified mean-field dynamo models,
where strong simplifying assumptions are used and various physical processes are parametrized. These models are nevertheless solar-like in the sense
 that  they produce reversals of the large-scale magnetic field and butterfly diagrams resembling the observations.
 Sequential data assimilation has been implemented for the first time by \citet{Kitiashvili08} in such a mean-field dynamo
model evolving jointly (in one spatial dimension) the three components of the magnetic field
 and a measure of magnetic helicity. The assimilated observations were the annually smoothed Wolf sunspot
 number for the period 1857-2007, in a sequential EnKF framework. The 1D model used in that study was a
standard $\alpha-\Omega$ dynamo model in which the toroidal field owes its origin to the differential
rotation (the $\Omega$-effect) and the poloidal field is created by helical turbulence within the solar convection
zone (the $\alpha$-effect). The prediction for the maximum sunspot number of Cycle 24 was $80$ in~2013.
This is remarkably close to the observed value of 82 for the amplitude of the cycle, and too early by 1 year for
the timing of cycle maximum. Further \cite{CameronSchuessler07} advocate that such predictions
done within 3 years of the minimum are easier, as simple correlations reach up to $80\%$ or so of success when
applied to past cycles.  More recently, data assimilation was performed with
a more complete 2D mean-field $\alpha\Omega$ Cartesian model using the 4D-Var variational approach \citep{JouveAssimi11};
it was shown that the latitude-dependent profile of the $\alpha$-effect could be reconstructed from the assimilation of
 synthetic magnetic data and that the method was versatile  and robust, which encouraged us to improve our model and method.

Such an improvement in the modeling can take the form of a flux-transport Babcock-Leighton model in which the poloidal field
is generated by the decay of active regions emerging at the solar
surface \citep{Babcock61,Leighton69}, and where a large-scale meridional flow
$\myvect{v}_p$
acts to advect the magnetic field in the whole convection zone.
 The main strength of such models is that they incorporate physical processes which have observable counterparts,
namely the active region evolution at the solar surface and the meridional circulation amplitude and pattern.
For the latter however, the observational constraints are limited. Most of our knowledge of $\myvect{v}_p$ is provided by
local helioseismology techniques which produce reliable measurements down to about 20~{\rm Mm} to 40~{\rm Mm} below the surface \citep{Haber02, Zhao04}.
We now know that the surface meridional flow is poleward and that the horizontal
velocity amplitudes are between 10 and 20 ${\rm ms^{-1}}$. Inferences from the advection
of super granules down to a depth of 70 {\rm Mm}  \citep{Hathaway12},  from $p$-mode
frequencies \citep{Mitra07} and improved time-distance analysis \citep{Zhao13,2015ApJ...805..133J} suggest a complex structure in the
convection zone, organized in multiple cells, as also confirmed by recent global helioseismic methods \citep{Schad13}.

Given the strong dependence of the magnetic cycle on the meridional flow amplitude
and profile in flux-transport Babcock-Leighton dynamo models \citep{JouveMC07}, the idea we pursue is to use data assimilation
to better constrain this flow. More specifically, our goal is to use time-dependent observations
of the magnetic field to find the optimal meridional flow $\myvect{v}_p$ (in structure and amplitude) that
 minimizes the misfit between the observations and the values predicted by the model.
 To this end, preliminary studies have been performed to first characterize the sensitivity of the
 magnetic field evolution to changes in $\myvect{v}_p$ \citep{Nandy11,Dikpati12} and to assess
 the so-called forecast horizon, i.e. the time interval (for $t>t_e$) over which reliable predictions
 can be achieved \citep{SanchezBLLongterm14}. These studies concluded that a modification of the amplitude of
 $\myvect{v}_p$ would show large changes in the evolution of the magnetic field in a time much shorter than
  the typical circulation time. \cite{SanchezBLLongterm14} provided an estimate of the exponential growth rate
 of an initial perturbation on the model trajectory (the error growth rate),
 and found a corresponding $e$-folding time of 2.76 solar cycles,  which is likely an over-estimate of the practical
 horizon of predictability for the Sun.
  These results are promising for the possible use of magnetic field data to infer the meridional
 flow profile and quite encouraging for possible future predictions of solar activity. Now that some key properties of
 this dynamical system (associated with this particular Babcock-Leighton dynamo model) are known, we can further consider to
 apply data assimilation techniques to it. In a recent paper, \cite{Dikpati14} resorted to the EnKF in order
 to reconstruct the meridional flow speed at the solar surface (having fixed the meridional flow profile to one large circulation
 cell per hemisphere) from synthetic observations of the magnetic field.  \cite{Dikpati14} found that the best reconstruction of
 this unique, time-dependent, parameter could be obtained
 provided at least 10 observations were used with a time between two analyses
  of 15 days and an observational error of less than $10$\%. We propose here to work along the same philosophical line,
 aiming this time at estimating not only the amplitude of $\myvect{v}_p$ at a particular location, but also its structure
  throughout the whole convection zone. To do so, we use a variational data assimilation (4D-Var) technique akin
 to that developed by \cite{JouveAssimi11}. We intend to demonstrate in the following the successful
 application  of 4D-Var to fully characterize the otherwise poorly constrained solar internal meridional flow $\myvect{v}_p$,
 by resorting to proof-of-concept experiments based on a particular flux-transport dynamo model.
  Proof-of-concept experiments rely on data generated by a free run of the model (a run unconstrained by real data); those
  synthetic data are subsequently used to verify and test the efficacy of any given assimilation scheme.  \par
     After presenting the details of the model in Sec.~\ref{sec:modeleq}, we introduce our implementation of 4D-Var in Sec.~\ref{sec:var}.
  The results of proof-of-concept  assimilation experiments with perfect data (i.e. without noise)
  are presented in Sec.~\ref{sec:twinnonoise}, while results for noised data are given in Sec.~\ref{sec:twinnoise}.
  A discussion of the consequences of this study is next given as a conclusion in Sec.~\ref{sec:conclu}.

\section{One major ingredient: the meridional circulation}
\label{sec:modeleq}

In this work, we choose to test our data assimilation technique with a spherical, axisymmetric mean-field dynamo model. 
We adopt the widely used flux-transport Babcock-Leighton model for which the meridional flow $\myvect{v}_p$ profile and amplitude is a 
 major ingredient \citep{Dikpati99,JouveMC07}. The model equations are presented in Appendix \ref{sec:blmodel} for reference. 
In this section, the meridional circulation, which is to be optimized by the assimilation, is described in detail. 

The meridional circulation is  henceforth expressed as the curl of a stream function, to ensure a divergence-free velocity field:
\begin{equation}\label{eq:curlvp}
 \myvect{v}_p= \nabla \times (\psi \myhat{e_{\phi}}),
\end{equation}
and the stream function is expanded as
\begin{equation}\label{eq:MC}
\begin{split}
& \psi (r,\theta)= -\frac{2(r-r_{mc})^2}{\pi(1-r_{mc})} \\
& \times
  \begin{cases}
     \sum\limits_{i=1}^{m} \sum\limits_{j=1}^{n} d_{i,j} \sin \left[ \frac{i\pi (r-r_{mc})}{1-r_{mc}} \right] 
 P_j^1 (-\cos \theta) & \text{if }r_{mc} \leq r \leq 1 \\
     0 & \text{if } r_{bot} \leq r<r_{mc},
  \end{cases}
\end{split}
\end{equation}
where the length is normalized with the solar radius $R_s$, the normalized radius $r$ varies between $r_{bot}=0.6$ and 
$r_{top}=1$ and the polar angle $\theta\in[0,\pi]$. The meridional flow is allowed to penetrate to a radius $r_{mc}=0.65$, i.e. slightly below the base of the convection zone located at $\secrevii{r_c}=0.7$. There is no flow between $r_{bot}$ and $r_{mc}$, which models a stationary layer of transition from the convection zone to the radiative zone. The magnetic diffusion time $R_s^2/\eta_t$ is chosen as the characteristic time scale, where $\eta_t$ is a typical value of the turbulent magnetic diffusivity (of order $10^{12}\mathrm{cm^2s^{-1}}$) in the convective envelope.

\begin{table*}[htbp]
 \caption{Coefficients of the 3 models characterizing the meridional flow based on expression \eqref{eq:MC} (see text 
 for details). Coefficients that do not 
  appear in this table are set to zero. }
 \label{tab:modeldij}
 \begin{tabular}{c*{5}{c}}
 \hline \hline
 Case     &$d_{1,2}$            & $d_{2,1}$             & $d_{2,2}$               & $d_{2,3}$             &  $d_{2,4}$   \\
 \hline
 1        & $3.33\times10^{-1}$ & $0.00$                & $0.00$                  & $0.00$                &  $0.00$    \\
 2        & $0.00$              & $0.00$                & $0.00$                  & $0.00$                &  $9.47 \times 10^{-2}$   \\
 3        & $0.00$              & $-5.74\times 10^{-2}$ & $-8.75 \times 10^{-2}$  & $-3.83 \times 10^{-2}$& $5.47 \times 10^{-2}$   \\
 \hline
 \end{tabular}%
 \end{table*}

\begin{figure}
\includegraphics[width=0.9\columnwidth]{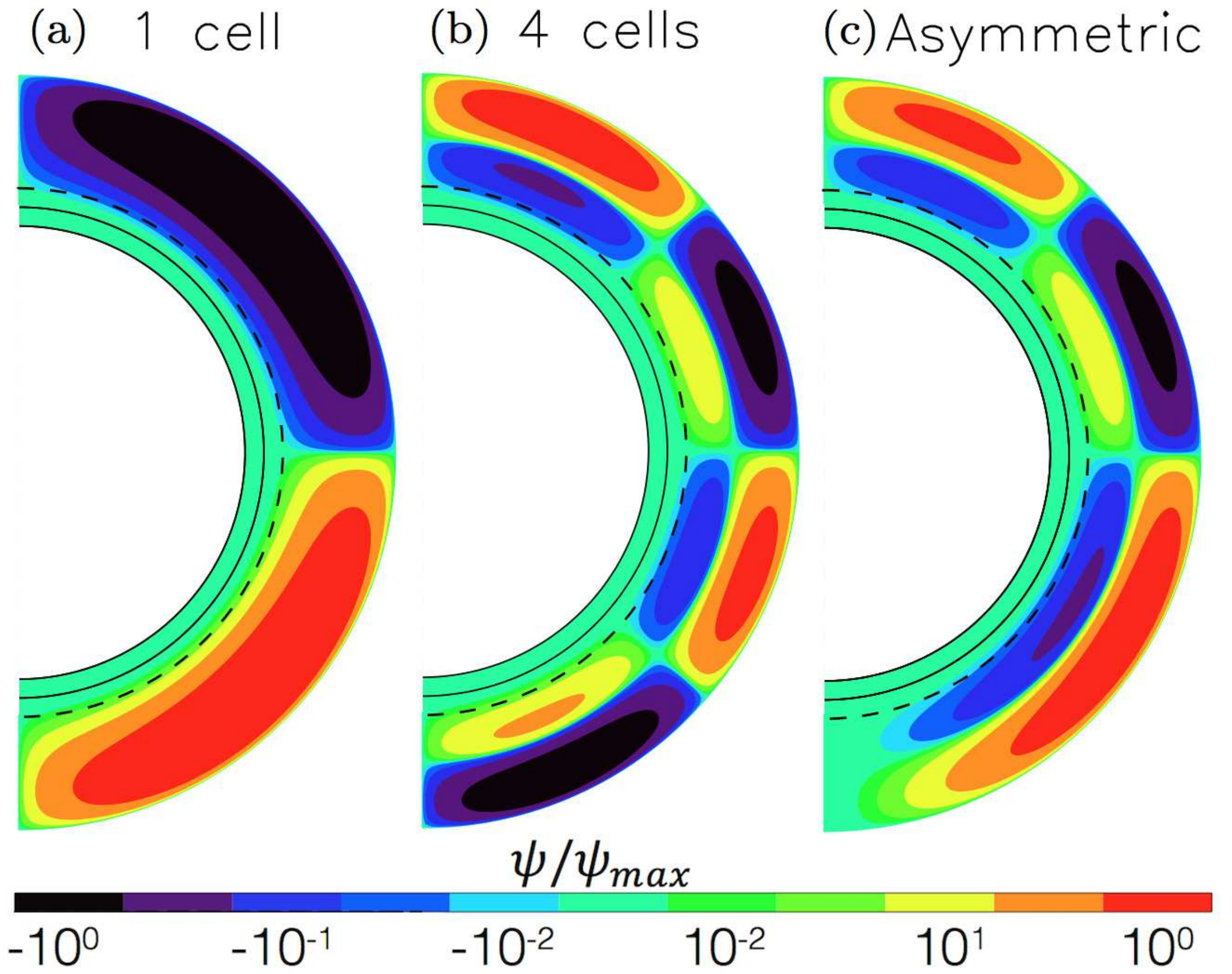}
\caption{Stream functions for the 3 cases studied [(a), (b) and (c) for cases 1, 2 and 3 respectively]. Case 1 is the unicellular model, case 2 is the 4-cell model, and case 3 is the asymmetric model. We indicate the base of the convection zone at $r=0.7R_s$ with a broken line in each plot.}
\label{fig:MCpsi}
\end{figure}

\floatsetup[figure]{style=plain,subcapbesideposition=top}
\begin{figure}[!h]
\sidesubfloat[]{\label{fig:btfy1}\includegraphics[width=0.95\columnwidth]{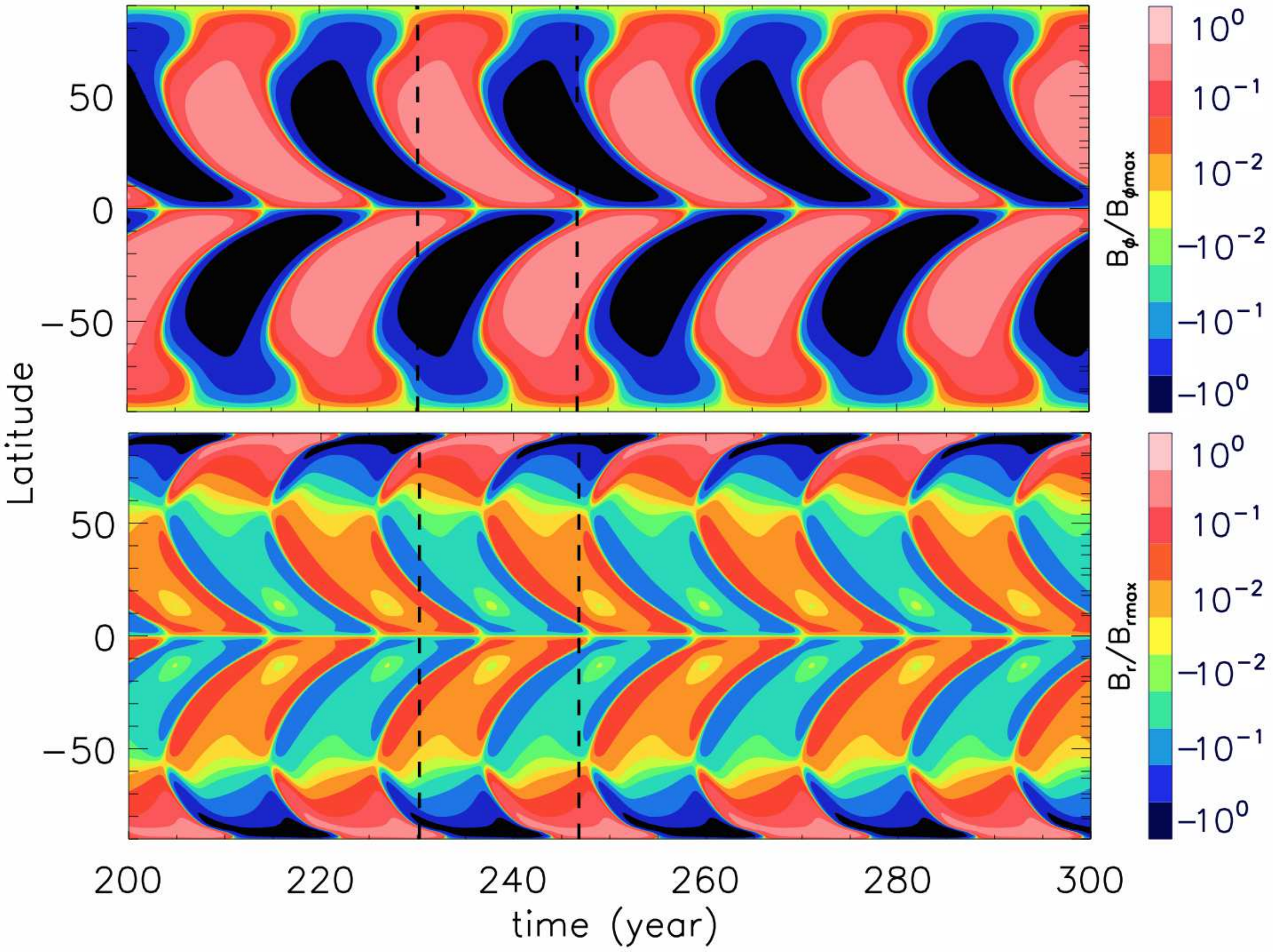}}\quad
\sidesubfloat[]{\label{fig:btfy2}\includegraphics[width=0.95\columnwidth]{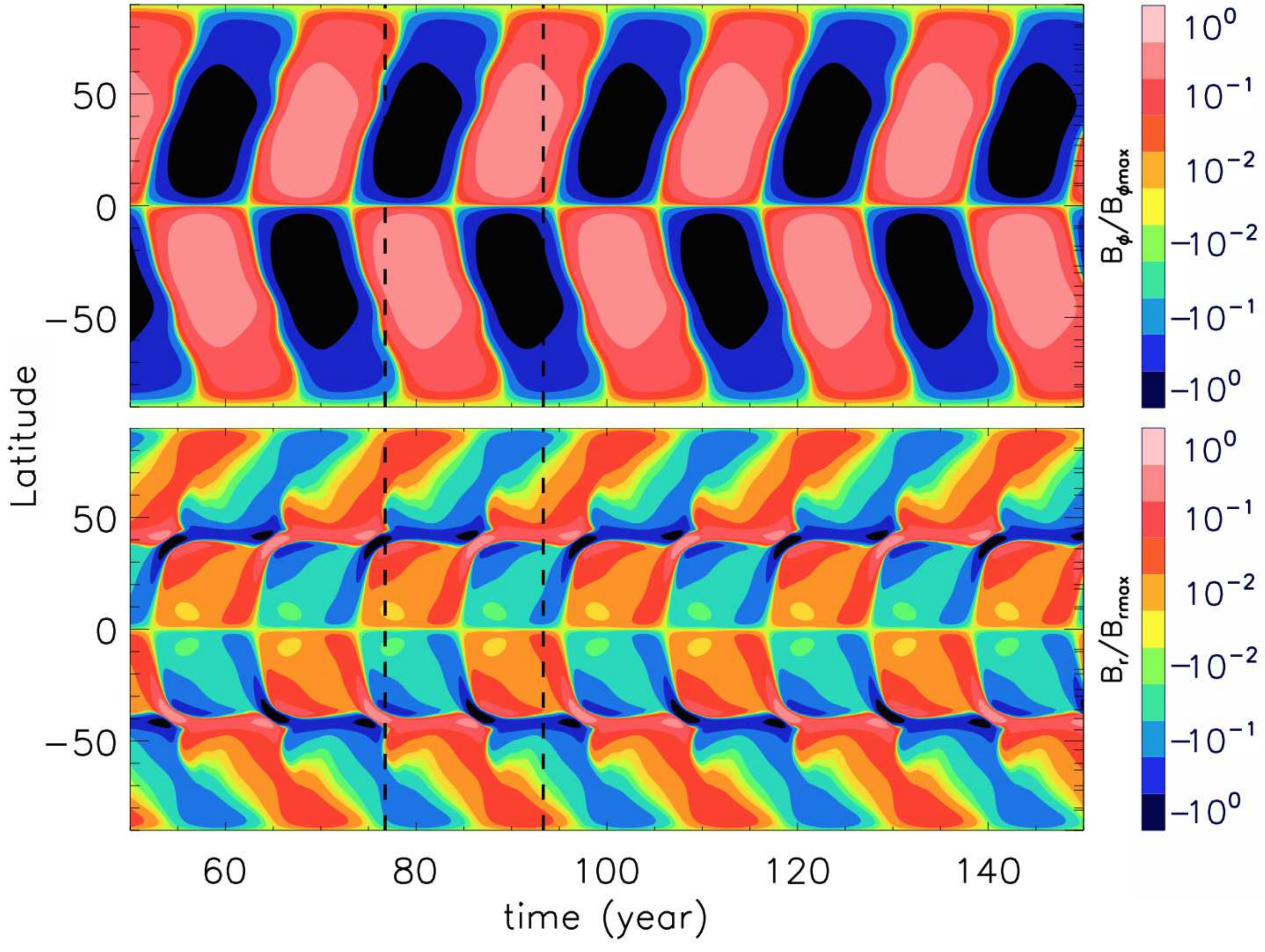}}\quad
\sidesubfloat[]{\label{fig:btfy3}\includegraphics[width=0.95\columnwidth]{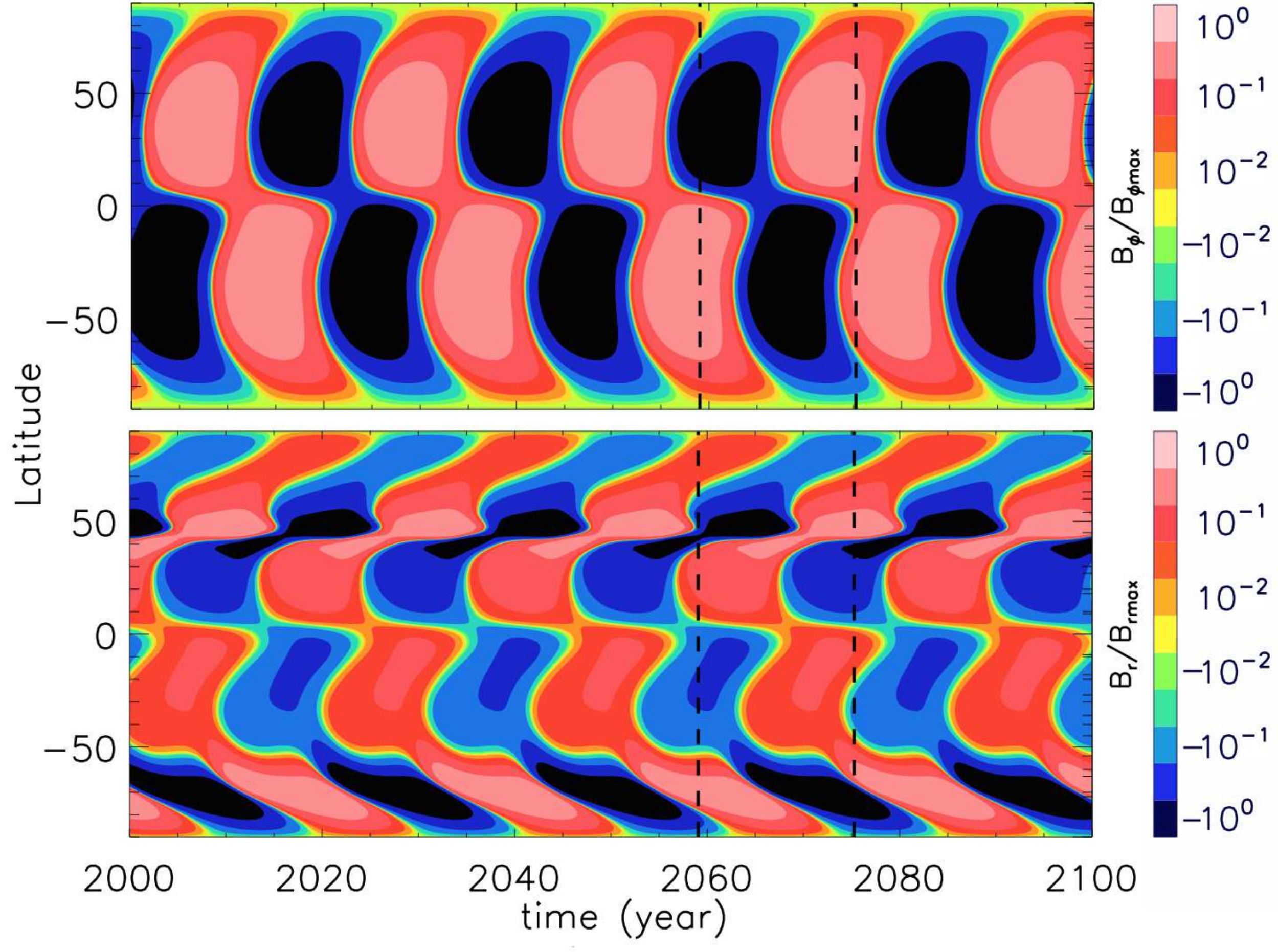}}\quad
\caption{Butterfly diagrams for the 3 cases studied [(a), (b) and (c) for case 1, 2 and 3 respectively], with toroidal field at the base of the convective zone (top panel) and surface radial field (bottom panel), and are normalized with their maximum respectively. Case 1 is from the simplest unicellular model, case 2 is from the 4-cell model, where the surface radial field shows the imprint of the 2 counter-cells. Also note the asymmetric nature of the butterfly diagram of case 3, while still maintaining a mostly anti-symmetric (dipole-like) solution. We mark a typical sampling window of width $1.5$ solar cycles in each case with dashed lines.}
\label{fig:btfy}
\end{figure}

Three different profiles of the meridional flow are constructed using 3 
different sets of expansion coefficients $\{d_{i,j}\}_{i\in{1,m},j\in{1,n}}$ in Eq.~\ref{eq:MC}. 
The size of the parameter space used in the following sections is $m=2$, $n=4$, which 
is an 8D parameter space. Those coefficients are given in~Table~\ref{tab:modeldij}. The non-listed coefficients are set to zero. We note that one can rewrite the expansion of the flow field [e.g. eq. \eqref{eq:MC}] using separable radial and latitudinal parts. The control vector would then be the expansion coefficients of the radial part and latitudinal part, resulting in a parameter space of lower dimension ($m+n$ vs $mn$). We however find that this results in a larger misfit in the assimilation procedure. \par
Case 1 is a unicellular stream function (in each hemisphere), case 2 is a 4-cell stream function with 2 cells in radius and 2 in latitude (in each hemisphere), case 3 is a more general model with 4 cells in the Northern hemisphere and only 2 cells in radius in the Southern hemisphere. Figure~\ref{fig:MCpsi} shows the contour plots of the stream functions of the meridional circulation of Cases 1, 2 and 3. The other physical parameters of the 3 cases are given in Appendix~\ref{sec:blmodel}. From surface observations of the horizontal flows done by \cite{2010ApJ...725..658U} and after projecting the data on associated Legendre polynomials $P_{\ell}^1(\theta)$, we note that the dominant mode is $\ell=1$, and that $\ell$ modes beyond $\ell=4$ are at least 3 orders of magnitudes smaller. We thus consider that stopping our latitudinal expansion of the streamfunction to $n=4$ is providing a fair reconstruction of the flow field. 
  Figure~\ref{fig:btfy} shows the resulting butterfly diagrams for the toroidal field at the base of the convection zone and the surface radial field. Those aspects are further documented in Appendix~\ref{sec:expstream}.

\section{Setting up the assimilation procedure}
\label{sec:var}

The physical model has now been described. The control vector, which is adjusted to fit the observational data, has been chosen to be the expansion of the meridional flow profile on particular radial and latitudinal functions (Eq \ref{eq:MC}). 
 The idea of this work is to apply a variational data assimilation technique [or 4D-VAR, see \citet{Talagrand87} for details] 
to this model, in which one seeks to minimize the misfit between the observations and the outputs of the model (characterized by an objective function $\mathcal{J}$) within a certain time interval. As a first step, we wish to proceed with twin (closed-loop)
 experiments where the magnetic data are produced by a free run of the model, as described for instance in \citet{JouveAssimi11}. The assimilation will be considered successful when the true state (the value of the control vector which was used to produce the observations) is recovered (to a certain accuracy) as a result of the minimization of the objective function. In this section, the setup of these twin experiments is described: the generation of magnetic data, the choice of the objective function and the minimization algorithm, the choice of initial guess and finally the diagnostics to assess the quality of the assimilation technique.

\subsection{Generating synthetic observational data}
\label{subsec:genobsdata}

In our twin experiments, the synthetic observations are generated by the direct Babcock-Leighton dynamo model governed by eq.\eqref{eq:Adyn} and \eqref{eq:Bdyn}, with the expansion coefficients of the meridional circulation eq.\eqref{eq:MC} given by Table~\ref{tab:modeldij}. The other parameters as well as the grid size are given in 
Table~\ref{tab:modelpara} of Appendix~\ref{sec:blmodel}. In each case the synthetic observations are the toroidal field at the tachocline and the vector potential of the surface poloidal field. These are taken from the reference trajectories of the magnetic field of the 3 cases described in section~\ref{sec:modeleq}, which are magnetic fields as a function of space and time, with dipole field as initial conditions, recorded when the periodic regime
 has been reached. The cycle period is $\sim 22$ years for all three cases (see Table~\ref{tab:modelpara}). Our choice of synthetic data is motivated by the future application to real solar observations: the toroidal field at the tachocline is indeed thought to be a proxy of the sunspot distribution and the vector potential of the surface poloidal field should be a good proxy for the observed surface radial magnetic field, since the radial field is the spatial derivative of the vector potential. Our synthetic magnetic data are thus chosen to mimic what real solar observations provide (sunspot number, polar cap field, butterfly diagrams, etc...).\par
\begin{figure}[!ht]
\includegraphics[width=\columnwidth]{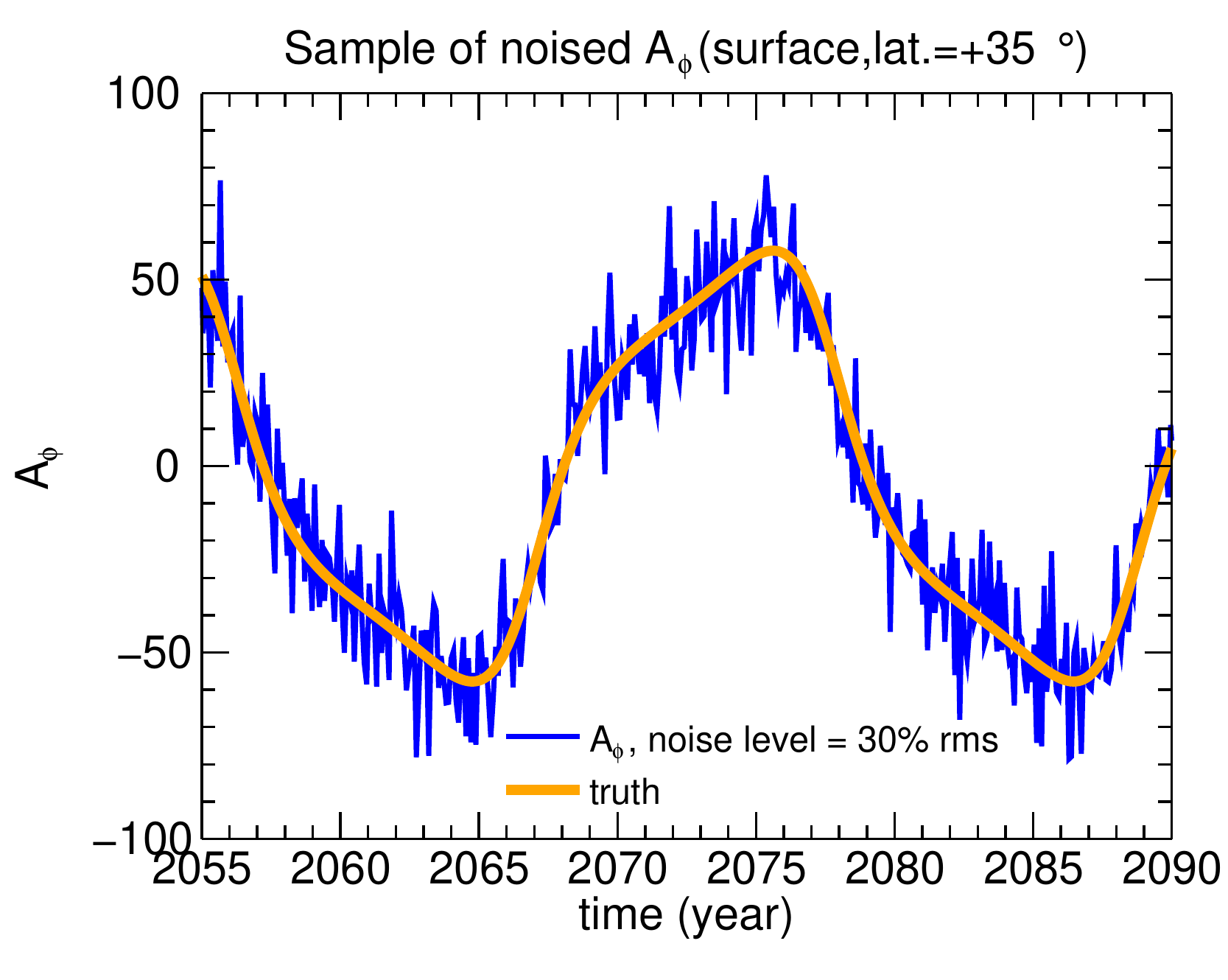}
\caption{Typical synthetic data generated in case 3, here we show the surface vector potential of the poloidal field 
at a fixed latitudinal location as a function of time.}
\label{fig:sample}
\end{figure}

 
For the purpose of our twin experiments, we first introduce our synthetic observations into our assimilation code, and analyze the converging behavior and performance of the code at various temporal observation windows as well as various spatial sampling in latitudes. Secondly, we add 
random noise to the synthetic observations and study the effects on the performance and accuracy 
of the optimization. The noise added is centered, normally distributed, with standard deviation 
being a fraction of the root mean square (r.m.s.) of the magnetic field components produced by 
the dynamo model. Examples of synthetic observations for case 3 (error-free and noised with a noise level of 30\% of r.m.s.) 
are shown in Figure \ref{fig:sample}, and the counterparts of cases 1 and 2 are similar thus not shown here.

\subsection{Choice of the objective function and minimization algorithm}
\label{subsec:choiceofobj}
As stated above, the idea of the variational method is to minimize a well defined objective function which measures the misfit between the 
observed quantities and corresponding outputs from the model.
 The objective functions being studied are the misfit of the toroidal field at the tachocline ($r_c = 0.7 R_s$),
\begin{equation}\label{eq:jb}
\mathcal{J}_B=\sum_{i=1}^{N_t^o} \sum_{j=1}^{N_\theta^o} \frac{\left[B_{\phi}(r_c,\theta_j,t_i)-B_{\phi}^o(r_c,\theta_j,t_i)\right]^2}%
{\sigmab^2(r_c,\theta_j)},
\end{equation}
and similarly the misfit of the poloidal field vector potential at the solar surface, 

\begin{equation}\label{eq:ja}
\mathcal{J}_A=\sum_{i=1}^{N_t^o} \sum_{j=1}^{N_\theta^o} \frac{\left[A_{\phi}(R_s,\theta_j,t_i)-A_{\phi}^o(R_s,\theta_j,t_i)\right]^2}%
{\sigmaa^2(R_s,\theta_j)},
\end{equation}
and their sum, $\mathcal{J}_A+\mathcal{J}_B$,

Here $B_{\phi}$ is the toroidal field predicted by our mean field dynamo model, and $B_{\phi}^o$ is the synthetic observation. 
The weight $\sigmab$ is the r.m.s. of the error-free 
toroidal field from the observations, and similarly for $A_{\phi}$, $A_{\phi}^o$ and $\sigmaa$. In real observations, $\sigma$ can also be time as well as space dependent and can be adjusted to give more weight to some observations which are more reliable than others. For example, introduction of more advanced instruments would produce higher accuracy in more recent observations. The expression is summed over all the spatial observations in latitude ($N_{\theta}^o$) and temporal observations ($N_t^o$), so that the total number of observations is $N^o=N_t^o N_\theta^o$.  \par

The objective function is minimized with a quasi-Newton method, which requires the evaluation of the objective function ${\mathcal J}$ and its 
gradient $\nabla {\mathcal J}$ with respect to the control vector, but does not require the exact computation of the Hessian,
 which is instead approximated iteratively by the Broyden-Fletcher-Goldfarb-Shanno formula. In our calculations, 
we use the minimization routine m1qn3 developed by J. Gilbert and C. Lemar\'{e}chal \citep{Gilbert_themodule} based on this algorithm. 
 ${\mathcal J}$ and $\nabla {\mathcal J}$ are computed following the integration of the forward and adjoint models, 
 respectively, details of which can be found in Appendix~\ref{sec:blmodel} and Appendix~\ref{sec:app_adj}, respectively. At the optimum, 
 $\nabla {\mathcal J}$ should be zero. Therefore, in the assimilation procedure, the stopping criterion is defined by $|\nabla \mathcal{J} |/|\nabla \mathcal{J}_0|$, which is the ratio between the magnitude of the gradient of the objective function (with respect to the control vector) after each iteration and the magnitude of the gradient at the initial guess ($\nabla \mathcal{J}_0$). We call this ratio the convergence criterion. In our twin experiments, the assimilation will terminate when the criterion $|\nabla \mathcal{J} |/|\nabla \mathcal{J}_0|$ drops below $10^{-6}$.  \par

Unless otherwise stated, we present the results and analysis adopting the objective function as the 
sum of the misfit on toroidal field at the tachocline and on the surface poloidal potential,  
${\mathcal J} = \mathcal{J}_B+\mathcal{J}_A$. Most of the tests conducted with this choice give reasonable estimate of the meridional flow so that we can discuss any possible trends based on these results. The effect of choosing only $\mathcal{J}_A$ or $\mathcal{J}_B$ as our objective function has also been investigated and is discussed in Section \ref{subsec:conv_char}. In general, the performance is less satisfactory in terms of both the misfit and the recovered flow pattern if only $A_{\phi}^o$ or $B_{\phi}^o$ are chosen for assimilation. Note that we are here using the toroidal field at the tachocline which in a realistic situation is not directly available unless one develops an operator that relates surface field observations to the field located at the base of the convective envelope. Such a relationship will be the focus of future work; suffice it to say here that one could for instance resort to 
 the three-halves law proposed by \cite{1988MNRAS.230..535B}. 

\subsection{Choice of initial guess for $\myvect{v}_p$ and initial condition for $\myvect{B}$}
\label{sec:initialguess}

The assimilation code requires an initial guess for both the meridional flow $\myvect{v}_p$ 
and the magnetic field $\myvect{B}$ as a starting point of the minimization. 
 Our initial guess for $\myvect{v}_p$ is always that of a unicellular flow, as we anticipate
  that this is the guess we will most likely make in an operational setting (when dealing
  with real data).  
  For case 1 (the unicellular
 case, recall Figure~\ref{fig:MCpsi}a), our initial guess for $\myvect{v}_p$
is therefore a unicellular flow, but one  which yields
 a magnetic cycle of 44 years  (we 
 thereby avoid considering an initial $\myvect{v}_p$ too close to the true $\myvect{v}_p$). 
 For cases 2 and 3,  we pick  a unicellular 
 $\myvect{v}_p$ producing a 22-yr cycle. 
 Let us stress that the performance of the assimilation method is in all 3 cases stable with respect to the period of 
 the cycle determined by the choice of initial unicellular $\myvect{v}_p$, up to a certain margin. 
 This margin shrinks as the complexity of the true 
 $\myvect{v}_p$ increases. This statement is further illustrated by some examples in Appendix~\ref{sec:initialperi}.

How do we set the initial condition (at $t=t_0$, say) for $\myvect{B}$? This is crucial since this choice
 determines the phase difference between the modeled field
and the observed one, once the modeled dynamo has entered its periodic regime. 
 Too large a phase lag can be detrimental to the optimization, to the point where it can
 simply fail, in particular if the true $\myvect{v}_p$ is substantially different 
 from the initial guess we just described (which is what happens for cases 2 and 3). 
 A possibility is to add $\myvect{B}_0\equiv\myvect{B}(t=t_0)$ to the control vector, and perform 
 an optimization of ${\mathcal J}$ by adjusting both $\myvect{v}_p$ and $\myvect{B}_0$. 
 That amounts to modifying the adjoint model (described in its current form 
 in Appendix~\ref{sec:app_adj}) in order to take the extra sensitivity to $\myvect{B}_0$
 into account. The strategy that we choose for this study is slightly different, and
   takes advantage of the periodic nature of the system we are interested in. 
  We take different trials for  $\myvect{B}_0$ by sampling regularly 
  a magnetic cycle produced by the integration 
  of the dynamo model based on the initial guess of $\myvect{v}_p$. 
  In this framework, the best $\myvect{B}_0$ is the one leading
 to the most successful optimization of $\myvect{v}_p$, following the
  diagnostics described in the next subsection. 
In practice this involves multiple trials with an ensemble of initial conditions for assimilating 
a single set of observations, and requires considerable computer time. 
 As we will demonstrate in the upcoming sections,  using this approach we have always successfully 
 found an initial $\myvect{B}_0$ that leads to  a good final estimation of the meridional circulation
 $\myvect{v}_p$. We encourage the interested reader to read Appendix~\ref{sec:initialcon} where we give more details on the 
 strategy used to initialize the data assimilation algorithm. One should also note that when trying to do a forecast, the initial condition 
 is usually taken from the previous assimilation cycle, and is consequently much closer to 
  the sought one than in the worst-case scenario that we consider here. 
  

\subsection{Diagnostics to assess the quality of the assimilation}
\label{subsec:paramspace}

The success of the assimilation depends on whether the estimated expansion coefficients 
 $d_{i,j}$ converge satisfactorily to their true value. If 
convergence is achieved, we assess the quality of the assimilation procedure by studying the 
number of iterations required to achieve a given accuracy and the value of the misfit at the end of 
minimization. In general, for a given set of observations, the assimilation halts when the magnitude of 
the gradient of the objective function decreases by a preset factor. This indicates that the misfit is 
minimized but from this alone there is no information about the accuracy and uniqueness of the optimal 
solution. However, in twin experiments, as observations are artificially generated by a known 
true state, the accuracy of the assimilation can be studied quantitatively by defining the following (relative) discrepancy:

\begin{equation}
\frac{\Delta p}{p} = \sqrt {\frac{\sum\limits_{i=1}^{m} \sum\limits_{j=1}^{n} \left(d_{i,j}-{d_{i,j}}_{true}
\right)^2}{\sum\limits_{i=1}^{m} \sum\limits_{j=1}^{n}{d^2_{i,j}}_{true}}},
\label{eq:dp}
\end{equation} 
here the coefficients with subscript {\it true} are the ones used to generate the synthetic observations 
(see Table~\ref{tab:modeldij}). This discrepancy is the Euclidean metric between the estimated control vector and 
its true counterpart, normalized by the norm of the true vector. 

In Sec.~\ref{sec:twinnoise}, when our synthetic observations are noised, we can measure the performance of the assimilation technique by the normalized misfit defined by 
\begin{equation}
\mathcal{J}_{norm}=\frac{1}{\epsilon}\sqrt{\frac{\mathcal{J}} {N}},
\label{eq:normmisfit}
\end{equation}
where $\epsilon$ is the level of noise introduced (as a fraction of r.m.s. of the field trajectory) 
and $N$ is the total number of observations. The fraction $\epsilon$ is used instead of the 
absolute standard deviation because from Eqs. \eqref{eq:jb} and \eqref{eq:ja}, the objective functions 
are already normalized with the r.m.s value. In our case, the observations are (unless otherwise noted) 
 the tachocline toroidal field and the surface potential vector, implying that $N=2N^o$. Statistically, an assimilation 
trial with ideal fitting of data will give $\mathcal{J}_{norm} \sim 1$, while $\mathcal{J}_{norm} \gg 1$ 
implies under fitting where the misfit is considered too large. On the other hand, $\mathcal{J}_{norm} \ll 1$ 
implies over fitting, given what is expected from the statistics. 

\section{Results using data without noise}\label{sec:twinnonoise}

In this section, we investigate the quality of convergence of the assimilation for cases 1, 2 and 3 when the observational data are perfect, i.e. 
when they are realizations of the direct model and are not perturbed. The results of this section will be used for a systematic 
analysis and testing of the assimilation method and will serve as reference for the following section where noise is added. We first 
show the efficiency of our assimilation technique on \firstrev{an ideal} case, with a reference distribution of observations. We then study 
the behavior of our technique when we vary the sampling of observations both in time and latitude. 
We focus on the effect of (i) the width of temporal windows over which observations are available, (ii) the temporal sampling, (iii) the latitudinal 
sampling. We investigate each of the above for case 1 and illustrate some typical examples for the 2 other cases for comparison.

\subsection{Recovery of true flow and observed magnetic field in \firstrev{an ideal} case}
\label{subsec:reference}

\floatsetup[figure]{style=plain,subcapbesideposition=top}
\begin{figure*}[!ht]
  \Large {$ \text{Initial guess}\xrightarrow{\makebox[0.65\columnwidth]{\text{\Large{Optimization}}}} \text{Final estimate}$}
  \sidesubfloat[]{\includegraphics[width=\columnwidth]{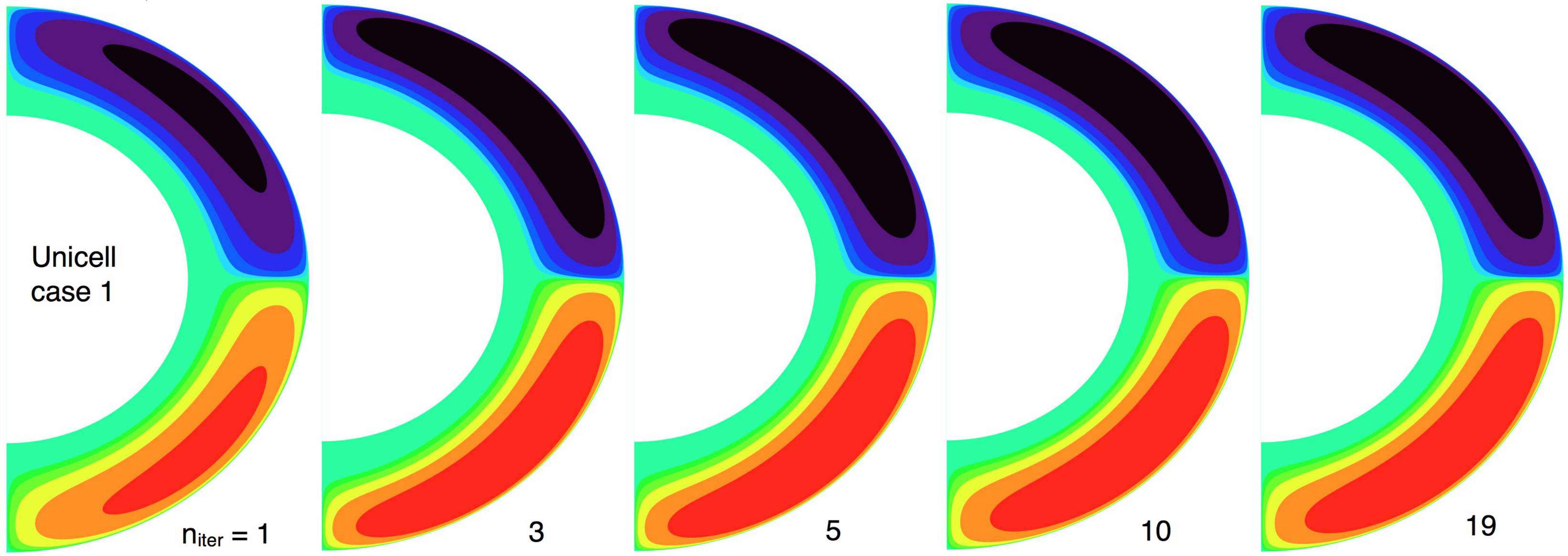}\label{fig:cellitercase1}}\quad%
  \sidesubfloat[]{\includegraphics[width=\columnwidth]{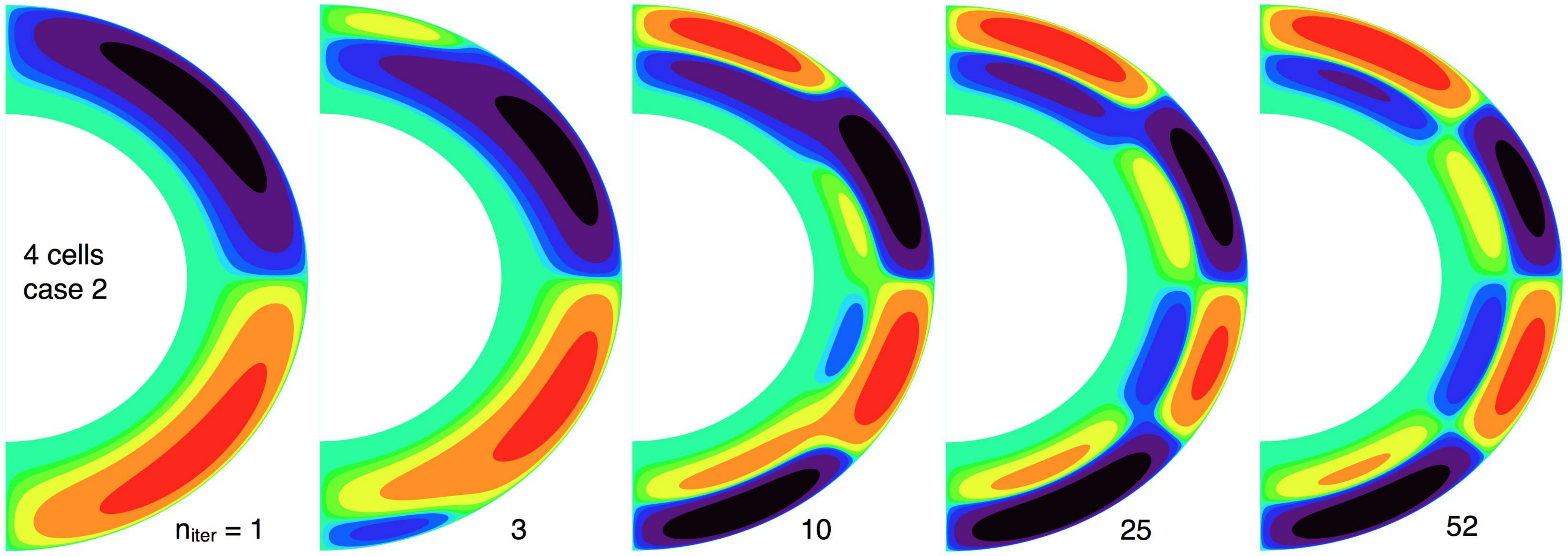}\label{fig:cellitercase2}}\quad%
  \sidesubfloat[]{\includegraphics[width=\columnwidth]{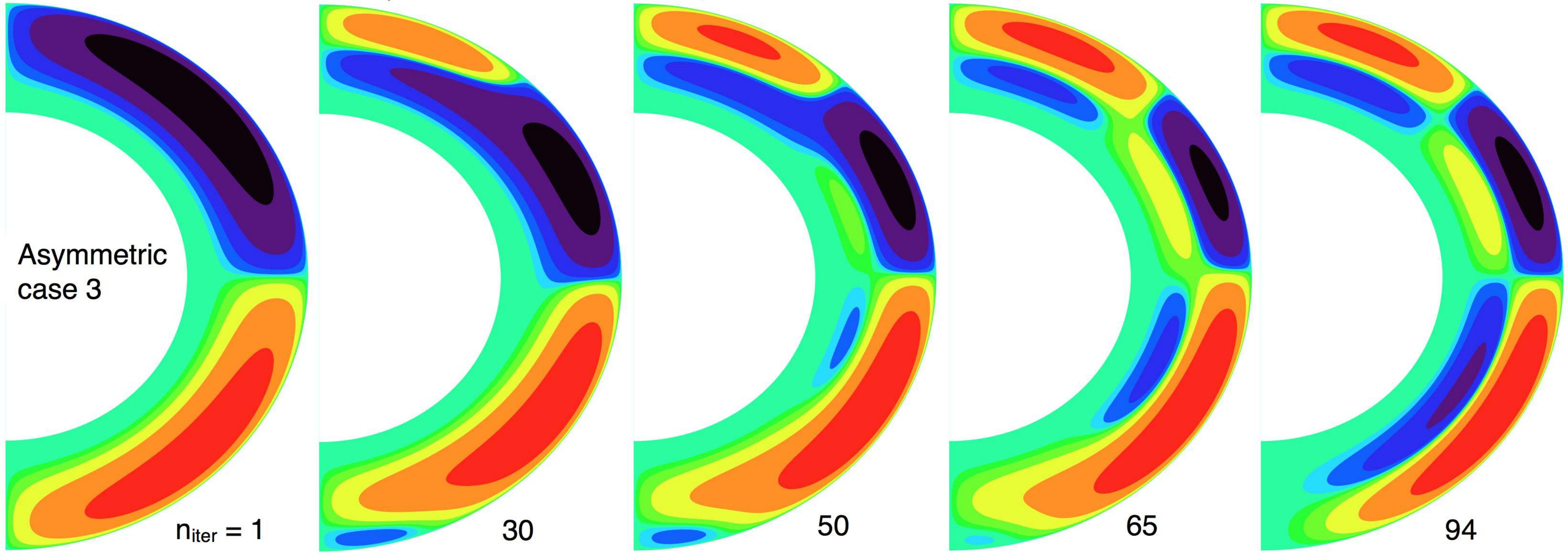}\label{fig:cellitercase3}}\quad%
\caption{Evolution in iteration count of assimilation of the meridional circulation for (a) case 1, (b) case 2 and (c) case 3. 
In case 1, from left to right, we show the stream function at the 1st, 3rd, 5th, 10th, 19th iteration. Note 
the change in the color mapping which shows the evolution of the model. For case 2, from left to right, we show the 
stream function at the 1st, 3rd, 10th, 25th, 52nd iteration, and for case 3, from left to right, 
we show the 1st, 30th, 50th, 65th, 94th iteration. For cases 2 and 3, note that the stream function eventually
evolves from the unicellular initial guess to the appropriate multi-cellular configuration. }
\label{fig:cellitercase123}
\end{figure*}

An illustrative example is considered here, for the three cases. 
The distribution of observations is uniform in latitude ($\Delta \theta =2.83^\mathrm{o}$) and time ($\Delta t= 33$ days) and the observational window 
has a width $t_e-t_s=1.5$ cycles. We show how the stream functions evolve from the unicellular initial guess to 
the final structure after assimilation. Figures~\ref{fig:cellitercase123} (a), (b), and (c) show 
the evolution for cases 1, 2, and 3 respectively. We can see that in cases 2 and 3, new cells appear 
at the expense of shrinking existing cells, and that 
it takes more iterations for the multi-cellular asymmetric structure to develop in case 3 
compared to the relatively simpler case 2. Note that in cases 1 and 2, the intermediate \secrev{states} of meridional 
circulation as the minimization proceeds are always antisymmetric with respect to the equator, since the 
symmetry of predicted values is preserved throughout the algorithm. For case 3, the synthetic data are asymmetric, 
and the meridional flow slowly shifts from the symmetric prior to a more adequate asymmetric structure, 
as shown in Figure~\ref{fig:cellitercase123} (c). In all cases, the basic structure of the meridional 
circulation is rapidly and nicely recovered. In the last 10 iterations, it comes to fine adjustments to further 
\secrev{decrease} the data misfit. This confirms that a unicellular prior is an appropriate initial guess, and it 
also demonstrates the potential of our assimilation method that can recover meridional circulation flow 
with multi-cellular and asymmetric profiles with respect to the equator as well as its deeper inner structure.

\begin{figure}[!ht]
\includegraphics[width=\columnwidth]{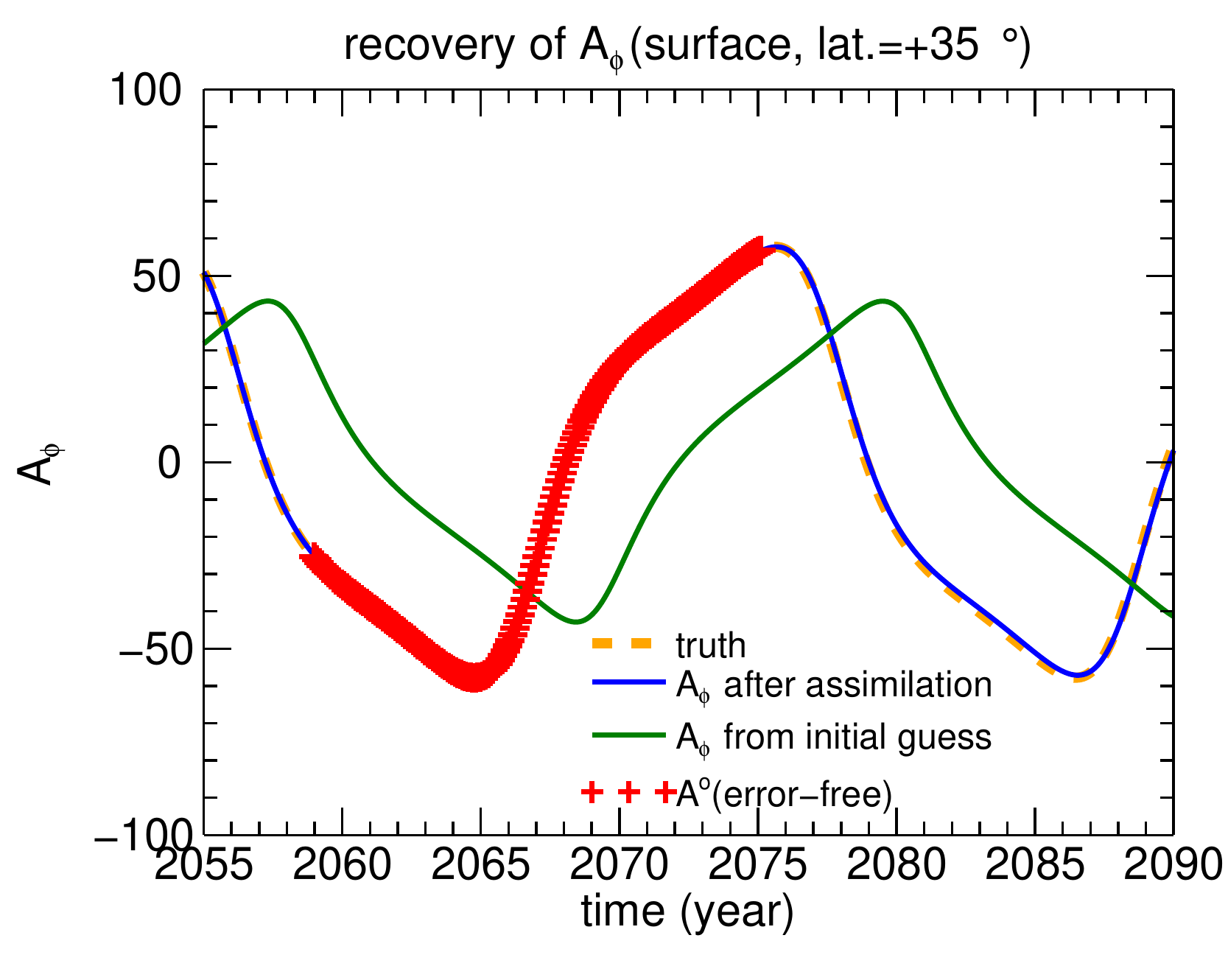}
\caption{
Vector potential of poloidal field at a given location as a function of time for case 3 (asymmetric meridional circulation).
Shown are the time series of the reference (dashed orange line), the initial guess before assimilation
(in green), the final estimate after assimilation (in blue), and the synthetic observations made at   
 that location (red crosses).}
\label{fig:recoveredsample}
\end{figure}

We now focus on the associated magnetic fields we get from the recovered meridional flow at the end of 
the assimilation procedure. In Figure~\ref{fig:recoveredsample}, we show for case 3, the magnetic fields at 
latitude $+35^{\circ}$ as a function of time (any other latitude yields similar results), (i) from the 
dynamo run with the initial guess for the meridional flow (green curve), (ii) from the dynamo run using the 
true meridional flow (orange curve), (iii) from the dynamo run with the final estimate for the meridional 
flow (blue curve), and (iv) from the synthetic observations (red crosses). The field trajectory from the true 
meridional circulation nearly overlaps with the trajectory from the final estimate and they are hardly 
distinguishable. We thus successfully recover the magnetic field trajectory from the free dynamo run with a 
minimal misfit for case 3, and similarly for all other cases showing convergence. \par
 
By assimilating synthetic observations with varying $N^o$, $N_{\theta}^o$, $N_t^o$, 
we find that a minimum $N^o$ is required to estimate a $\myvect{v}_p$ which effectively 
minimizes ${\mathcal J}$, and that the  accuracy of the estimate 
decreases when the width of the assimilation window $t_e-t_s$ is too short. 
The unicellular case is the most tolerable, with an estimate of $\myvect{v}_p$  accurate down to $\Delta p/p \sim 10^{-6}$ 
with $N^o=9$ and $t_e-t_s=0.5$ solar cycle only. For the 4-cell case 2, the lowest possible observations ingested in the tests leading to success is $N^o=84$, 
with $t_e-t_s=1.5$ solar cycles.  Fewer observations results in optimization failure and a decrease of $t_e-t_s$ causes a drop of 
accuracy of the estimate. For the most complicated asymmetric case 3, the smallest $N^o$  is found to be $N^o=966$ for a $t_e-t_s=1.5$~solar cycles. 
Decreasing $N^o$ and $t_e-t_s$ have similar consequences as for case 2.  In general, we also find that increasing $N_{\theta}^o$ is more efficient 
in improving the convergence rate than  increasing $N_t^o$.

\subsection{Convergence behavior vs observational window width}
\label{subsec:samwin}

\floatsetup[figure]{style=plain,subcapbesideposition=top}
\begin{figure*}[!ht]
  \sidesubfloat[]{\includegraphics[width=0.4\columnwidth]{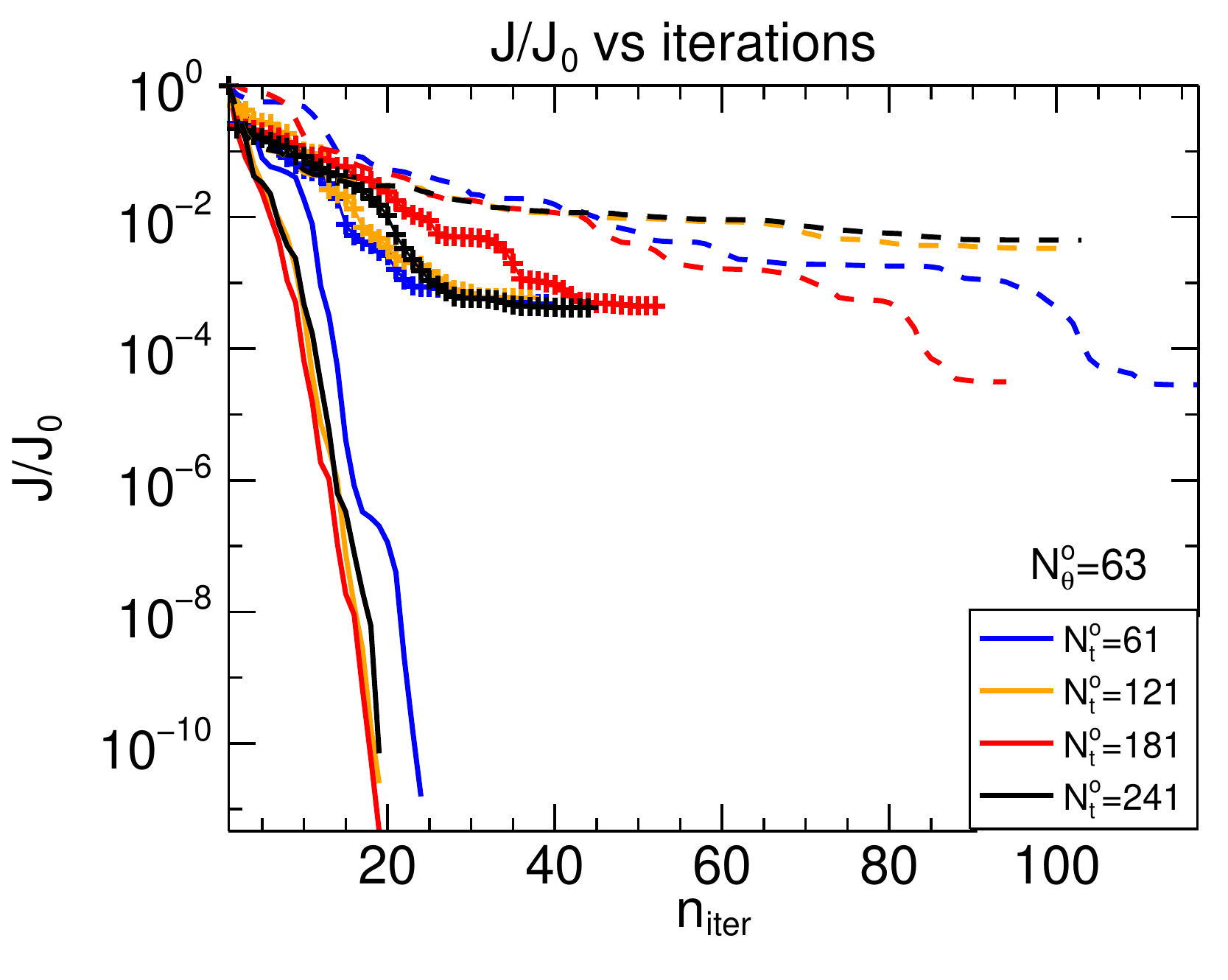}\label{fig:wj}}\quad%
  \sidesubfloat[]{\includegraphics[width=0.4\columnwidth]{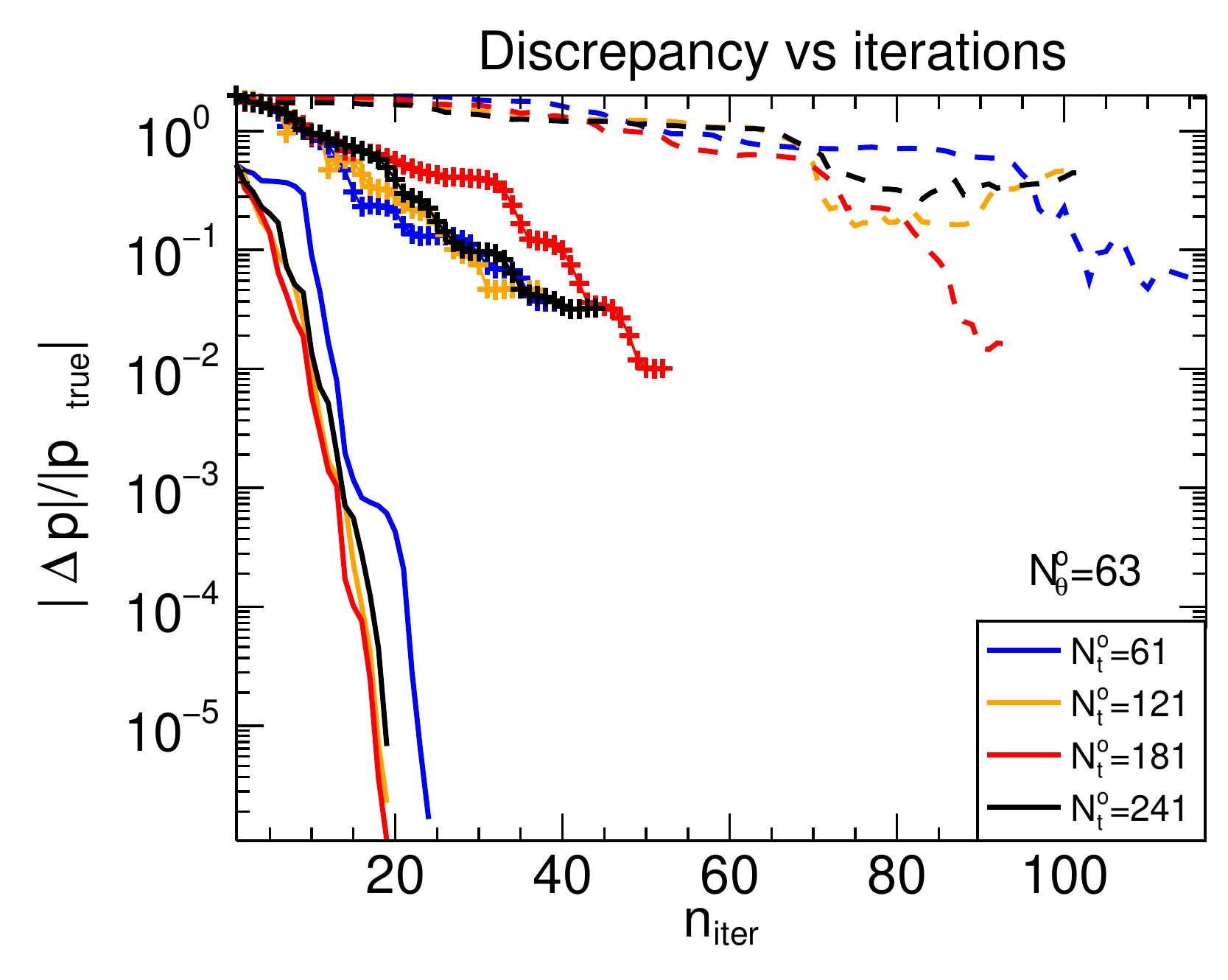}\label{fig:wdp}}\quad%
   \caption{Optimization of objective function (left) and increase in accuracy (right) as the iterative
   minimization proceeds, at a fixed sampling frequency of $\sim$ 1 month (33 days), for varying
   assimilation window widths.  
   $N_t^o=61,121,181\text{ and }241$ correspond to sampling windows of width 0.5, 1.0, 1.5 and 2.0 solar cycle(s), respectively. 
   Here and on the plots which follow, we indicate case 1, case 2 and case 3 with solid lines, crosses and broken lines respectively. 
   The number of observation points in latitude is the same for all cases ($N_{\theta}^o=63$). 
   The convergence rate increases with the number of iterations, i.e., when the predicted state gets closer to the true state. }
   \label{fig:ConvMetime}
\end{figure*}

\begin{figure}[!ht]
  \includegraphics[width=\columnwidth]{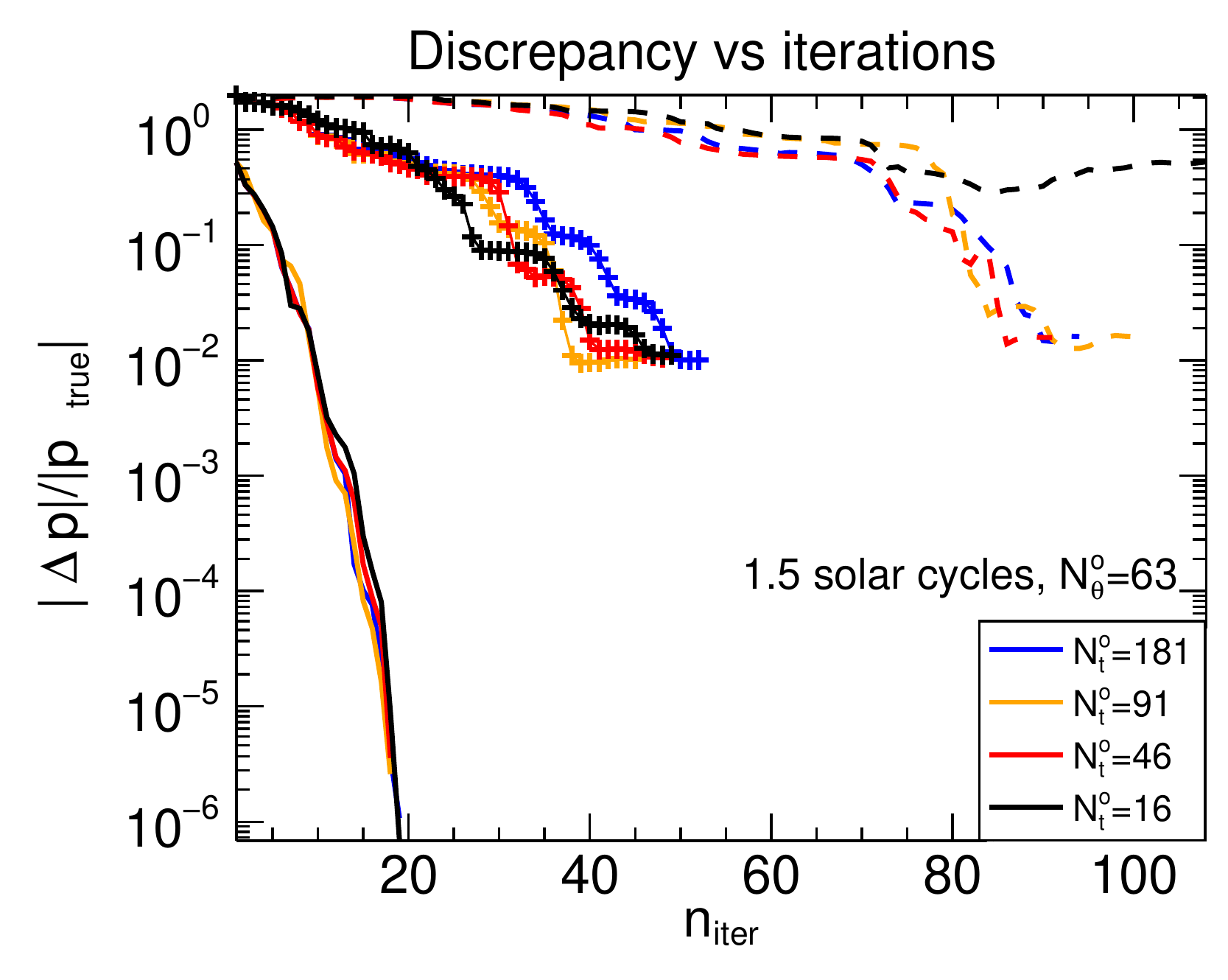}
   \caption{Increase of accuracy as the minimization proceeds, at various temporal sampling frequencies,
   for a fixed assimilation window of width 1.5 solar cycle and a fixed number
 of observation points in latitude ($N_{\theta}^o=63$). The effect of the sampling frequency on the convergence 
 is less significant than that of the temporal window width, except for the most complex asymmetric case with the sparsest temporal sampling.}
\label{fig:ConvMefreqcaseall}
\end{figure}

First, we investigate the convergence behavior of the twin experiment, at a fixed sampling frequency of $\sim$ 1 
month (33~days) of observations, i.e., $\Delta t$ is kept constant, and further $N_{\theta}^o$ is constant, 
with 63 observations evenly distributed in colatitude from $\theta=$ 0 to $\pi$, i.e., $\Delta \theta=2.83^{\circ}$. 
Observations are taken regularly in time so the width of the sampling window $t_e-t_s$ increases as $N_t^o$ increases. 
In this study the temporal window width $t_e-t_s$ varies from 0.5 solar cycle to about 2 cycles. In each assimilation trial, we investigate the 
relation of  $\mathcal{J}/{\mathcal{J}}_0$ and ${\Delta p}/p$ with the corresponding iteration count respectively (${\mathcal{J}}_0$ is the 
value of objective function at the initial guess of meridional circulation). The results are shown in 
Figure~\ref{fig:ConvMetime}. In all cases the \firstrev{nomralized} objective functions $\firstrevii{\mathcal{J}/\mathcal{J}_0}$ \firstrev{[panel (a)]} and the discrepancies  $\firstrevii{dp/p}$ \firstrev{[panel (b)]} 
diminish as the iteration count \secrev{increases}, showing that the optimization method is giving an improved 
estimate of meridional circulation after each iteration, thereby reducing the misfit between the observations and the field predicted by the dynamo model. Note that the objective function and discrepancy can get to extremely small values ($\Delta p/p<10^{-5}$) in case 1 compared to cases 2 and 3. This is because the initial condition of the assimilation is always the dynamo field from the unicellular flow model, and the synthetic observations in case 1 are also from the same \secrev{unicellular profile} (but the strength of the flow is not the same so the assimilation is not trivial), so the model prediction is essentially the same as the synthetic observations when the true meridional circulation is recovered. 
\secrev{While in cases 2 and 3, assimilation is initialized using a 1-cell configuration which is different 
from the 4-cell and asymmetric profiles used to generate the synthetic observations, resulting 
in a more challenging minimization, 
making it more difficult to reach the absolute minimum.}
As shown, the latter is of the order $>10^{-2}$, even when the synthetic observations are noise-free. Nonetheless, we are still able to reconstruct the true flow and the trajectory of the field (which will be shown in the following sections). 
The number of iterations required (for a fixed convergence criterion) increases as the complexity of the model increases: case 1 requires fewest iterations and case 3 requires most iterations. In case 3, only the trial with a sampling window of width 1.5 cycles can give a discrepancy of $10^{-2}$ (others are higher), as the objective function for such a complex meridional flow is also complicated and difficult to be optimized. In some situations, the assimilation algorithm terminates at a local minimum of the objective function in the parameter space, like the trials in case 3 with temporal windows of 1.0 cycle and 2.0 cycles. In those cases, the performance in terms of minimization of misfit are lower than that of the 1.5 cycles trial.  \par
For all cases, a short sampling window of 0.5 cycle lowers the accuracy of the results (see the blue curves in Figure~\ref{fig:ConvMetime}), as the short observation window may not provide enough constraints on the meridional circulation. \par
Another feature of the assimilation method is that the accuracy increases slowly at the beginning of the iteration, and speeds up eventually as the improved forecast gets closer to the true state, where the objective function is close to quadratic. This is a characteristic of the assimilation algorithm. As the meridional circulation becomes more complicated, it takes more iterations for the estimated control vector
 to reach the region of quadratic convergence. \par
In this particular study of the impact of the sampling window width on the 
 quality of the assimilation, we find that a window width of 1.5 cycles gives on average the best a posteriori 
fit to the observations.  

\floatsetup[figure]{style=plain,subcapbesideposition=top}
\begin{figure}[!ht]
  \sidesubfloat[]{\includegraphics[width=0.77\columnwidth]{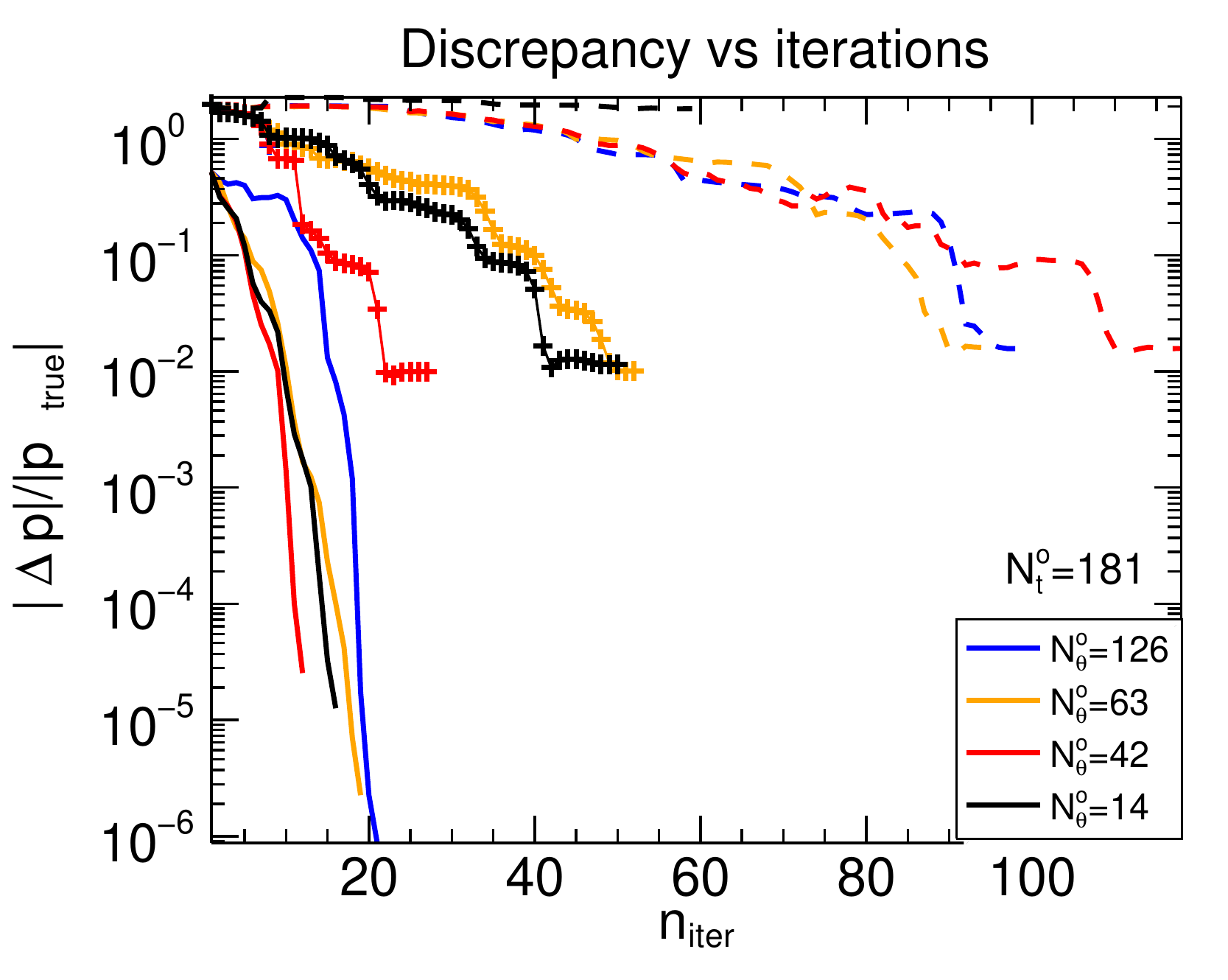}\label{fig:udp1}}\quad%
  \sidesubfloat[]{\includegraphics[width=0.77\columnwidth]{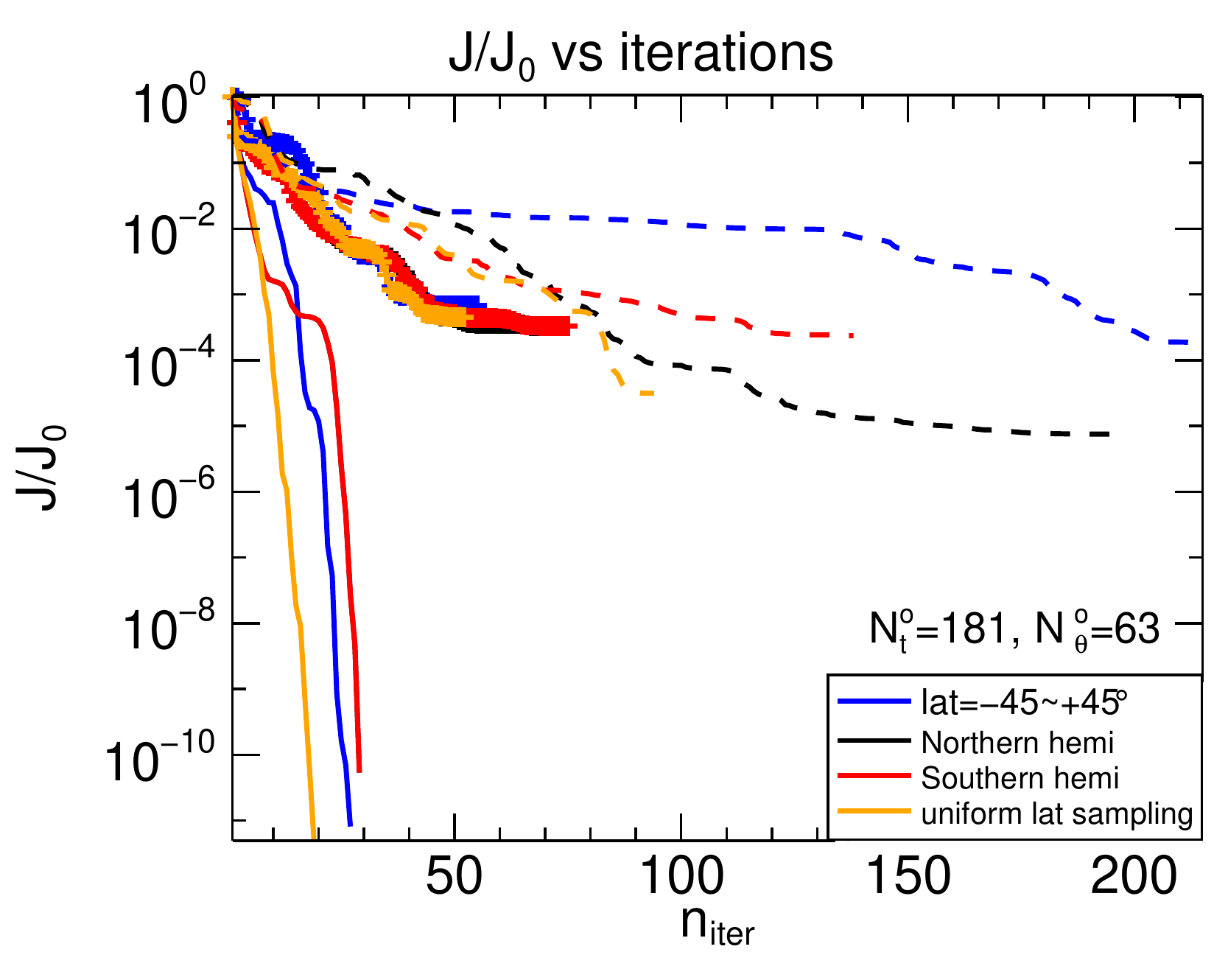}\label{fig:hj1}}\quad%
  \sidesubfloat[]{\includegraphics[width=0.77\columnwidth]{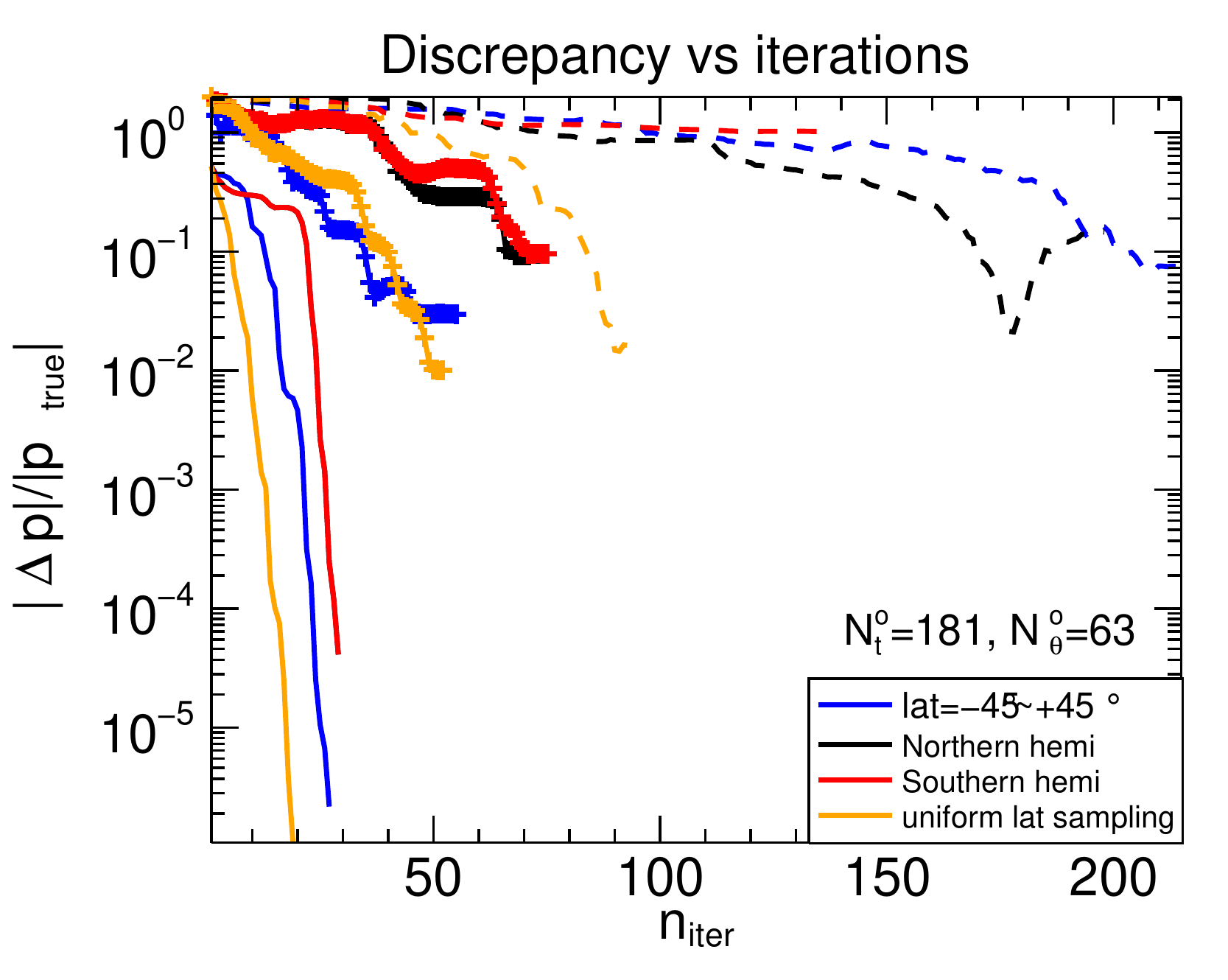}\label{fig:hdp1}}\quad%
   \caption{Evolution of $\Delta p/p$ as minimization proceeds 
 for various uniform spatial samplings [panel (a)] and of $\mathcal{J}/{\mathcal{J}}_0$ and ${\Delta p}/p$ 
 for nonuniform sampling in latitude [panels (b) and (c)]. 
  Blue, black, and red refer to sampling restricted to the active band, the Northern hemisphere and the Southern hemisphere, 
  respectively. 
  In the 2 last panels, a case with uniform sampling is also shown for comparison.}
   \label{fig:ConvMeuhlat}
\end{figure}

\subsection{Convergence behavior at different sampling frequencies}
\label{subsec:samfreq}

From the previous section, we identified that the optimal observational window width $t_e-t_s$ for almost all cases was 1.5 cycles. 
We thus now fix this temporal window width and investigate the effect of changing the sampling frequency. Spatial 
sampling is held constant and evenly distributed ($\Delta \theta=2.83^{\circ}$). The results are 
shown in Figure~\ref{fig:ConvMefreqcaseall}. We see that the convergence behavior is 
relatively insensitive to the change in sampling frequencies, with only one exception 
for case 3. The highest sampling frequency shown at $\Delta t = 33$ days corresponds 
to $N_t^o=181$ , while  $N_t^o=16$ for the sparsest sampling, $\Delta t=396$ days. The sparsest sampling 
consists of a total of $N^o=1008$ observations, which is sufficient to estimate the 8 expansion coefficients 
of the meridional flow, provided that the sampling window is wide enough to characterize the flow. We thus find 
that in all these cases where $N^o \gg 8$, having more temporal observations within a fixed 
sampling window does not introduce any new characteristics to the objective function during the 
assimilation process. However, the trial with the sparsest sampling in case 3 is an exception as the 
true meridional flow is the most complex. This probably requires more frequent observations to correctly estimate the 
flow structure. Overall, a sampling frequency of one month to a trimester seems adequate when considering the perspective of using real solar data.

\floatsetup[figure]{style=plain,subcapbesideposition=top}
\begin{figure*}[!ht]
  \sidesubfloat[]{\includegraphics[width=0.23\columnwidth]{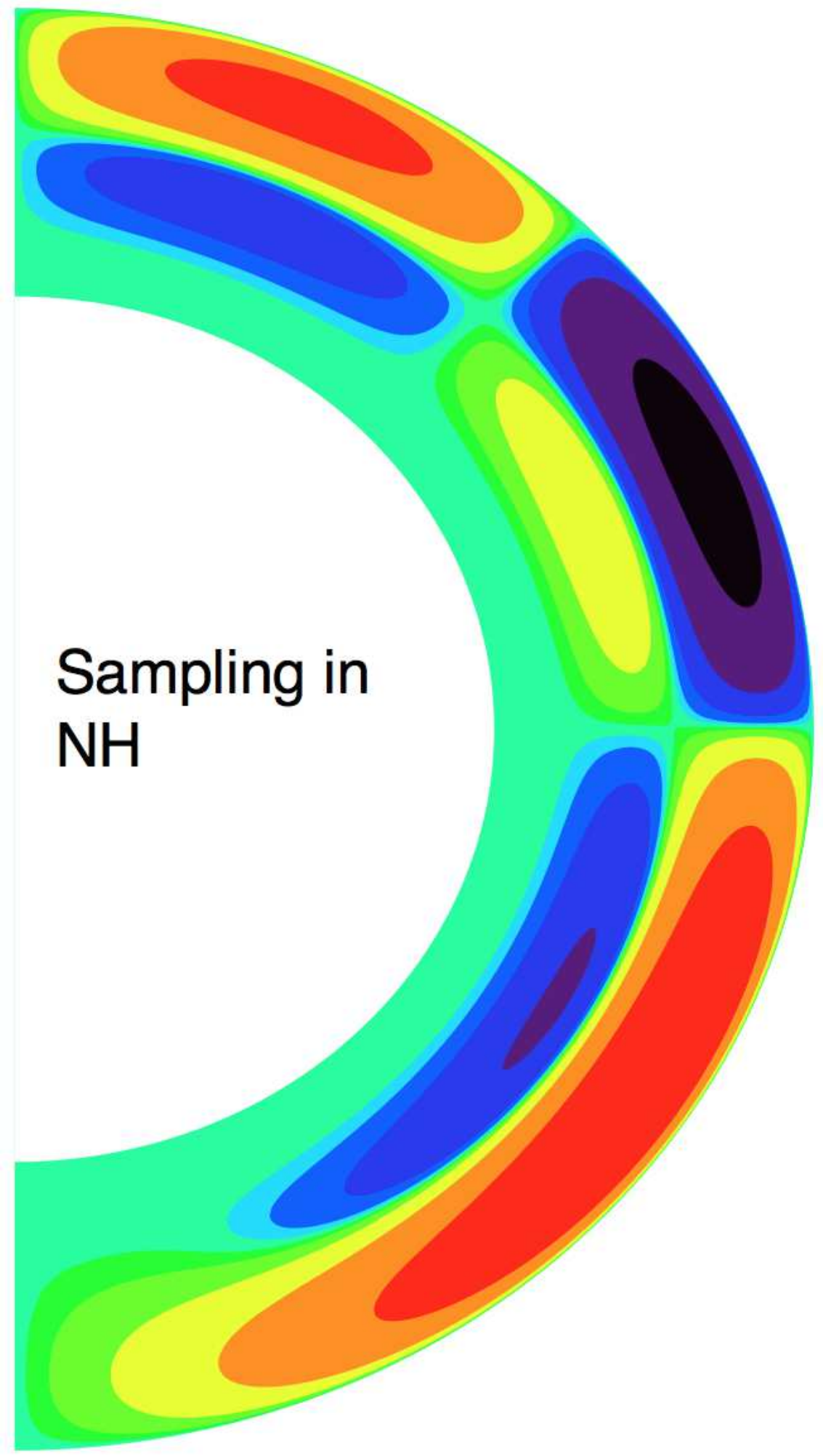}\label{fig:flowNHcase3dr}}\quad%
  \sidesubfloat[]{\includegraphics[width=0.23\columnwidth]{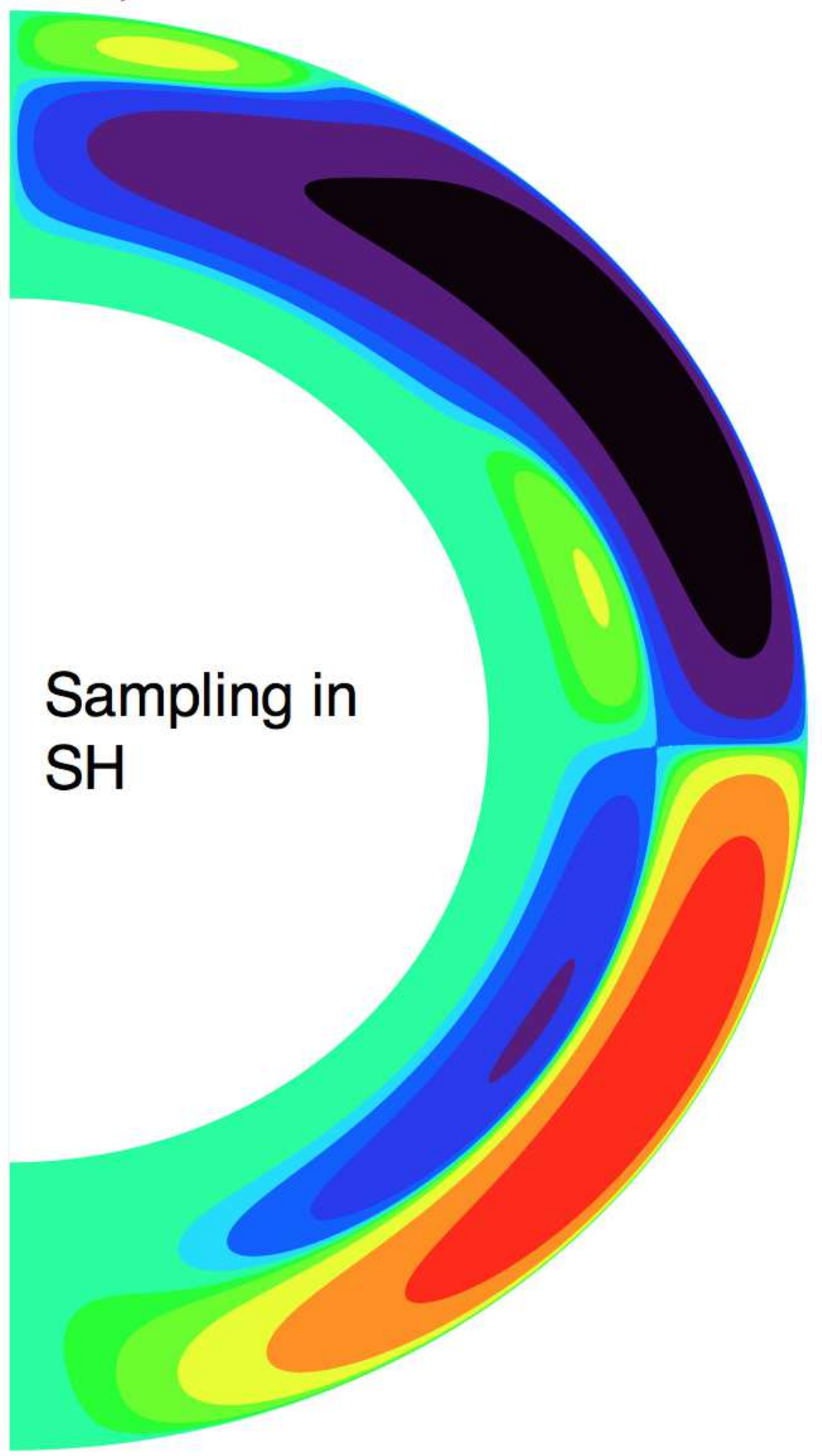}\label{fig:flowSHcase3dr}}\quad%
  \sidesubfloat[]{\includegraphics[width=0.23\columnwidth]{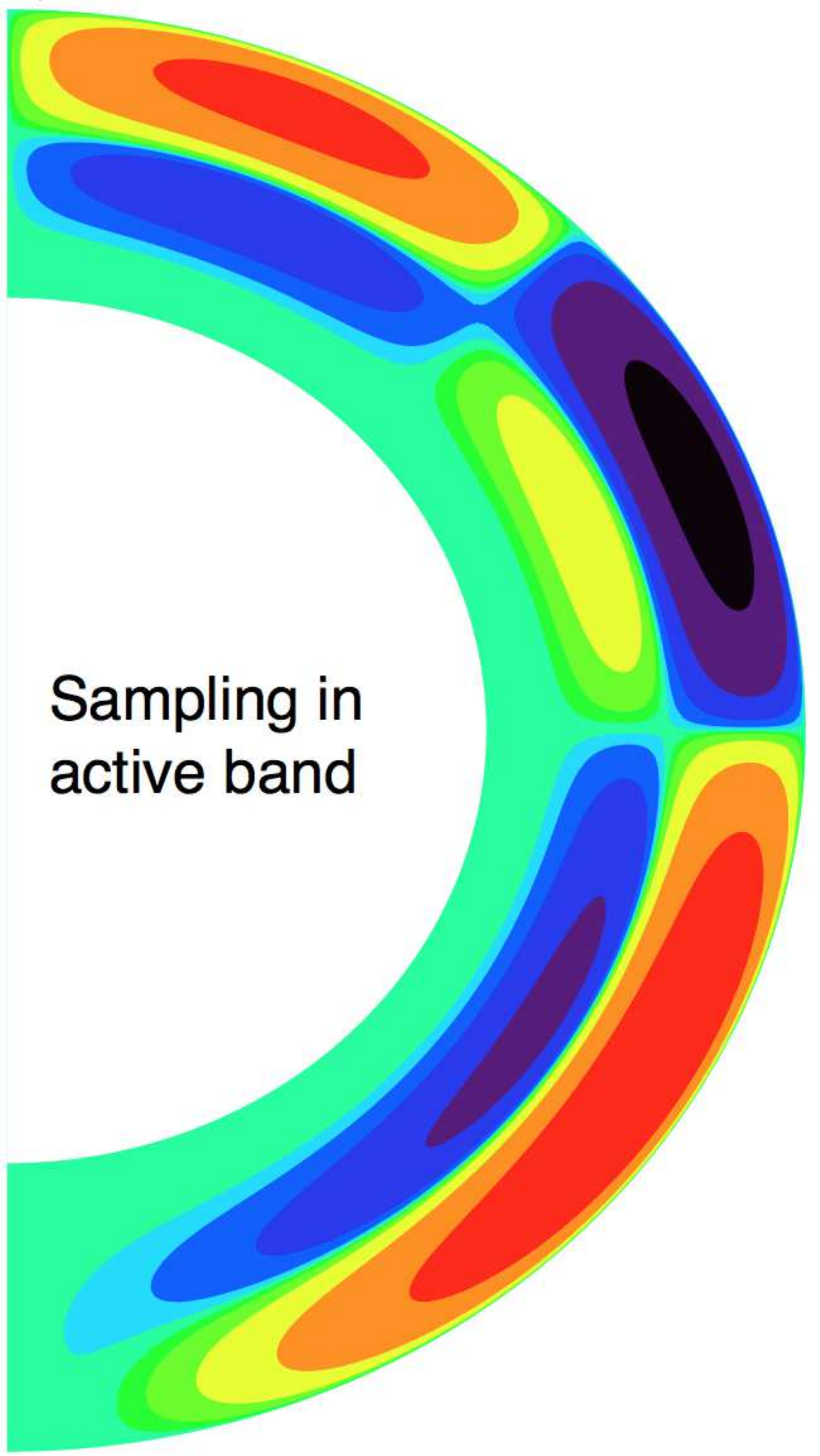}\label{fig:flowactbandcase3dr}}\quad%
   \caption{Stream function of the estimated meridional flow of case 3 at the end of assimilation, with
   a latitudinal sampling restricted to  (a) the Northern hemisphere, (b) the Southern hemisphere and (c) the activity band. 
   Observations are made every month for 1.5 solar cycles. This shows that observing in one hemisphere leaves the estimation of 
   the meridional circulation in the other hemisphere less ideal. In (c), the synthetic observations are noised with $\epsilon=1\%$
    (see text for details).}
   \label{fig:flowforecastcase3drNSH}
\end{figure*}

\begin{figure}[!ht]
\includegraphics[width=0.9\columnwidth]{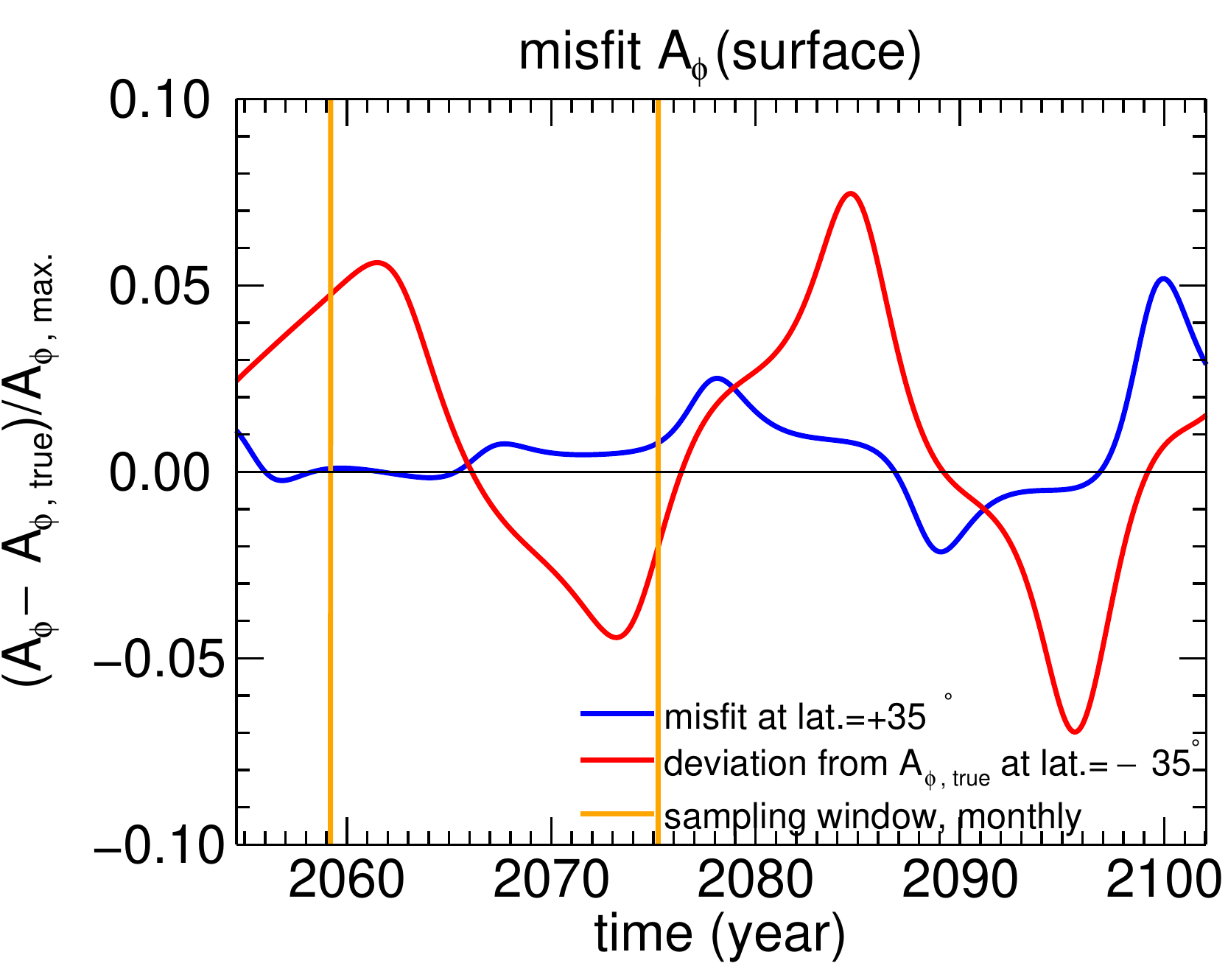}
\caption{Normalized deviation of the predicted field 
from the true field at $35^{\circ}$ (blue) and at $-35^{\circ}$ (red), for case 3 
when magnetic observations are restricted to the Northern hemisphere. 
 The vertical orange lines indicate the limits of the observational window.
 Note that outside the sampling window, the deviation starts to grow.}
\label{fig:NSmisfit}
\end{figure}

\subsection{Convergence behavior at different latitudinal samplings}

\label{subsec:nonoiselat}

In this section, we fix the temporal window to 1.5 cycles and sampling frequency to 1 month. We now investigate the results of the assimilation procedure when the distribution of observations in latitude is varied. We show the results with (I) uniform sampling in latitude, (II) nonuniform sampling: sampling in one hemisphere only and sampling in the activity band only (latitude $-45^o$ to $+45^o$).
The results for (I) and (II) are shown in Figure~\ref{fig:ConvMeuhlat}. For (I) [panel (a)], there is no systematic trend relating the sampling density and convergence behavior. Unlike the situation of changing sampling frequencies, the objective function is more sensitive to a change of sampling density in latitudes, and a denser spatial sampling does not necessarily result in faster convergence. A latitudinal sampling of $\Delta \theta =4.25^{\circ}$ requires the least iterations for convergence for cases 1 and 2, but the $\Delta \theta =2.83^{\circ}$ sampling results in higher accuracy. For case 3, the sparsest spatial sampling fail to converge and assimilation stops eventually without significant optimization, showing that for a flux-transport dynamo with a complex flow, more spatial observations are needed to estimate the flow structure. 

 In (II) [panels (b) and (c)], a uniform sampling always gives better convergence than other nonuniform patterns for the same number of observations. Again, case 1 gives the smallest misfit and discrepancy. For sampling in one hemisphere only, notice the following features: 

(i) In case 1, the convergence paths shown for sampling in the Northern or Southern hemisphere only coincide. Indeed, since the data and prior are both antisymmetric about the equator, the resulting \firstrev{normalized} objective functions $\firstrevii{\mathcal{J}/\mathcal{J}_0}$ \firstrev{[panel (b)]} are exactly the same. Note however that for case 2, which is also antisymmetric, the results for the Northern and Southern hemisphere are close but not exactly the same. This is due to the fact that the initial condition for the assimilation is unicellular and the system has thus to undergo a transient, resulting in not exactly the same steady states for both trials. This is not true for case 3 where the model is asymmetric.

(ii) Although the discrepancy from the true state is significantly higher than in the case with uniform sampling, the meridional flow in the whole domain can be recovered by sampling in one hemisphere only, even for the asymmetric case~3. This can be explained by the fact that we do not try to reconstruct the point-wise meridional flow but look for the coefficients of a particular expansion (eq.\ref{eq:MC}), which implies certain symmetries. Also, this shows that the magnetic observations in one hemisphere give information for the meridional circulation in the other hemisphere by various physical processes like hemispheric coupling. It thus demonstrates the global structure of the magnetic field produced in these models.

(iii) Let us now focus on case 3 with observations in the Northern hemisphere only [black dashed line in Figure~\ref{fig:ConvMeuhlat}(c)]. As there is no constraint on the hemisphere where the field is not sampled, the optimization algorithm gives a \secrev{state} which is slightly different from the true flow. This is illustrated in Figure~\ref{fig:flowforecastcase3drNSH}(a) where the result for the meridional flow is shown at the end of the assimilation. Compared to Figure~\ref{fig:MCpsi}(c), we see that the Northern hemisphere where observations are available is much better recovered than the Southern hemisphere, although the recovery is still quite satisfactory in the Southern hemisphere (as stated before, the initial guess was indeed a unicellular flow). To have more quantitative estimates of the quality of the recovery of the solution in this case, we plot the deviation between the forecast and the true trajectory for the magnetic poloidal potential at some selected latitudes in Figure~\ref{fig:NSmisfit}. We find that during the observational window, the misfit on $A_\phi$ in the Northern hemisphere is of the order of half a percent when it is about 10 times higher for the Southern hemisphere (which is still a low value given the fact we have no observations in this hemisphere). This figure also illustrates that the deviations start to grow after the end of the observational window, with a faster growth in the Southern hemisphere. This indicates that the reliability of the prediction decreases quite quickly when observations are no longer available. Typically here in about 1 or 2 cycles, the benefit of data assimilation is lost. \par
For the other latitudinal samplings considered for case 3, namely in the Southern hemisphere only and activity band only, we find similar results. The recovery of the flow is shown in Figure~\ref{fig:flowforecastcase3drNSH}(b) and (c). We note that when the observations are available in the hemisphere where the flow is less complex [panel (b)], the structure of the flow in the other hemisphere is quite different from the true one. Indeed, in this case, the algorithm probably stopped in a local minimum and the final value of the discrepancy stays very high ($\Delta p/p\sim 1$), as seen in Figure~\ref{fig:ConvMeuhlat}(c) (red dashed line). However, when observations are available in the activity belt only [panel (c) of Figure~\ref{fig:flowforecastcase3drNSH}], the structure of the flow is much better recovered and the discrepancy drops to a value of less than $10\%$. 

From the above systematic study of the sampling patterns, a temporal observational window of 1.5 solar cycles with monthly sampling and uniform in latitude (with $\Delta \theta =2.83^{\circ}$) gives relatively robust convergence among all the trials (though not necessary giving the fastest convergence in all 3 cases). Therefore, for illustration purposes, we will use this reference sampling in most of the analysis and demonstrations below.

\subsection{Other characteristics of the convergence}

\label{subsec:conv_char}

We here study some additional properties of the minimization procedure. We first focus on the choice of the objective function. 
We can choose to assimilate the tachocline toroidal field or surface potential vector only, i.e. using the objective functions \eqref{eq:jb} or \eqref{eq:ja} for optimization. For case 1, data assimilation is effective in terms of minimizing the misfit and recovering the true meridional circulation even if we use \eqref{eq:jb} or \eqref{eq:ja} only. For the more complex cases 2 and 3, using only one of the objective functions lowers the performance of optimization, i.e., more iterations are required to recover the meridional circulation to the same accuracy, compared to the case where the sum of both objective functions is used. For example, for case 3, it takes 94 iterations to assimilate both components but 141 iterations are needed if only the toroidal field is given, to reach a discrepancy of $\sim 2\%$. Also, in most cases, the misfit at the end of the assimilation is higher than when both fields are considered. We also showed that when one component of the field only is observed, the decline of performance cannot be compensated by a longer sampling time, e.g. 2 solar cycles (one complete magnetic cycle) instead of 1.5. Overall, both $A_{\phi}^o(R_s,\theta,t)$ and $B_{\phi}^o(r_{c},\theta,t)$ are necessary for the assimilation procedures to capture a meridional circulation which can optimize the misfit, especially when the true meridional circulation is far more complex than one cell per hemisphere. For the next section, we will thus use the synthetic observations both on $A_\phi$ and $B_\phi$ to perform the assimilation. As noted above in a real situation we will not have direct access to the toroidal field at the base of the convection zone, but instead to
surface field (via for instance sunspot observations) that we will have to relate to the deep toroidal field via an adequately defined
 operator. We have tested the influence on the convergence of shifting the location of the toroidal sampling and found that in the unicellular case it has little influence. For the multicellular cases 2 \& 3, as long as the sampled depth \secrev{probes} the deeper secondary cell, the data assimilation procedure behaves the same. Nevertheless in the Babcock-Leighton framework it is natural to relate the sunspot number to the field strength at the base of the convection zone, as it mimics the rise of toroidal flux tubes to the surface. Defining the corresponding
  operator will be the subject of our next study.

\begin{figure*}[!ht]
  \sidesubfloat[]{\includegraphics[width=0.4\columnwidth]{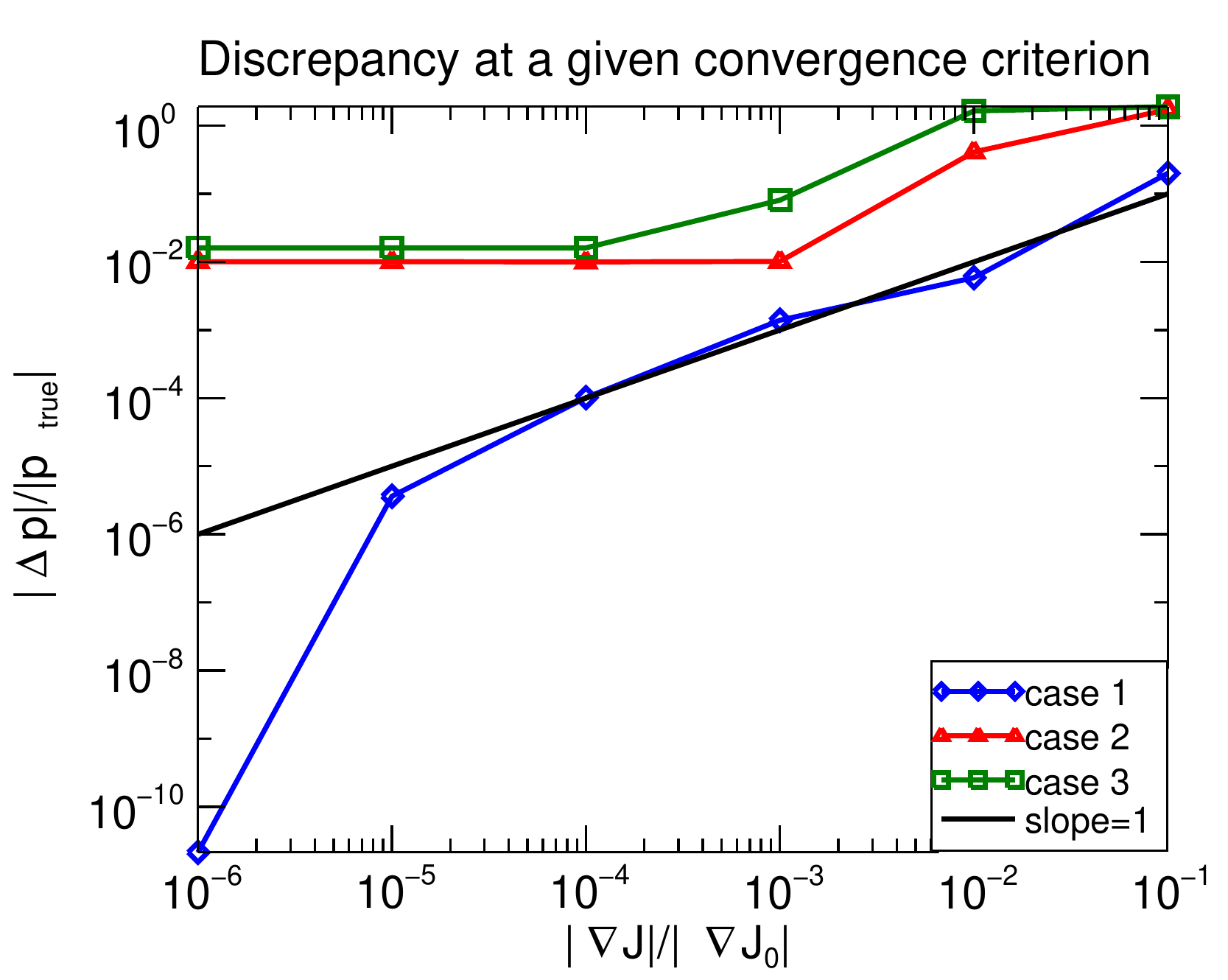}\label{fig:graddp}}\quad%
  \sidesubfloat[]{\includegraphics[width=0.4\columnwidth]{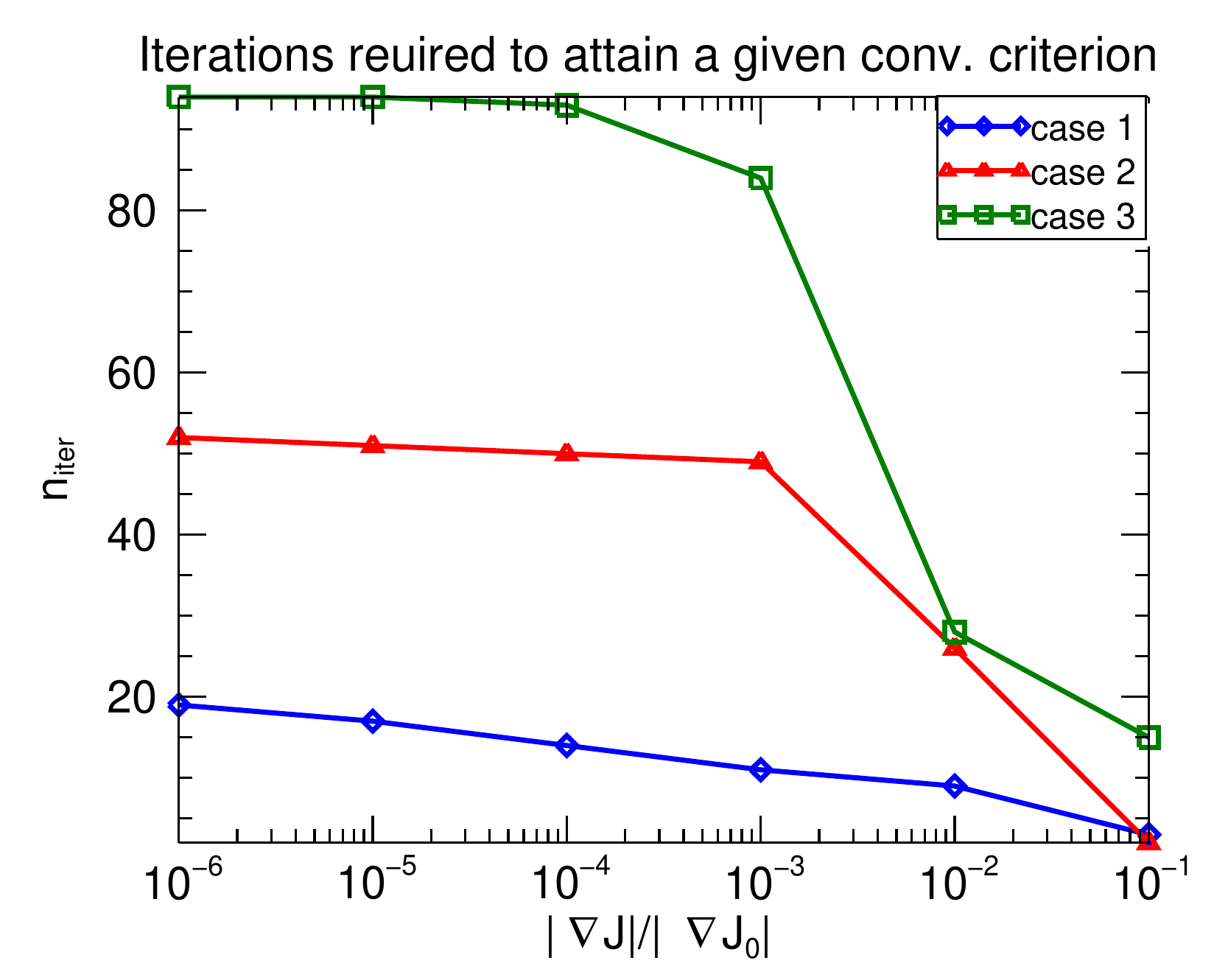}\label{fig:gradjiter}}\quad%
   \caption{Accuracy of the estimated parameters (panel (a)) and required number of iterations (panel (b)) at different convergence criteria. The discrepancy $\Delta p/p$ decreases with the criteria $|\nabla \mathcal{J}|/|\nabla \mathcal{J}_0|$ (the line of slope 1 is shown for reference).  On panel (b), note the region of quadratic convergence as the criterion drops below $10^{-3}$.
}
\label{fig:dpvsgradJcase}
\end{figure*}

We also investigate here the relationship between the accuracy of the forecast parameters $\secrevii{{d_{ij}}}$, defined in equation \eqref{eq:dp} and the convergence criterion. \secrev{Plots of $\Delta p/p$ against $|\nabla \mathcal{J} |/|\nabla \mathcal{J}_0|$ are shown in Figure~\ref{fig:dpvsgradJcase}.} 
In all cases, $\Delta p/p$ decreases as the preset criteria is lowered, although in cases 2 and 3, the discrepancy cannot get lower then $\sim 1\%$. Indeed, as mentioned above, \secrev{the true flow structure in cases 2 and 3 are more complicated 
than in case1, making the minimization more difficult.} 
The number of iterations required for various criteria are shown in Figure~\ref{fig:dpvsgradJcase}(b). It shows that when  $|\nabla \mathcal{J} |/|\nabla \mathcal{J}_0|<10^{-3}$ convergence becomes quadratic and a few further iterations can decrease the gradient by 2 to 3 orders of magnitude.

\floatsetup[figure]{style=plain,subcapbesideposition=top}

\section{Results using data with noise}
\label{sec:twinnoise}

In this section, we move to a more realistic situation where we carry out the same twin experiments but with noised data. The synthetic observations are perturbed by a normally distributed random noise, with standard deviation being a factor of the r.m.s. of the magnetic field/vector potential at steady state, denoted by $\epsilon$ in the preceding sections. 

\subsection{Dependence of convergence behavior on the noise level}
\label{subsec:noiselevel}

\floatsetup[figure]{style=plain,subcapbesideposition=top}
\begin{figure*}[!ht]
  \sidesubfloat[]{\includegraphics[width=0.4\columnwidth]{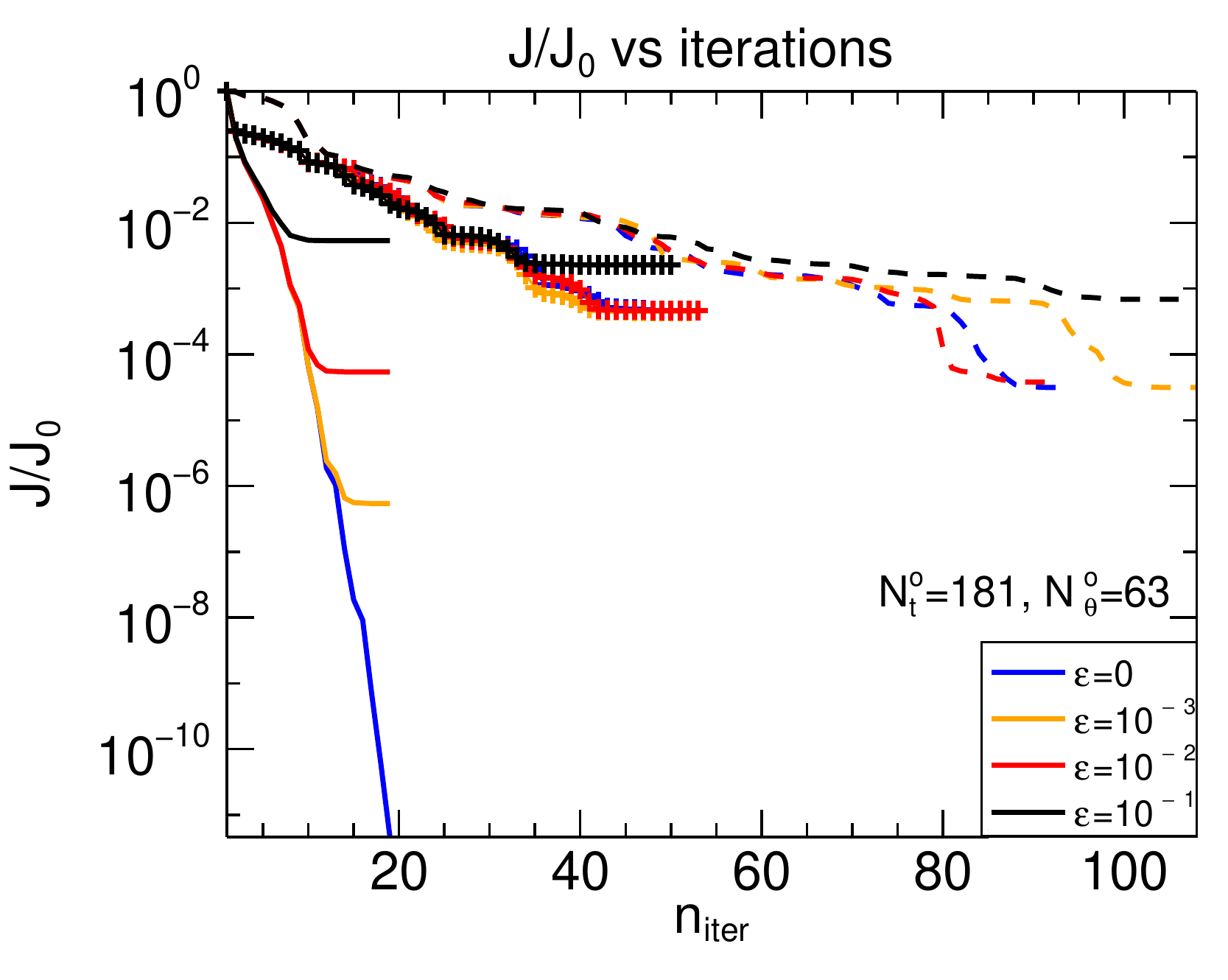}\label{fig:Jvsiter}}\quad%
  \sidesubfloat[]{\includegraphics[width=0.4\columnwidth]{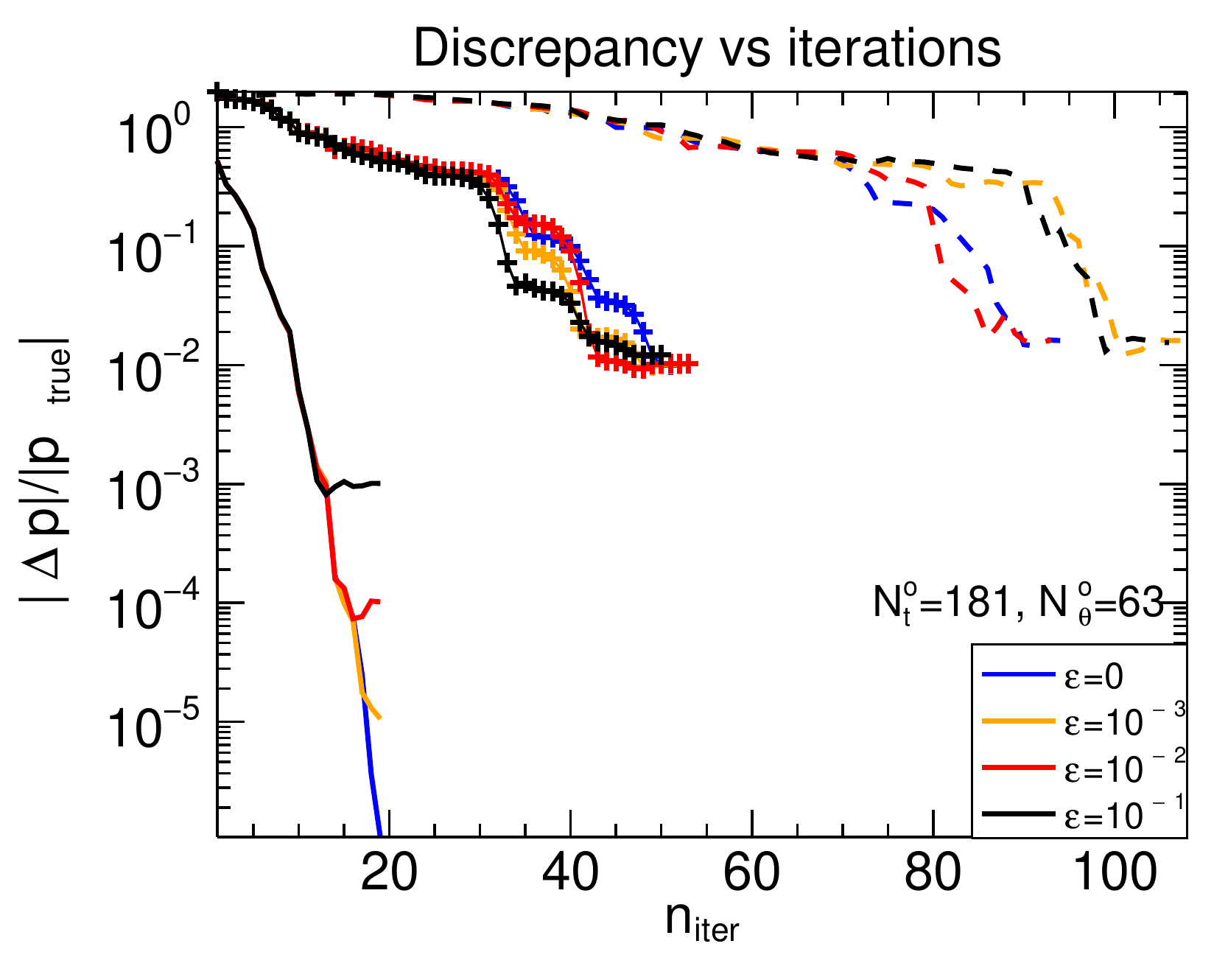}}\label{fig:dpvsiter}\quad%
   \caption{Objective function (left) and  $\Delta p/p$ (right) as the minimization evolves, with the reference sampling, at different noise levels. 
Note that in case 1 when the noise increases, the curves of objective function and $\Delta p/p$ starts to level off at higher values.}
   \label{fig:Jdpvsiter}
\end{figure*}

For our reference sampling, we carry out assimilation with synthetic observations noised with different values of $\epsilon$ in cases 1, 2 and 3. The evolution of the objective function and the discrepancy as the minimization proceeds are shown in Figure~\ref{fig:Jdpvsiter}. The \firstrev{normalized} objective function $\firstrevii{\mathcal{J}/\mathcal{J}_0}$ \firstrev{[panel (a)]} at the optimal parameters is no longer exactly zero, but is positive, increasing as the noise level increases. Similarly, $\Delta p/p$ at the optimum
 also increases with the noise level. The number of iterations required is slightly more than most of the corresponding 
perfect (error-free) situations, i.e., it takes $\sim 20$ iterations for the simplest unicellular case to converge, about $\sim 60$ for case 2 and $\sim 110$ for case 3 where the true meridional flow becomes more complex. 

\secrev{Note that for case 1, as the noise level $\epsilon$ increases, the optimization remains successful: the normalized objective function reaches a value proportional to $\epsilon^2$, as it should in a pure linear situation. For cases 2 and 3, the effect of noise on assimilation is less straightforward, for the reasons outlined above in the perfect data case (see Sec. \ref{subsec:samwin}). 
}
This contributes to the misfit of data at the end of the assimilation in addition to the artificially added noise.
Another feature is the relationship between the residual discrepancy $\Delta p/p$ and the noise level $\epsilon$. For case 1, it is visible in Figure~\ref{fig:Jdpvsiter}(b) that $\Delta p/p$ increases linearly with the noise level, i.e., the accuracy decreases linearly as the noise level is increased. The linear relationship implies that the change in magnetic field with respect to that of the norm of the control vector \{$d_{i,j}$\} is linear. Moreover, the discrepancy when no noise is added, is nonzero. This is because the optimization terminates when the preset convergence criterion is reached, and as the criteria is finite, there is a finite residue at the end of the assimilation. However, for cases 2 and 3, such linear relationship is not obvious and the discrepancy increases very slowly with $\epsilon$ [barely observable in Figure~\ref{fig:Jdpvsiter}(b)], since, as stated before, \secrev{the minimization for more complicated flow profiles is more challenging using a simple unicellular flow initial guess.} 
Therefore, in cases 2 and 3, such a linear relationship rooted from the property of the model is hindered. 

\subsection{Distributions of the deviations of the field}
\label{subsec:distdev}

In this subsection, we investigate how the predicted $A_\phi$ deviates from
 the true one based either on the
 initial guess of $\myvect{v}_p$ or on its final estimate after assimilation, 
 considering the $N^o$ observations that are made. 
  We shall refer to these differences as the innovations in the former case 
  and to the residuals in the latter case. 

\begin{figure}[!ht]
\includegraphics[width=\columnwidth]{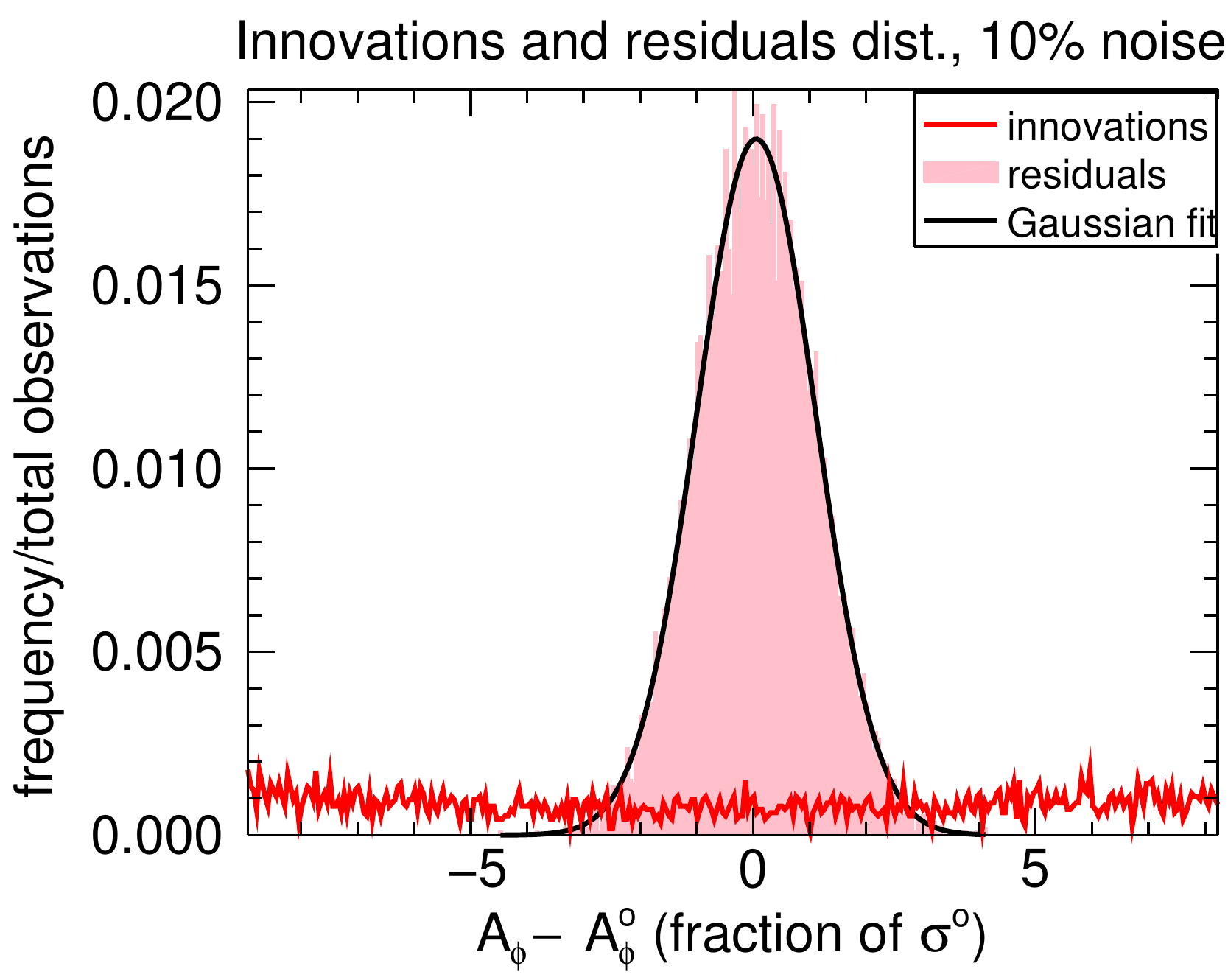}
 \caption{Distribution of the misfit (both innovations and residuals) with noise level of $10\%$ r.m.s. in case 3. The distribution is fitted with a Gaussian (with least square fit, shown in black line), the corresponding average and standard deviation are listed in Table~\ref{tab:avestd}.}
\label{fig:priorresidue3}
\end{figure}

As in the previous sections, the prior (or initial guess) in case 1 is a unicellular flow producing a magnetic cycle of $\sim 44$ years while for cases 2 and 3 the prior is also a unicellular flow but producing a 22-yr cycle. Again, the reference sampling (1.5 solar cycles, sampling every month) is used. The distribution of the deviations $A_{\phi}-A_{\phi}^o$ for the most complicated case 3 is shown in Figure~\ref{fig:priorresidue3}. The noise added to the data is $\epsilon=10\%$ for this example. Initially, the innovations are broadly distributed, showing the large discrepancy between the true trajectory and the initial state. After assimilation, the residuals show a peak which resembles a Gaussian (verified with a least square fit shown in black line in the figure), with a kurtosis of $3.05$ (a value of 3 is expected for a perfect Gaussian) and a skewness of $1.04\times 10^{-2}$ (zero is expected for unbiased distribution). The corresponding average and standard deviation are consistent with the settings of our twin experiment, i.e., zero mean and $\sigma = 10\%$ r.m.s. (Table~\ref{tab:avestd}). The normalized misfit is $1.03$. In this example, we illustrate that the assimilation algorithm not only minimizes the misfit and gives an estimate of the meridional circulation close to the true one (Sec.~\ref{subsec:noiselevel}), but also correctly recovers the normal distribution of the synthetic noise, with a normalized misfit being statistically consistent (i.e., $\sim 1$).

\begin{table*}[!ht]
\caption{The average and standard deviation of $A_{\phi}-A_{\phi}^o$, for an assimilation 
with the reference sampling. Shown are the statistics of the innovations (prior to assimilation)
 and those of the residuals (after completion of assimilation). Averages and
 standard deviations are normalized by the root mean square of $A_{\phi}$ of the reference solution. 
The values for the residuals are consistent with the noise added to the data (i.e., zero average, $\sigma = 10\%$ r.m.s.) and the normalized misfits are close to 1.}\label{tab:avestd}
\begin{tabular}{crrrrrr}
\hline \hline
 Case &            Mean &  Stand. dev.    & Norm. misfit        &  Mean residual & Stand. dev.          &   Norm. misfit \\
      &      innovation &  of innovations & before assim.       &                &  of  residuals       &   after assim. \\
\hline                                        
  1   &  $   0.72$ &      $152$  \% & $13.6$                  & $-1.89$ $10^{-4}$&    $9.86\%$   &     $0.995$     \\
\hline                                        
  2   &  $  -1.61 $&     $164$ \% &   $23.2$                & $2.65$  $10^{-2}$&    $11.2\%$    &     $1.11 $  \\
\hline                                       
  3   & $  -0.99$  &   $524$ \% &     $39.3$             & $9.27$  $10^{-3}$&     $10.4\%$&     $1.03$       \\
\hline                                        

\end{tabular}
\end{table*}

\floatsetup[figure]{style=plain,subcapbesideposition=top}
\begin{figure}[!ht]
\includegraphics[width=\columnwidth]{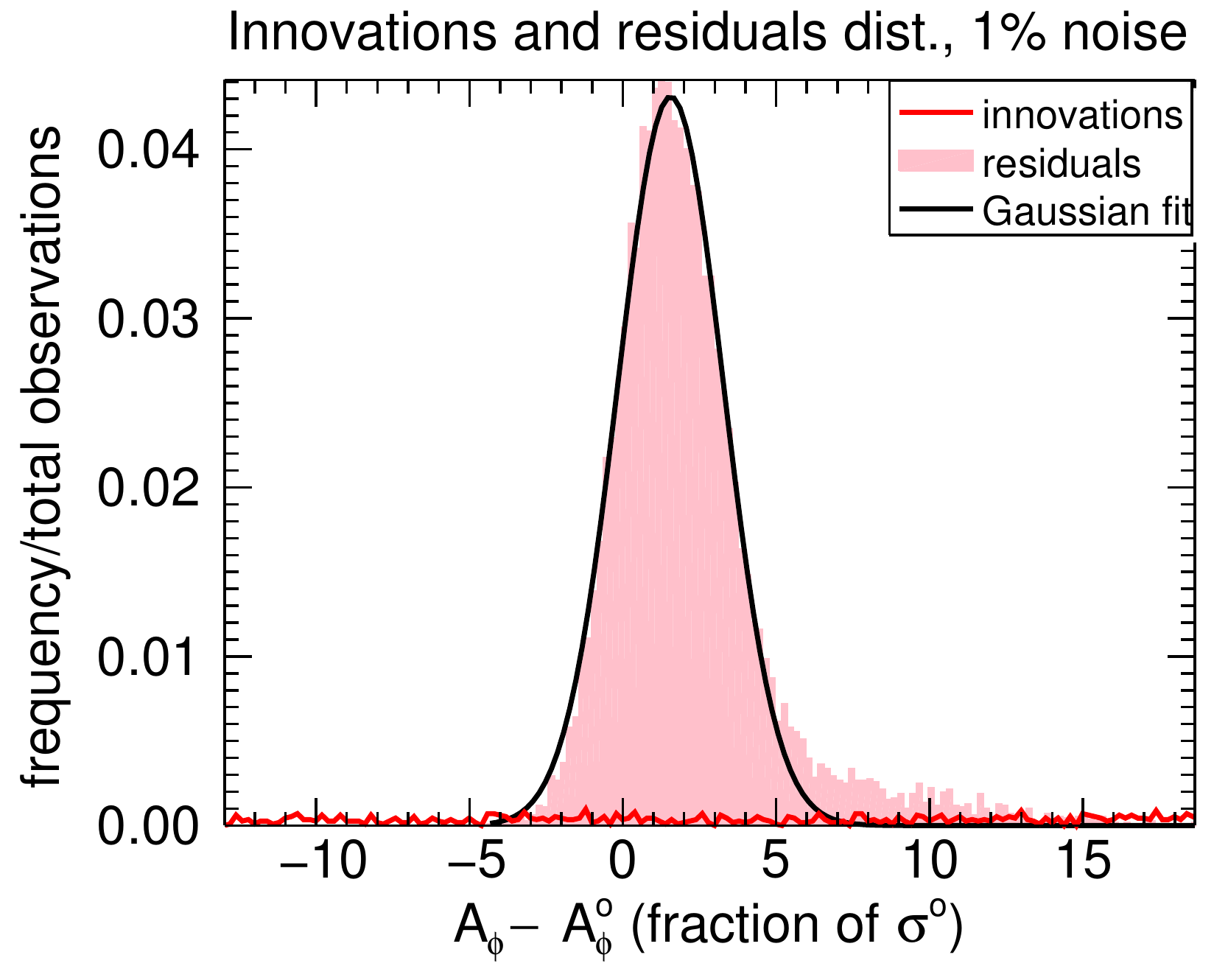}
\caption{Same as figure \ref{fig:priorresidue3}, for a latitudinal sampling restricted to  the activity band only
 and a noise level of $\epsilon=1\%$. Note that the distribution of residuals slightly deviates from a Gaussian distribution.}
\label{fig:distactband}
\end{figure}

We now consider a more realistic situation, where the field in the activity band only is observed (keeping the temporal sampling the same as the reference), for cases 2 and 3. We plot the distribution in Figure~\ref{fig:distactband} for case 3 (case 2 is similar and thus not shown), 
with a noise level of $\epsilon=1\%$. Starting from a broadly distributed innovations as in the previous case, 
the residuals crowd again around zero. However, the standard deviations are now \secrev{larger} than that 
of the synthetic noise: $2.86$ for case 2 and $2.32$ for case 3 (where it should be 1 for an ideal situation). 
Moreover, the distributions depart somewhat from pure Gaussians and are biased, with a kurtosis $=8.09$ indicative 
of extended wings and a skewness $=1.79$ for the case 3 shown in Figure~\ref{fig:distactband}.  The corresponding 
normalized misfits are $3.01$ and $2.71$ respectively, which is an indication of under fitting. The estimated 
meridional circulation [shown in Figure~\ref{fig:flowforecastcase3drNSH}(c)] is nevertheless still close 
to the truth. Therefore, it is still possible to estimate a complex multi-cellular flow in this more realistic 
example of nonuniform sampling and noised data, even if the statistics of the residuals clearly indicate that the recovery is not perfect.

\begin{figure}[!ht]
\includegraphics[width=\columnwidth]{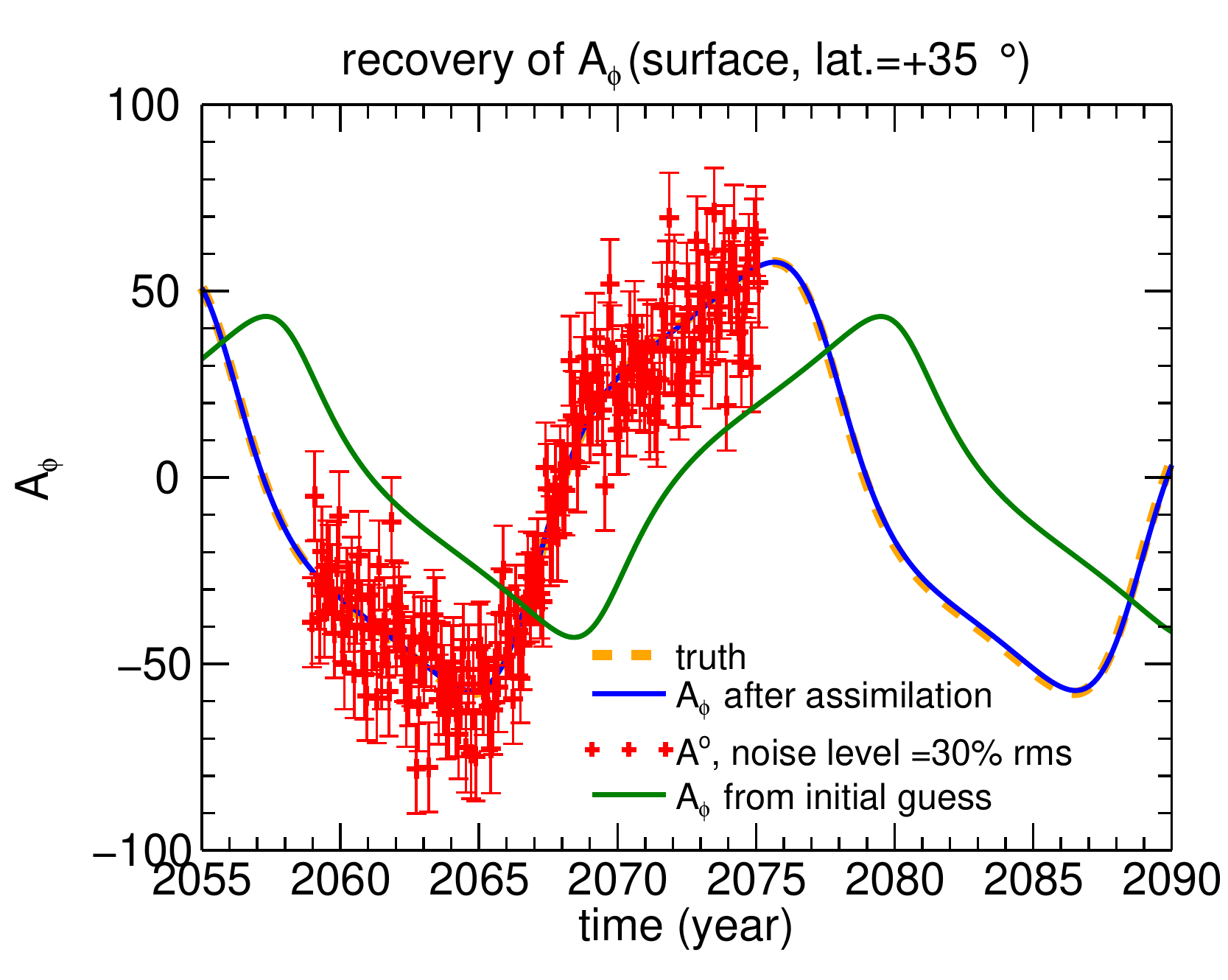}
\caption{Same as Fig.\ref{fig:recoveredsample}, save that synthetic data are noised 
with a noise level of $30\%$ relative to the r.m.s.. Note that the deviation grows slowly as time evolves.}
\label{fig:recoveredsamplenoise}
\end{figure}

Finally, we show in Figure~\ref{fig:recoveredsamplenoise} the recovered field of the assimilation for case 3, when the data files are noised with $\epsilon=30\%$. The trajectory obtained from the initial guess is shown in green, the final forecast after assimilation in blue and the true trajectory in orange. Note that the forecast trajectory only deviates by a small amount from the true trajectory as time evolves.

\bigskip 

\bigskip 

\section{Discussion and Conclusions}
\label{sec:conclu}

In this study, a first step towards predicting future solar activity using data assimilation has been presented. A variational data assimilation technique has been applied to a mean-field axisymmetric flux-transport Babcock-Leighton dynamo model, widely used in the community to reproduce key properties of the real solar cycle. As a proof of concept, we have focused our study on the estimation of the meridional circulation by assimilating magnetic proxies into our data assimilation procedure. We have successfully adjusted a control vector representing the expansion coefficients of the meridional flow onto radial and latitudinal functions such as to minimize the deviations (misfit) of the outputs of the model to synthetic magnetic observations.

Using twin experiments where observations are produced by the model itself (but where noise can be added to perturb the data), we show that with adapted sampling of different components of the magnetic field, and starting from a classical unicellular meridional circulation, we are able to minimize the misfit to the observations and recover a complex flow profile, with multiple cells and asymmetry with respect to the equator. By performing a systematic study of the effects of the spatial and temporal sampling patterns on the efficiency of the assimilation method, we find an optimal sampling with an observational window of width 1.5 cycles, uniform sampling in latitude with $\Delta \theta \sim 3^\mathrm{o}$ and monthly observations for almost all cases considered. An interesting aspect of this systematic study however, is that observing in one hemisphere only or in the activity belt only \firstrev{can produce flow reconstructions} of reasonably good agreement with the true state, even for a complex flow structure. When noise is added to the observational data, a normalized misfit close to $1$ is found in the optimal case, showing that the true state is very well recovered. When an even more realistic case is considered (complex flow and $30\%$ noise or activity band sampling), minimization remains effective and the estimated flow is still reasonable, though the residuals deviate slightly from Gaussian distribution, with a higher normalized misfit. Overall, proof is made that the assimilation method is successful throughout our systematic studies.

The variational technique can be already very useful at testing the sensitivity of various outputs of the models to poorly constrained input parameters, such as the meridional flow profile. Indeed, below $\sim 40$Mm, inversions based on helioseismic techniques such as ring diagram or time distance analysis do not seem to give consistent results \citep{Zhao13,2015ApJ...805..133J}. We showed that the high sensitivity of the components of the magnetic field to the flow structure and amplitude in flux-transport Babcock-Leighton models makes it possible to estimate the meridional circulation profile by assimilating magnetic field observations. If such a model is a good representation of what actually happens in the Sun, this technique is extremely promising to be able to much better constrain this internal flow.

As far as predictions of future solar activity is concerned, our present studies are still limited at a stage of proof of concept based on twin experiments where the assimilated data is not real solar observations. However, we used magnetic outputs which are believed to be good proxies for real magnetic measurements at the solar surface like sunspot distributions, polar fields or structure of the butterfly diagram. 
Note that we did use a direct measure of the toroidal field at the base of the convective envelope and that in reality one would have to construct an operator that adequately relates the observed surface field to the deeply anchored and hidden magnetic field. Defining such an 
operator will be the subject of our next study. 

At this stage, the meridional circulation in our model is expressed as an expansion on basis functions, which defines the flow globally in the whole convection zone. This basis was limited to 2 sinusoidal radial functions and 4 Legendre polynomials in latitude in the present analysis for practical purpose as nothing prevents us to go beyond. This global definition obviously implies some symmetries and possibly artificial coupling between hemispheres, which would probably not be the case with a point-wise definition. The latter would of course involve much more coefficients to recover and would require to impose a constraint that the recovered flow is  divergence-free. When projecting surface meridional circulation onto associated Legendre polynomials, as we have done using data from  \cite{2010ApJ...725..658U}, we find that small velocity structures contain little power compared to the global low order modes. Hence even though our \secrev{strong formulation} to recover the meridional circulation has its own limitation, we do capture 
the dominant components, which encourages us to apply it to real solar data. Note also that in the present work 
we have deliberately chosen to initialize our assimilation procedure with a magnetic field which was only weakly related to the solution that
we were looking for (recall Sec.~\ref{sec:initialguess} and Appendix~\ref{sec:initialcon}). 
 In an operational forecasting procedure, one would use instead the magnetic field obtained from the 
previous assimilation cycle, which would presumably give rise to a faster convergence to the optimum.

In this work, only synthetic observations with constant cycle period were used. This deviates from the activity of the real Sun, which presents strong modulation of its magnetic cycle, both in duration and amplitude. 
If we want to be able to take into account the temporal modulation of 
the solar cycle (which is obviously needed if we want to make predictions), the next step is 
to produce a model with such modulations. This is possible in the mean-field dynamo framework if stochastic 
fluctuations of the dynamo coefficients or on the meridional flow are introduced \citep[see for example][]{Charbonneau00,Ossendrijver02} or 
if the back-reaction of the magnetic field on the flow is considered through the Malkus-Proctor 
effect \citep[e.g.][]{Malkus75,Tobias97,Moss00,Bushby06,2006ApJ...647..662R}. The idea would then be to assimilate synthetic data 
based on a model with a time dependent meridional flow, such that the auto-correlation of the modeled 
flow is similar to that of the observed flow in the real Sun. The predictive skills, predictability limit and sensitivity 
of these models to perturbations of the input parameters have to be studied, to gain some insights 
 on the reliability of predictions 
which could be provided. Incorporating real solar data coming from instruments onboard satellites like \emph{Hinode, Stereo, SDO} 
and the future \emph{Solar Orbiter} mission into a well-suited model is then the ultimate goal of our work. If flux-transport models are 
valid to explain the large-scale solar magnetic activity, they should enable us to produce a quantitative estimate of the 
meridional flow in the solar interior. Based on such an estimate, predictions of the timing, amplitude and shape of the next 
solar cycle will be made and compared with that provided by other existing methods 
relying on geomagnetic precursors or other statistical estimates \citep[see for instance,][]{Hathaway10,Petrovay10}.

\appendix
\section{The Babcock-Leighton mean-field dynamo model}\label{sec:blmodel}
In this section, we present the Babcock-Leighton dynamo model and the corresponding physical ingredients we adopted for the assimilation for reader's reference. \par

We start from the induction equation to model the solar dynamo, describing the evolution of large scale magnetic field $\myvect{B}$,
\begin{equation}
\label{eq:MHDKD}
\partial_t \myvect{B}= \nabla \times (\myvect{v} \times \myvect{B} ) - \nabla \times (\eta \nabla \times \myvect{B}),
\end{equation}
where $\eta$ is the effective magnetic diffusivity. We adopt a kinematic formulation, i.e. the velocity field $\myvect{v}$ is prescribed instead of being a dynamical variable.
By introducing the spherical coordinate system and assuming axisymmetry, we rewrite the magnetic field and velocity field as a sum of poloidal and toroidal components:
\begin{equation}
\label{magcom}
\myvect{B}=B_{\phi} \myhat{e_{\phi}} + \nabla \times (A_{\phi} \myhat{e_{\phi}}) \text{ and }
\myvect{v} (r,\theta)=\myvect{v}_p(r,\theta)  + r \sin \theta \Omega(r,\theta) \myhat{e_{\phi}},
\end{equation}
where $B_{\phi}$ is the toroidal field and $A_{\phi}$ is the poloidal potential and
 $\Omega(r,\theta)$ is the differential rotation. \par

Then the induction equation \eqref{eq:MHDKD} can be rewritten in poloidal and toroidal components as

\begin{equation}\label{eq:Adyn}
\begin{split}
\partial_t A_{\phi} =\frac{\eta}{\eta_t}  \left(\nabla^2 - \frac{1}{\varpi^2}\right) A_{\phi} 
-Re \frac{\myvect{v}_p}{\varpi}\cdot \nabla (\varpi A_{\phi})  
 + C_sS(r, \theta, B_{\phi}), 
\end{split}
\end{equation}

\begin{equation}\label{eq:Bdyn}
\begin{split}
\partial_t B_{\phi} =\frac{\eta}{\eta_t}  \left(\nabla^2 - \frac{1}{\varpi^2}\right) B_{\phi} 
+\frac{1}{\varpi} \frac{\partial (\varpi B_{\phi})}{\partial r} \frac{\partial (\eta/\eta_t)}{\partial r} 
-Re \varpi \myvect{v}_p\cdot \nabla \left( \frac{B_{\phi}}{\varpi }\right) 
- Re B_{\phi} \nabla \cdot \myvect{v}_p 
+ C_{\Omega} \varpi \left[ \nabla \times (A_{\phi} \myhat{e_{\phi}}) \right] \cdot \nabla \Omega  , 
\end{split}
\end{equation} 
where $\varpi=r \sin \theta$. The domain is $r \in [0.6,1]$ and $\theta \in [0,\pi]$ as specified in Sec.~\ref{sec:modeleq}. The toroidal field $B_{\phi}=0$ at the boundary of the domain, and for $A_{\phi}$, we impose the pure radial field approximation at the surface, i.e., $\partial_r (r A_{\phi})=0$ at $r=1$, and $A_{\phi}=0$ on all the other boundaries. Here and in the following, the length is normalized with solar radius $R_s$, time is normalized with the diffusive time scale $R_s^2/{\eta}_t$ where ${\eta}_t$ is the envelope diffusivity. Using this normalization, we introduce 3 dimensionless parameters, namely the Reynolds number based on the meridional flow speed $Re=R_su_o/\eta_t$, the strength of the Babcock-Leighton source $C_s=R_ss_o/\eta_t$ and the strength of the $\Omega$-effect $C_{\Omega}=\Omega_oR_s^2/\eta_t$, with $u_o$ and $s_o$ given in Table~\ref{tab:modelpara} and $\Omega_o = 2\pi \times 456\mathrm{nHz}$.

\begin{table}[H]
\caption{Parameters of the 3 models being studied. Unicellular (case 1), 4 cells (case 2), asymmetric (case 3).}
\label{tab:modelpara}
 \begin{tabular}{c*{6}{c}}
 \hline \hline
 Case     & Resolution             & Time step  & $u_o$       & $\eta_t$         & $s_o  $  &  Cycle period   \\
          & $n_r \times n_{\theta}$&            & $(cms^{-1})$& $(cm^2s^{-1})$   & $(cm s^{-1})$& (yrs)   \\
 \hline
 1        & $128^2$                  & $10^{-6}$    & 690         & $10^{11}$          & 50 & 22.0   \\
 2        & $128^2$                  & $10^{-6}$    & 1379        & $2 \times 10^{11}$ & 201& 21.6   \\
 3        & $128^2$                  & $10^{-6}$    & 1034        & $2.4 \times 10^{11}$ & 17.2& 21.7   \\
 \hline
 \end{tabular}%

\end{table}

The physical ingredients for the model \secrev{include} a differential rotation which generates the toroidal magnetic field from the poloidal field:
\begin{equation}
\label{eq:diffrot}
\Omega(r,\theta)=\Omega_c + \frac{1}{2}(1-\Omega_c-c_2\cos^2\theta)\left[ 1+\tanh\left(\frac{r-r_c}{d_1} \right) \right],
\end{equation}
with $\firstrevii{d_1=0.016}$, $\firstrevii{r_c=0.7}$ \firstrev{(base of the convection zone)}, $\Omega_c=0.92$ and $c_2=0.2$,

the Babcock-Leighton source of poloidal field, with a quenching term to prevent the magnetic energy from growing exponentially:
\begin{equation}
\label{eq:BLS}
\begin{split}
S(r,\theta,B_{\phi})=\frac{1}{2}\left[ 1+\tanh\left(\frac{r-r_2}{d_2} \right) \right]\left[ 1-\tanh\left(\frac{r-1}{d_2} \right) \right]\cos\theta \sin\theta 
                      \left\{ 1+ \left[ \frac{B_{\phi}(r_c,\theta,t)}{B_0} \right]^2\right\}^{-1}B_{\phi}(r_c,\theta,t), 
\end{split}
\end{equation}
with $d_2=0.008$, $r_2=0.95$ and $B_0=10^4$. Note the dependence of toroidal field at the base of the convection zone results in a nonlocal source term,

and the magnetic diffusivity which is given by
\begin{equation}
\label{eq:etaprofile}
\frac{\eta}{\eta_t}=\frac{\eta_c}{2\eta_t}\left[1-\tanh\left(\frac{r-r_c}{d_1}\right)\right]+\frac{1}{2}\left[1+\tanh\left(\frac{r-r_c}{d_1}\right)\right],
\end{equation}
where $\eta_c=10^9$ cm$^2$~s$^{-1}$.

Another important ingredient is the meridional flow $\myvect{v}_p$ which advects the magnetic field in the meridian plane. Since this ingredient is at the center of this present study, it is specified in the main body of the text, in Sec.~\ref{sec:modeleq}.


\section{Derivation of the adjoint Babcock-Leighton model}\label{sec:variapp}
\label{sec:app_adj}

In this appendix, we follow and adapt the procedure described by \cite{Talagrand03} in order to derive 
the adjoint dynamo model needed to express efficiently the sensitivity of the 
 objective function to its control vector. The novelty here with respect to the previous derivation by \cite{JouveAssimi11}
 stands in the fact that we are operating in spherical geometry with a Babcock-Leighton flux transport dynamo model, as opposed to 
  in Cartesian geometry with a simpler $\alpha-\Omega$ dynamo model. 
  Let us consider the coupled induction equations \eqref{eq:Adyn} 
and \eqref{eq:Bdyn} for the fields $A_{\phi}$ and $B_{\phi}$. We look for solutions over the 
domain $D = [r_{bot}, r_{top}] \times [0, \pi] \times [0, 2\pi\firstrevii{]} \times [t_s,t_e]$ in the $(r, \theta, \phi, t)$ -space. These equations are first order in time and second order in space. \par

Consider $A_{\phi}^o(\myvect{r}, t)$ and $B_{\phi}^o(\myvect{r}, t)$ as our observations over the domain $D$. 
Since we assimilate data on the toroidal and poloidal fields, our objective function 
\begin{equation}\label{eq:jvol}
\mathcal{J}=
\frac{1}{2} \int  \mathrm{d}^3\myvect{r} \int \mathrm{d}t \left[
(B_{\phi}-B_{\phi}^o)^2+(A_{\phi}-A_{\phi}^o)^2
\right]
,
\end{equation}
where the spatial integration is over the domain of the dynamo model described in Sec.~\ref{sec:blmodel}. The system considered 
is axisymmetric, so that integration with respect to $\phi$ is equivalent to multiplication by a factor of $2\pi$.

We aim to express the variations of the objective function $\mathcal{J}$ subject to variations of $A_{\phi}$ and $B_{\phi}$ for all points in space at the initial time $t = t_s$, as well as to variations in the meridional flow $\myvect{v}_p$. 
 Such variations are constrained by the dynamo equations. 
Let us linearize and differentiate equations \eqref{eq:Adyn} and \eqref{eq:Bdyn} with respect to $A_{\phi}$, $B_{\phi}$ and $\myvect{v}_p$. The corresponding 
 equations write 
\begin{equation}\label{eq:Atl}
\begin{split}
\partial_t \delta A_{\phi} 
=&
\frac{\eta}{\eta_t}  \left(\nabla^2 - \frac{1}{\varpi^2}\right) \delta A_{\phi}
-Re \frac{\delta \myvect{v}_p}{\varpi}\cdot \nabla (\varpi A_{\phi})
-Re \frac{\myvect{v}_p}{\varpi}\cdot \nabla (\varpi \delta A_{\phi})
\\&
+ C_s f(r) g(\theta) \frac{1-B_{\phi}^2(r_c,\theta,t)/B_0^2}{\left[1+ B_{\phi}^2(r_c,\theta,t)/B_0^2\right]^2}\delta B_{\phi}(r_c,\theta,t),
\end{split}
\end{equation}

\begin{equation}\label{eq:Btl}
\begin{split}
\partial_t \delta B_{\phi} 
=&
\frac{\eta}{\eta_t}  \left(\nabla^2 - \frac{1}{\varpi^2}\right) \delta B_{\phi} 
+\frac{1}{\varpi} \frac{\partial (\varpi \delta B_{\phi})}{\partial r} \frac{\partial (\eta/\eta_t)}{\partial r} 
  -Re \varpi \delta \myvect{v}_p\cdot \nabla \left( \frac{B_{\phi}}{\varpi }\right)  
\\&  
- Re \delta B_{\phi} \nabla \cdot \myvect{v}_p 
- Re B_{\phi} \nabla \cdot \delta \myvect{v}_p 
+ C_{\Omega} \varpi \left[ \nabla \times (\delta A_{\phi} \myhat{e_{\phi}}) \right] \cdot \nabla \Omega, 
\end{split}
\end{equation} 
where we use the explicit expression of the Babcock-Leighton source \eqref{eq:BLS}, and the variation is up to first-order 
in $\delta B_{\phi}$. \par
The variations of ${\mathcal J}$ are subject to the constraints defined by these last two equations. 
 Consequently, we introduce the Lagrange multipliers $-A_{\phi}^*(\myvect{r},t)$ and $-B_{\phi}^*(\myvect{r},t)$, respectively, 
 for equations \eqref{eq:Atl} and \eqref{eq:Btl}. The notations $A_{\phi}^*$ and $B_{\phi}^*$ are used so that  when 
 the derivation proceeds, we can identify them as defining the adjoint 
 magnetic field ($A_{\phi}^*$ and $B_{\phi}^*$ being the adjoint poloidal potential 
and toroidal field, respectively). We get 
\begin{equation}\label{eq:varJ}
\begin{split}
\delta \mathcal{J}
=&
\int \mathrm{d}^3\myvect{r} \int \mathrm{d}t \Bigg\{    
(B_{\phi}-B_{\phi}^o)\delta B_{\phi} + (A_{\phi}-A_{\phi}^o)\delta A_{\phi}
\\&-A_{\phi}^*\Bigg[
\partial_t \delta A_{\phi} 
-\frac{\eta}{\eta_t}  \left(\nabla^2 - \frac{1}{\varpi^2}\right) \delta A_{\phi}
+Re \frac{\delta \myvect{v}_p}{\varpi}\cdot \nabla (\varpi A_{\phi})
+Re \frac{\myvect{v}_p}{\varpi}\cdot \nabla (\varpi \delta A_{\phi})
\\&- C_s f(r)g(\theta) \frac{1-B_{\phi}^2(r_c,\theta,t)/B_0^2}{\left[1+ B_{\phi}^2(r_c,\theta,t)/B_0^2\right]^2}\delta B_{\phi}(r_c,\theta,t)
\Bigg]
\\&-B_{\phi}^*\Bigg[
\partial_t \delta B_{\phi} 
-\frac{\eta}{\eta_t}  \left(\nabla^2 - \frac{1}{\varpi^2}\right) \delta B_{\phi} 
-\frac{1}{\varpi} \frac{\partial (\varpi \delta B_{\phi})}{\partial r} \frac{\partial (\eta/\eta_t)}{\partial r} 
  +Re \varpi \delta \myvect{v}_p\cdot \nabla \left( \frac{B_{\phi}}{\varpi }\right)  
\\&  +Re \varpi \myvect{v}_p\cdot \nabla \left( \frac{\delta B_{\phi}}{\varpi }\right)  
+ Re \delta B_{\phi} \nabla \cdot \myvect{v}_p 
+ Re B_{\phi} \nabla \cdot \delta \myvect{v}_p 
- C_{\Omega} \varpi \left[ \nabla \times (\delta A_{\phi} \myhat{e_{\phi}}) \right] \cdot \nabla \Omega 
\Bigg]
\Bigg\}. 
\end{split}
\end{equation}
The differential operators acting
 on the variations of $A_{\phi}$, $B_{\phi}$, and $\myvect{v}_p$ can be removed via 
integration by parts, at the expense of introducing boundary integrals, either over the surface $\partial V$ of the spatial domain 
 (we will remove the notation $\partial V$ for clarity after its first introduction), or at the end-points in time $t_s$ and $t_e$. 

For example, for the time derivative and diffusion of $A_{\phi}$, one gets 
\begin{equation}
\label{eq:dtAbypart}
-\int \mathrm{d}^3\myvect{r} \int \mathrm{d}t A_{\phi}^* \partial_t \delta A_{\phi}
=\int \mathrm{d}^3\myvect{r} \int \mathrm{d}t \delta A_{\phi}  \partial_t A_{\phi}^* 
\left. -\int \mathrm{d}^3\myvect{r}  A_{\phi}^* \delta A_{\phi} \right|_{t_s}^{t_e},
\end{equation}
and 
\begin{equation}\label{eq:lapAbypart}
\begin{split}
\int \mathrm{d}^3\myvect{r} \int \mathrm{d}t \frac{A_{\phi}^*\eta}{\eta_t}  \left(\nabla^2 - \frac{1}{\varpi^2}\right) \delta A_{\phi}
=& \int \mathrm{d}^3\myvect{r} \int \mathrm{d}t \delta A_{\phi}  \left(\nabla^2 - \frac{1}{\varpi^2}\right) \frac{A_{\phi}^*\eta}{\eta_t}
+\int \mathrm{d}t \int_{\partial V} \mathrm{d}\myvect{a} \cdot \left( 
\left. 
\frac{A_{\phi}^*\eta}{\eta_t} \nabla \delta A_{\phi} -  \delta A_{\phi} \nabla \frac{A_{\phi}^*\eta}{\eta_t}  
\right)
\right|_{\partial V},
\end{split}
\end{equation}
respectively. In addition, the rearrangements for the nonlocal Babcock-Leighton source term (a novelty of this study) write 
\begin{equation}\label{eq:BLrearr}
\begin{split}
&\int \mathrm{d}^3\myvect{r} \int \mathrm{d}t A_{\phi}^* C_s f(r)g(\theta) \frac{1-B_{\phi}^2(r_c,\theta,t)/B_0^2}
{\left[1+ B_{\phi}^2(r_c,\theta,t)/B_0^2\right]^2}\delta B_{\phi}(r_c,\theta,t)
\\&=\int \mathrm{d}^3\myvect{r} \int \mathrm{d}t A_{\phi}^* C_s f(r)g(\theta) \int \mathrm{d}r' \delta(r'-r_c)  
\frac{1-B_{\phi}^2(r',\theta,t)/B_0^2}{\left[1+ B_{\phi}^2(r',\theta,t)/B_0^2\right]^2}\delta B_{\phi}(r',\theta,t)
\\&=\int \sin \theta \mathrm{d}\theta \int \mathrm{d}\phi \int r'^2 \mathrm{d}r' \int \mathrm{d}t A_{\phi}^* C_s f(r')g(\theta) 
\int \mathrm{d}r \delta(r-r_c)  \frac{1-B_{\phi}^2(r,\theta,t)/B_0^2}
{\left[1+ B_{\phi}^2(r,\theta,t)/B_0^2\right]^2}\delta B_{\phi}(r,\theta,t)
\\&=\int \mathrm{d}^3\myvect{r} \int \mathrm{d}t \,\,  C_s g(\theta)\frac{\delta(r-r_c)}{r^2}  
\frac{1-B_{\phi}^2(r,\theta,t)/B_0^2}{\left[1+ B_{\phi}^2(r,\theta,t)/B_0^2\right]^2}
\left[\int \mathrm{d}r' r'^2 A_{\phi}^*(r',\theta,t) f(r') \right] \delta B_{\phi}(r,\theta,t).
\end{split}
\end{equation}
Note the introduction of the $\delta$-function $\delta (r'-r_c)$ in the first right-hand side 
 of equation \eqref{eq:BLrearr}, which should be distinguished from the $\delta$ symbol used to represent variations of field variables. \par

By grouping the terms by variations, we get the following equation for $\delta \mathcal{J}$: 
\begin{equation}\label{eq:rearrdelJ}
\begin{split}
\delta \mathcal{J}
=& \int \mathrm{d}^3\myvect{r} \int \mathrm{d}t \Bigg\{
\Bigg[
\partial_t A_{\phi}^* + \left(\nabla^2 - \frac{1}{\varpi^2}\right) \frac{\eta A_{\phi}^*}{\eta_t} 
+ Re \varpi \nabla \cdot \frac{A_{\phi}^*\myvect{v}_p}{\varpi} 
+ C_{\Omega} \myhat{e_{\phi}} \cdot \nabla (\varpi B_{\phi}^*) \times \nabla \Omega 
+(A_{\phi}-A_{\phi}^o)
\Bigg] \delta A_{\phi}
\\&+\Bigg[
\partial_t B_{\phi}^* + \left(\nabla^2 - \frac{1}{\varpi^2}\right) \frac{\eta B_{\phi}^*}{\eta_t}
+Re \frac{1}{\varpi}  \nabla \cdot (\varpi B_{\phi}^* \myvect{v}_p) 
-\frac{1}{r}\partial_r\left(r B_{\phi}^* \partial_r \frac{\eta}{\eta_t}\right)
-Re B_{\phi}^* \nabla \cdot \myvect{v}_p
\\&+C_s g(\theta) \frac{\delta(r-r_c)}{r^2}  \frac{1-B_{\phi}^2/B_0^2}{(1+ B_{\phi}^2/B_0^2)^2} \int \mathrm{d}r' r'^2 A_{\phi}^*(r',\theta,t) f(r')
+(B_{\phi}-B_{\phi}^o)
\Bigg] \delta B_{\phi}
\\&+Re\bigg[
\nabla (B_{\phi}^* B_{\phi})
-A_{\phi}^* \frac{1}{\varpi} \nabla (\varpi A_{\phi}) - \varpi B_{\phi}^* \nabla \frac{B_{\phi}}{\varpi }
\bigg] \cdot \delta \myvect{v}_p
\Bigg\}
\\& \left. -\int \mathrm{d}^3\myvect{r}  A_{\phi}^* \delta A_{\phi} \right|_{t_s}^{t_e}
+\int \mathrm{d}t \int \mathrm{d}\myvect{a} \cdot \left( 
\left. 
\frac{A_{\phi}^*\eta}{\eta_t} \nabla \delta A_{\phi} -  \delta A_{\phi} \nabla \frac{A_{\phi}^*\eta}{\eta_t}  
-\myvect{v}_p A_{\phi}^* Re  \delta A_{\phi}
+ C_{\Omega} \myhat{e_{\phi}} \times \nabla \Omega B_{\phi}^* \varpi \delta A_{\phi}
\right)
\right|_{\partial V}
\\& \left. -\int \mathrm{d}^3\myvect{r}  B_{\phi}^* \delta B_{\phi} \right|_{t_s}^{t_e}
+\int \mathrm{d}t \int \mathrm{d}\myvect{a} \cdot \left( 
\left.
\frac{B_{\phi}^*\eta}{\eta_t} \nabla \delta B_{\phi} -  \delta B_{\phi} \nabla \frac{B_{\phi}^*\eta}{\eta_t}  
-\myvect{v}_p B_{\phi}^* Re  \delta B_{\phi}
+\myhat{e_r} B_{\phi}^* \delta B_{\phi} \frac{\partial \eta/\eta_t}{\partial r}
-Re  \delta \myvect{v}_p B_{\phi}^* B_{\phi} 
\right)
\right|_{\partial V}.
\end{split}
\end{equation}

The expression is valid for any $A_{\phi}^*(\myvect{r},t)$ and  $B_{\phi}^*(\myvect{r},t)$. The  first
 three lines of the above equations vanish if we require $A_{\phi}^*$ and $B_{\phi}^*$ to 
 satisfy the following partial differential equations: 
\begin{equation}\label{eq:lambdaeqt}
-\partial_t A_{\phi}^* = \left(\nabla^2 - \frac{1}{\varpi^2}\right) \frac{\eta A_{\phi}^*}{\eta_t} 
+ Re \varpi \nabla \cdot \frac{A_{\phi}^* \myvect{v}_p}{\varpi} 
+ C_{\Omega} \myhat{e_{\phi}} \cdot \nabla ( \varpi B_{\phi}^*) \times \nabla \Omega 
+(A_{\phi}-A_{\phi}^o)
,
\end{equation}
and 
\begin{equation}\label{eq:gammaeqt}
\begin{split}
-\partial_t B_{\phi}^* =& \left(\nabla^2 - \frac{1}{\varpi^2}\right) \frac{\eta B_{\phi}^*}{\eta_t}
+Re \frac{1}{\varpi}  \nabla \cdot (\varpi B_{\phi}^* \myvect{v}_p) 
-\frac{1}{r}\partial_r\left( r B_{\phi}^* \partial_r \frac{\eta}{\eta_t}\right)
-Re B_{\phi}^* \nabla \cdot \myvect{v}_p
\\&+C_s g(\theta) \frac{\delta(r-r_c)}{r^2}  \frac{1-B_{\phi}^2/B_0^2}{(1+ B_{\phi}^2/B_0^2)^2} \int \mathrm{d}r' r'^2 A_{\phi}^*(r',\theta,t) f(r')
+(B_{\phi}-B_{\phi}^o)
.
\end{split}
\end{equation}
 
Now we can identify equations \eqref{eq:lambdaeqt} and \eqref{eq:gammaeqt} as 
the adjoint equations of the forward dynamo model \eqref{eq:Adyn} and \eqref{eq:Bdyn}, respectively, 
with the adjoint field variables $A_{\phi}^*$ and $B_{\phi}^*$. Note that the nonlocality of the 
(forward) Babcock-Leighton effect results in a nonlocality of its adjoint and 
in the introduction of the $\delta$-function. Interestingly, in these adjoint
 equations, differential rotation now acts upon $A_{\phi}^*$, while the 
 Babcock-Leighton effect has an imprint on  $B_{\phi}^*$. 
 This is contrary to the `forward' situation and illustrates nicely 
 the general mechanism of `transposition' that forms the backbone of
 any adjoint-based variational approach \citep{Talagrand03}.

Let us now inspect the boundary terms (in space). The boundary conditions of the forward dynamo model
 are that 
\begin{eqnarray}
\partial_r(r A_{\phi}) &=&0\mbox{ at } r=r_{top} ,\\
 A_{\phi}&=&0 \mbox{ at } r=r_{bot}, \mbox{ and} \\ 
B_{\phi}&=&0 \mbox{ at } r=r_{bot} \mbox{ and } r=r_{top}.  
\end{eqnarray}
 The respective variations of these quantities must consequently vanish. We also have
 that $\mathrm{d}\myvect{a}\cdot \myvect{v}_p=0$ on $\partial V$, which suppresses some of the components of the boundary terms 
 appearing in the last two lines of the above equation. 
Now, if we further require that  
\begin{eqnarray}
\partial_r(A_{\phi}^* r \eta/\eta_t)&=&0\mbox{ at } r=r_{top} ,\\
 A_{\phi}^*&=&0 \mbox{ at } r=r_{bot}, \mbox{ and} \\ 
B_{\phi}^*&=&0 \mbox{ at } r=r_{bot} \mbox{ and } r=r_{top}
\end{eqnarray}
at all times, the surface integrals in  \eqref{eq:rearrdelJ} identically vanish. These conditions are the adjoint boundary conditions
 that are naturally associated with the adjoint problem. 
At this stage, the variations of ${\mathcal J}$ reduce to 
 
\begin{equation}\label{eq:intermediateJ}
\begin{split}
\delta \mathcal{J}
=& 
\int \mathrm{d}^3\myvect{r} \, Re\delta \myvect{v}_p \cdot \int \mathrm{d}t \bigg[
\nabla (B_{\phi}^* B_{\phi})
-A_{\phi}^* \frac{1}{\varpi} \nabla (\varpi A_{\phi}) - \varpi B_{\phi}^* \nabla \frac{B_{\phi}}{\varpi }
\bigg]
\left. -\int \mathrm{d}^3\myvect{r}  A_{\phi}^* \delta A_{\phi} \right|_{t_s}^{t_e}
\left. -\int \mathrm{d}^3\myvect{r}  B_{\phi}^* \delta B_{\phi} \right|_{t_s}^{t_e}. 
\end{split}
\end{equation}
The adjoint fields are auxiliary fields whose task is to help us compute the sensitivity 
of ${\mathcal J}$ to its control parameters, and we can conveniently ask them to 
satisfy the following terminal conditions
\begin{equation}
A_{\phi}^* = B_{\phi}^* = 0 \mbox{ at } t=t_e. 
\end{equation}
This leaves us with the following variation of ${\mathcal J}$
\begin{equation}\label{eq:reducerearrdelJ}
\begin{split}
\delta \mathcal{J}
=& 
\int \mathrm{d}^3\myvect{r} \, Re\delta \myvect{v}_p \cdot \int \mathrm{d}t \bigg[
\nabla (B_{\phi}^* B_{\phi})
-A_{\phi}^* \frac{1}{\varpi} \nabla (\varpi A_{\phi}) - \varpi B_{\phi}^* \nabla \frac{B_{\phi}}{\varpi }
\bigg]
+ \left. \left[\int \mathrm{d}^3\myvect{r}  \left( A_{\phi}^* \delta A_{\phi} + B_{\phi}^* \delta B_{\phi}\right)\right]\right|_{t=t_s}
.
\end{split}
\end{equation}
This expression shows that the partial derivatives of ${\mathcal J}$ with respect to $A_{\phi}(\myvect{r},t_s)$,
 $ B_{\phi}(\myvect{r},t_s)$ and $\myvect{v}_p (\myvect{r})$ write respectively 
\begin{equation}\label{eq:partialdelJ}
\begin{aligned}
&\frac {\partial \mathcal{J}}{\partial A_{\phi}} (\myvect{r},t_s)=  A_{\phi}^*(\myvect{r},t_s), \quad
\frac {\partial \mathcal{J}}{\partial B_{\phi}} (\myvect{r},t_s)=  B_{\phi}^*(\myvect{r},t_s),
\\
&\frac {\partial \mathcal{J}}{\partial \myvect{v}_p} (\myvect{r}) =  Re \int \mathrm{d}t \bigg[
\nabla (B_{\phi}^* B_{\phi})
-A_{\phi}^* \frac{1}{\varpi} \nabla (\varpi A_{\phi}) - \varpi B_{\phi}^* \nabla \frac{B_{\phi}}{\varpi }
\bigg]. 
\\
\\
\end{aligned}
\end{equation}
The actual calculations of these derivatives demand in particular
 the knowledge of $A_{\phi}^*(\myvect{r},t)$ and $B_{\phi}^*(\myvect{r},t)$. Those are obtained
 from the integration of the adjoint equations \eqref{eq:lambdaeqt} and \eqref{eq:gammaeqt} subject
 to the boundary and terminal conditions we just discussed. The integration is carried out
 backwards from $t_e$ to $t_s$, which is what makes it stable: the partial time derivative
 is preceded by a minus sign, and the Laplacian on the right-hand sign does not therefore
  lead to instabilities  \citep{Talagrand87}.

 In practice, 
instead of discretizing and numerically integrating the adjoint equations \eqref{eq:lambdaeqt} and \eqref{eq:gammaeqt}, we 
develop the adjoint model directly from the discretized forward model \citep{Talagrand91}.  This
ensures that the computation of the gradient is consistent between the forward and adjoint models. 
 Furthermore, as we observe the magnetic proxies at discrete times and positions, 
and with uncertainties, the driving terms $A_{\phi}-A_{\phi}^o$ and $B_{\phi}-B_{\phi}^o$ in the adjoint model are 
collections of delta functions in space and time, divided by the appropriate variances. 
 This is what we effectively implemented in our optimization routine for the twin experiments. \par 

In this study, as discussed in Sec.~\ref{sec:initialguess}, $ A_{\phi} (\myvect{r},t_s)$ and $ B_{\phi}(\myvect{r},t_s)$ are not 
 included in the control vector. We are therefore left with the sensitivity of ${\mathcal J}$ to the sole
 $\myvect{v}_p(\myvect{r})$. This steady flow 
is mathematically represented by a streamfunction, recall Eq.~\eqref{eq:MC}. It is thus divergence-free, 
i.e., the term $-ReB_{\phi}^*\nabla\cdot\myvect{v}_p$ in equation \eqref{eq:gammaeqt} 
and hence the term $\nabla(B_{\phi}^* B_{\phi})$ in the $2^\mathrm{nd}$ line of equation \eqref{eq:partialdelJ} need not be
 considered. We can then rewrite the variation of the meridional flow in terms of the streamfunction, 
\begin{equation}\label{eq:streambypart}
\begin{split}  
&\int \mathrm{d}^3\myvect{r} \int \mathrm{d}t (-Re)\bigg[
A_{\phi}^* \frac{1}{\varpi} \nabla (\varpi A_{\phi}) + \varpi B_{\phi}^* \nabla \frac{B_{\phi}}{\varpi }
\bigg] \cdot \nabla \times (\delta \psi \myhat{e_{\phi}})
\\=&-Re \int \mathrm{d}^3\myvect{r} \int \mathrm{d}t 
\bigg[
\nabla \frac{A_{\phi}^*}{\varpi} \times \nabla (\varpi A_{\phi}) + \nabla (\varpi B_{\phi}^*) \times \nabla \frac{B_{\phi}}{\varpi }
\bigg]
\cdot \myhat{e_{\phi}} \delta \psi
+ Re \int \mathrm{d}t \int \mathrm{d}\myvect{a} \cdot 
\left.
\bigg[
A_{\phi}^* \frac{1}{\varpi} \nabla (\varpi A_{\phi}) + \varpi B_{\phi}^* \nabla \frac{B_{\phi}}{\varpi }
\bigg]
\times  \myhat{e_{\phi}} \delta \psi
\right|_{\partial V}.
\end{split}
\end{equation}

The surface term again vanishes, and by substituting the variation of  
$$
 \psi (r,\theta)= -\frac{2(r-r_{mc})^2}{\pi(1-r_{mc})} \\
  \begin{cases}
     \sum\limits_{i=1}^{m} \sum\limits_{j=1}^{n} d_{i,j} \sin \left[ \frac{i\pi (r-r_{mc})}{1-r_{mc}} \right] 
 P_j^1 (-\cos \theta) & \text{if }r_{mc} \leq r \leq 1 \\
     0 & \text{if } r_{bot} \leq r<r_{mc},
  \end{cases}
$$
into Eq.~\eqref{eq:streambypart}, the partial derivative of the objective function with respect to each expansion coefficient $d_{i,j}$ writes 
\begin{equation}\label{eq:partialdij}
\begin{split}
 \frac{\partial \mathcal{J}}{\partial d_{i,j}}=
     Re 
\int \mathrm{d}^3 \myvect{r}
\left \{
\frac{2(r-r_{mc})^2}{\pi(1-r_{mc})}\sin \left[ \frac{i\pi (r-r_{mc})}{1-r_{mc}} \right]
 P_j^1 (-\cos \theta)
\int \mathrm{d}t  
\left[
\nabla \frac{A_{\phi}^*}{\varpi} \times \nabla (\varpi A_{\phi}) + \nabla (\varpi B_{\phi}^*) \times \nabla \frac{B_{\phi}}{\varpi }
\right]
\cdot \myhat{e_{\phi}}
\right \}.
\end{split}
\end{equation}
Again, the calculation of this derivative requires the integration of the adjoint equations for $A_{\phi}^*$ and $B_{\phi}^*$ subject
 to the boundary and terminal conditions already discussed above. Note also that the radial part of the 
 volumetric integration is now restricted to the domain $r_{mc}<r<r_{top}=1$.

\section{Assimilation model started from an ensemble of  initial conditions taken from a unicellular dynamo model}\label{sec:initialcon}
As discussed in Sec.~\ref{sec:initialguess}, 
the assimilation is carried out without the knowledge of the true meridional circulation $\myvect{v}_p$
and the true initial condition for $\myvect{B}$. 
In practice, the assimilation tests different 
initial conditions for $A_\phi$ and $B_\phi$, based on a collection of snapshots 
from a 22-yr periodic reference dynamo model whose variability is controlled by 
a unicellular meridional flow.   
Under favorable circumstances, there is potentially a time for which the predicted magnetic field 
can be almost in phase with the synthetic data, opening the way to a successful recovery
 of the meridional circulation; otherwise too large a phase difference leads to a the misfit remaining suboptimal,
 and an unsuccessful recovery.  
 For each trial, we let the forward model iterate through the transient regime. When the periodic regime is reached, 
 the the misfit between the synthetic observations and the predicted trajectory (i.e. the objective function) is evaluated. 
 Those multiple trials of assimilation for the same set of observations are performed in an embarrassingly parallel framework. 

Let us illustrate this further by considering a twin experiment for case 3. The synthetic observations are obtained using an asymmetric flow associated with a 22-yr magnetic cycle period and the other parameters given in Table~\ref{tab:modelpara} and ~\ref{tab:modeldij}. The observations are noised with $\epsilon=10\%$. The sampling window has a width of 1.5 solar cycles. The field is sampled monthly and the observations are uniformly distributed
  in latitude. We take as initial guess an ensemble of initial conditions from a 22-year reference dynamo model with a unicellular flow, evenly distributed within this period 22 years.
To label these snapshots in the reference model, we plot the time evolution of the toroidal field at the tachocline at the latitude $+35^{\circ}$ over a period. We define the instant when the field is zero with positive time derivative to be year zero. We assimilate the same set of data for each initial conditions and the objective function used is the sum $\mathcal{J}_A+\mathcal{J}_B$. The convergence criteria is $10^{-6}$. The result for an ensemble of 10 such snapshots is shown in Table~\ref{tab:init_case3dr}.

 The iteration converges to a minimum discrepancy of $\sim1.55\%$ when the snapshot of year 14.0 is chosen as the initial conditions for assimilation, and the corresponding normalized misfit is close to 1. This shows that our assimilation is robust with respect to the choice of initial conditions, provided that the choice is physical and the forward model is allowed to iterate through the transient regime. \par 

\begin{table}[!ht]
\caption{Assimilation results for case 3, showing the need to resort to an ensemble of initial magnetic conditions 
 in order to eventually achieve a good recovery of the meridional circulation.  
 The 10 equally-spaced snapshots defining those initial conditions are extracted from a 
  22 yr-cycle of a dynamo model driven by our initial guess of $\myvect{v}_p$. The sampling window for assimilation
 has a width of 1.5 solar cycles, with observations made every month and uniformly distributed in latitude. 
 A divergent iteration is marked with an `x's for the final discrepancy and normalized misfit. The assimilation is successful with minimum discrepancy of $1.55\%$ if the snapshot of year 14.0 is used as the initial condition for the assimilation.}
\resizebox{\columnwidth}{!}{%
\begin{tabular}{c*{11}{c}}
\hline \hline
 snapshot    & 1.07 & 3.22 & 5.37 & 7.52 &9.67   & 11.8 & 14.0                  & 16.1 & 18.3 & 20.4   \\
 epoch (year)     &      &      &      &      &       &      &                       &      &      &        \\
\hline                                        
 $n_{iter}$           &  9    & 8   &  142 & 9    &  75   & 93   & 106                   &  172 &  91  & 87      \\
\hline
 $\Delta p/p$         & x     &    x& 3.28 & 13.4 & 0.236 &0.544 & $1.55 \times 10^{-2}$ & 1.48 & 0.326& $8.65 \times 10^{-2}$ \\
\hline                                        
 $\mathcal{J}_{norm}$ & x     &    x& 9.90 & 10.2  & 2.34  & 2.57 & 1.03                 & 3.22 & 2.24 & 1.18    \\
\hline
\label{tab:init_case3dr}
\end{tabular}}
\end{table}

\section{Sensitivity of assimilation results to the period of unicellular initial guesses}\label{sec:initialperi}
In this appendix we show some examples demonstrating the stability of the performance of the assimilation method with respect to the choice of period of the dynamo model based on unicellular meridional circulations of various amplitude, in the vicinity of period 44 years for case 1, 22 years for cases 2 and 3, as mentioned in Sec.~\ref{subsec:choiceofobj}. The reference sampling is used, i.e., 1.5 cycles, monthly in time and uniform in space at $\Delta \theta =2.83^{\circ} $. We start with a unicellular stream function and record the performance of convergence. The data is noised at $\epsilon=10\%$. We tabulate the tested trials in Table~\ref{tab:perper}. The convergence criterion is fixed to $|\nabla \mathcal{J}|/|\nabla \mathcal{J}_0| < 10^{-6}$, we indicate a successful converging trial with the corresponding iterations required, and a divergent trial with an `x'. The results show that, the convergence is stable with respect to the change of period of the dynamo model with unicellular initial guess within a finite margin (in the vicinity of 44 years for case 1, 22 years for case 2 and 3). This also justifies our choice of initial guesses in the previous analysis. However, this margin of stability shrinks as the meridional flow gets more complex from the unicellular case to the most complicated asymmetric case.

\begin{table}[!ht]
\caption{Assimilation performance with respect to initial guesses of unicellular stream function with different strength, which makes the period of the dynamo vary. In case 1, the 22 years unicellular meridional flow case is the same model as the data file, so no essential assimilation is done, as a result it is not included in the test. A successful trial shows a normalized misfit close to 1, while a trial which is under fitted has a misfit $\gg 1$. The convergence is stable in the vicinity of the periods chosen to be the initial guesses for the analysis above, but the margin of the variation of the period shrinks as the model gets more complicated.}
\begin{tabular}{c*{6}{c}}
\hline \hline
 Case 1, guessed $d_{1,2}$& 0.13 & 0.145  & 0.16 & 0.19 & 0.22  \\
 corr. period (yrs)   & 54.2 &  48.3  & 43.5 & 36.6 & 31.6  \\
\hline                                        
 $n_{iter}$           &   19   & 20     &  19  & 16 &  16    \\
\hline                                        
 $\mathcal{J}_{norm}$& 1.00   & 1.00   & 1.00 &1.00 & 1.00   \\
\hline                                        
&&&&&&\\
\hline \hline
 Case 2, guessed $d_{1,2}$& 0.130& 0.150& 0.163 & 0.175 & 0.180   \\
 corr. period (yrs)   & 27.1 & 23.6  & 21.6  &  20.0  &  19.4   \\
\hline                                        
 $n_{iter}$           &  49  & 46   & 50    &  45   & 32     \\
\hline                                        
 $\mathcal{J}_{norm}$  & 1.11 & 1.11 & 1.11  &  1.11 & 11.3    \\
\hline                                        
&&&&&&\\
\hline \hline
 Case 3, guessed $d_{1,2}$& 0.180    & 0.185    & 0.190  & 0.195      & 0.200   \\
 corr. period (yrs)       & 23.1     & 22.6     & 22.2   & 21.8       & 21.4     \\
\hline                                        
 $n_{iter}$               &  76      & 83       &  106   & 95         & 94       \\
\hline                                        
 $\mathcal{J}_{norm}$    &  3.04    & 1.03     &  1.03   & 1.03      & 6.36       \\
\hline                                        
\label{tab:perper}
\end{tabular}
\end{table}

\section{Parameter space using a separable stream function}\label{sec:expstream}
In this appendix, we discuss the differences for the data assimilation procedure between using a separable stream function 
in the dynamo model or using the general linear combination defined in \eqref{eq:MC}. We also 
illustrate with a few examples how well we can recover the separable  $\{a_i, b_j\}$, $i=1,..,m$, $j=1,..,n$ coefficients. \par

 The key difference between the two mathematical structures of stream function is that the general expansion $\sum\limits_{i,j} d_{i,j} \sin \left[ \frac{i\pi (r-r_{mc})}{1-r_{mc}} \right] P_j^1 (-\cos \theta)$ constitutes a complete set in 2D physical space as $m,n$ approaches infinity, while expanding the radial and polar dependencies separately $\sum\limits_{i=1}^{m} a_i \sin \left[ \frac{i\pi (r-r_{mc})}{1-r_{mc}} \right] \sum\limits_{j=1}^{n} b_jP_j^1 (-\cos \theta)$ is only a subset of the general expansion in the 2D space. Therefore, theoretically, there is a trade off between two situations. The separable stream function model is neater and uses fewer parameters ($m+n$) to fit the observations compared with the general 2D expansion ($mn$) for constant $m, n$. However, the misfit of using the separable model of stream function in assimilation will always be essentially equal or greater than the general expansion, as there are more degrees of freedom to control for the latter case during optimization. Also, having more parameters to adjust in the general structure, it is easier for the assimilation algorithm to reach a region with lower gradient in the parameter space compared with the separable expansion.    \par
In operational sense, the dependencies of stream function (and so does the objective function and its gradient in the parameter space) on the parameters for the separable model are more complicated than those of the general expansion. For the general 2D expansion, the parameters to be estimated are $d_{i,j}$'s appear as a linear combination in the expansion of the stream function in \eqref{eq:MC}. While expanding the radial and polar dependencies separately, the parameters being estimated would be $a_i$'s and $b_j$'s, which are nonlinearly coupled in the stream function, and the expression must first be linearized (in $a_i$'s and $b_j$'s) to evaluate the adjoint during operation.

In the following examples, we still limit the parameter space to $m=2$, $n=4$, so that $m+n=6$ vs $mn=8$ in the general case. Since it is the product $a_ib_j$ characterizes the stream function, two different pairs of $a_i,b_j$ can describe the same meridional flow if the product $a_ib_j$ is the same. In the evaluation of $\Delta p/p$ using the separable model, we replace $d_{i,j}$ with $a_ib_j$ (the forecast) in \eqref{eq:dp}. \par

We compare the performance of one assimilation trial for cases 1, 2 and 3 using the reference sampling. We use the sum of both objective functions $\mathcal{J}_A + \mathcal{J}_B$ and a 1-cell flow as initial guess. The convergence criterion for both the separable and general expansion model are $|\nabla \mathcal{J}|/|\nabla \mathcal{J}_0| < 10^{-6}$ (note that as the parameter spaces for 2 stream function structures are different, the criteria may correspond to different accuracies as the gradients are taken with respect to different sets of variables). The synthetic observations are unnoised to rule out the possible effects of noise on the comparison. 

\begin{figure}[H]
  \includegraphics[width=0.5\columnwidth]{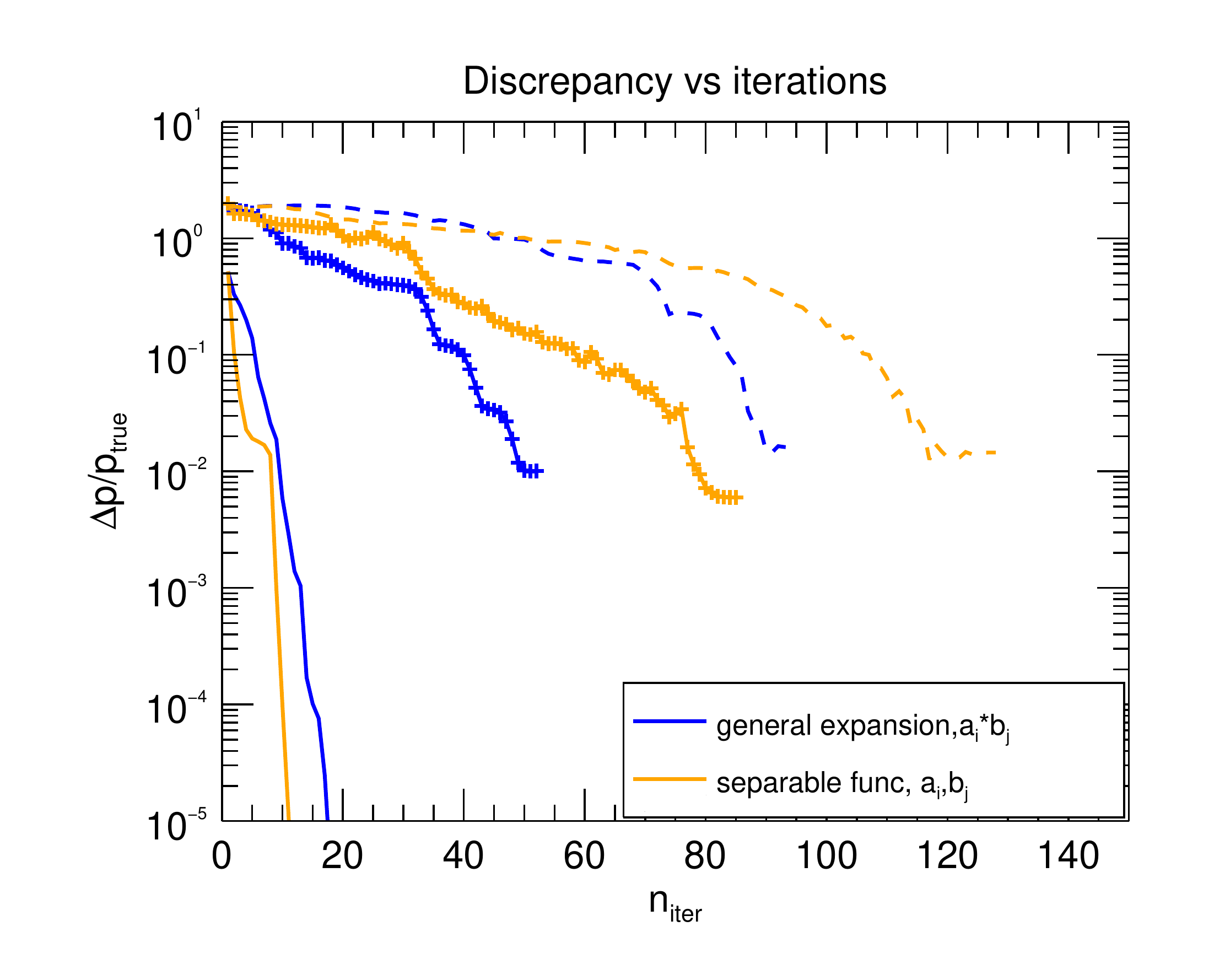}
   \caption{$\Delta p/p$ as the iteration evolves for the 3 cases, sampling monthly for 1.5 solar cycles, uniformly in latitudes. Note that the general expansion model is more efficient than the separable stream function expansion in this case 2 and 3.}
   \label{fig:effgenvssepcase123}
\end{figure}

The performance in terms of the convergence behavior as iteration evolves is plotted in Figure~\ref{fig:effgenvssepcase123}. The efficiency of assimilation using the separable model is slightly higher than that of the general expansion in case 1, but significantly lower than the latter in cases 2 and 3, when the flow becomes more complicated. We see that for cases 2 and 3, for the same accuracy, the separable model needs more iterations to converge than the general expansion, and for the same number of iterations, the former gives a greater discrepancy than the latter.

\acknowledgments
The authors gratefully acknowledge support via the IdEx Sorbonne-Paris-Cit\'{e} project DAMSE, 
the ERC PoC SolarPredict, CNES Solar Orbiter Grant and INSU/PNST support.
We are grateful to Roger Ulrich for giving us access to his surface meridional circulation measurements.
IPGP contribution \firstrev{3673}. 


\bibliography{mc_paper_info}
\bibliographystyle{./apj}

\end{document}